\documentstyle[12pt]{article}
\makeatletter
\@addtoreset{equation}{section}
\renewcommand{\theequation}{\thesection.\arabic{equation}}
\makeatother
\def\T{\displaystyle \mathop{T}}
\def\R{\displaystyle \mathop{R}}
\def\a{\displaystyle \mathop{a}}
\def\V{\displaystyle \mathop{V}}
\def\G1{\displaystyle \mathop{G}}
\def\e1{\displaystyle \mathop{e}}
\def\u1{\displaystyle \mathop{u}}

\def\T{\displaystyle \mathop{T}}
\def\n1{\displaystyle \mathop{\nu}}
\def\ps1{\displaystyle \mathop{\Psi}}

\def\p1{\displaystyle \mathop{p}}
\def\J1{\displaystyle \mathop{J}}
\def\ttt{\displaystyle \mathop{\theta}}
\def\O1{\displaystyle \mathop{O}}

\def\hG{\displaystyle \mathop{\hat{G}}}
\def\gam{\displaystyle \mathop{\gamma}}
\def\Gam{\displaystyle \mathop{\Gamma}}
\def\bGam{\displaystyle \mathop{\bar{\Gamma}}}
\def\pr{\displaystyle \mathop{\partial}}
\def\lpr{\displaystyle \mathop{\buildrel\leftarrow\over{\partial}}}
\def\hgam{\displaystyle \mathop{\hat{\gamma}}}
\def\hpr{\displaystyle \mathop{\hat{\partial}}}
\def\Dlt{\displaystyle \mathop{\Delta}}
\def\hN{\displaystyle \mathop{\hat{N}}}
\def\he{\displaystyle \mathop{\hat{e}}}
\def\hp1{\displaystyle \mathop{\hat{p}}}
\def\hps{\displaystyle \mathop{\hat{\Psi}}}
\def\bhp{\displaystyle \mathop{\bar{\hat{\Psi}}}}
\def\bp{\displaystyle \mathop{\bar{\Psi}}}

\def\tbp{\displaystyle \mathop{\widetilde{\bar{\Psi}}}}

\def\S{\displaystyle \sum}
\def\IIn{\displaystyle \int}
\def\FFr{\displaystyle \frac}
\def\Lm{\displaystyle \lim}
\def\pd{\displaystyle \prod}
\def\hb{\displaystyle \mathop{\hat{b}}}
\def\F{\displaystyle \mathop{F}}
\def\sig1{\displaystyle \mathop{\sigma}}

\def\hh{\displaystyle \mathop{\hat{h}}}
\def\HH{\displaystyle \mathop{\hat{H}}}
\def\th{\displaystyle \mathop{\widetilde{h}}}
\def\TH{\displaystyle \mathop{\widetilde{H}}}
\def\TG{\displaystyle \mathop{\widetilde{G}}}
\def\h1{\displaystyle \mathop{h}}
\def\H1{\displaystyle \mathop{H}}
\def\A1{\displaystyle \mathop{A}}

\def\D1{\displaystyle \mathop{D}}
\def\D1{\displaystyle \mathop{D}}
\def\g1{\displaystyle \mathop{g}}
\newtheorem{guess}{Conjecture}

\begin{document}
\begin{center}
{\Large {\bf OPERATOR MANIFOLD APPROACH TO GEOMETRY AND PARTICLE PHYSICS}}
\vskip 0.5truecm
{\normalsize G.T.Ter-Kazarian\footnote
{E-mail address:gago@bao.sci.am}}\\
{\small Byurakan Astrophysical Observatory, Armenia 378433}\\
{\small June 3, 1996}\\
\end{center}
\begin{abstract}
The question that guides our discussion is 
{\em how did the geometry and particles come into being?} To explore 
this query the present theory reveals primordial deeper structures 
underlying fundamental concepts of contemporary physics. 
We begin with a drastic revision of a role of local internal symmetries in 
physical concept of curved geometry. Under the reflection of fields and 
their dynamics from Minkowski space to Riemannian a standard gauge 
principle of local internal symmetries was generalized. The gravitation gauge 
group is proposed, which was generated by hidden local internal symmetries.
In all circumstances, it seemed to be of the most importance for understanding 
of physical nature of gravity. 
Last two parts of this paper address  
to the question of physical origin of geometry and basic concepts 
of particle physics such as the fundamental fields of quarks  
with the spins and various quantum numbers, internal symmetries and so
forth; also four basic principles of Relativity, Quantum, Gauge and Color 
Confinement, which are, as it was proven, all derivative and  come into being
simultaneously. The substance out of which the geometry and particles are 
made is a set of new physical structures - {\em the goyaks}
~\footnote{\normalsize{\em goyak}
in Armenian means {\em an existence (an existing structure)}. This term has 
been used in [1-5]}, which are involved 
into reciprocal linkage establishing processes.
The most promising aspect of our approach so far is the fact  that
many of the important anticipated properties, basic concepts and 
principles of particle physics are appeared quite naturally in the framework 
of suggested theory.
\end{abstract}
\section {Introduction}
\label {int}
Although a definite pattern for the theoretical description of particle
physics has emerged, which is attractive enough both theoretically and 
phenomenologically, but it could not be regarded as the final word in 
particle physics and many fundamental questions have yet to
be answered. In all previous studies the concepts 
and right symmetries, also basic principles of particle physics have 
been put into theory by hand. The difficulties associated with this step 
are notorious. In view of all this, we are led to consider the problem, 
as it was seen at the outset, how to find out the substance or basic
structures, something deeper, that underly both geometry and 
particles? It will be appropriate to turn to them as the 
primordial deeper structures. The absence of the vital physical theory, 
which will be able to reveal these structures and answer to the right 
questions mentioned above, imperatively stimulates the search for general 
constructive principles.
The guiding line framing our discussion throughout this
paper is the generalization and further expansion of ideas of the 
theory of distortion of space-time continuum [1-5],
wherein, as it is believed, the space-time is not pre-determined background 
on which physical processes take place, but a dynamical entity itself. 
We start with the formulation of general gauge principle, 
with in the scope of which we outlined the rules governing the general 
distortion of geometry. 
The last two parts of the theory are a revised and brief
version of [4,5].
In second part our task is to develop
and understand the conceptual foundations for our viewpoint.
We elaborate a mathematical framework in order to describe the
persistent processes of creation and annihilation of {\em regular goyaks}. 
It is, in fact, a still wider generalization of the familiar methods of 
secondary quantization with appropriate expansion over the geometric objects.
Hence, it gives rise to the formalism of {\em operator manifold}, which
yields the quantization of geometry, which
differs in principle from all earlier studies.
The geometry realization condition should be cleared up.
But it turned out to be a trivial one, which leads
to geometry without particles and interactions.
There is still an other choice of realization of geometry,
which subsequently yields the geometry with the particles and
interactions. Connected to it, we employ the wave functions of 
{\em distorted goyaks} to extend the knowledge here gained regarding the 
quark fields being introduced 
in the color space of internal degrees of freedom. The local distortion
rotations around each fixed axis yield the quarks or anti-quarks.
They obey exact color confinement, and the spectrum of hadrons would emerge
as the spectrum of the color singlet states.
Finally, it has been shown that the gauge principle holds for any
physical system which can be treated as a definite system of distorted goyaks.
The theory of goyaks predicts a class of possible models of internal 
symmetries, which utilize the idea of gauge symmetry and reproduce the 
known phenomenology of electromagnetic, weak and strong interactions. 
To save writing we guess it worthwhile to leave the other
concepts such as the flavors and so forth with associated aspects 
of particle physics for an other treatment. It will not concern us
here and must be further discussed. Surely this is an important
subject for separate research.
\vskip 0.5truecm
\begin{center}{\Large {\bf Part I. General Gauge Principle and Gravitation}}
\end {center}
\section {Formulation of Principle}
\label{Prin}
In spite of unrivaled by its simplicity and beautiful features, Einstein~'s
classical theory of gravitation clashes with some basic principle of field 
theory. This state of affairs has not much changed up to present and proposed 
abundant models of gravitation are not conductive to provide unartificial
and unique recipe for resolving controversial problems of energy-momentum 
conservation laws of gravitational interacting fields, localization of energy 
of gravitation waves and also severe problems involved in quantum gravity. 
It may seem foolhardy to think the rules governing such issues
in scope of the theory accounting for gravitation entirely in terms of
intricated Riemannian geometry. The difficulties associated with this step 
are notorious, however, these difficulties are technical. In the main,
they stem from the fact that Riemannian geometry, in general, has not
admitted a group of isometries. So, the Poincare 
transformations no longer act as isometries. For example, it is
not possible to define energy~-momentum as Noether currents related to exact 
symmetries and so on. On the other hand, the concepts of local internal 
symmetries and gauge fields [6-8] have became a powerful tool and successfully
utilized for the study of electroweak and strong interactions. The intensive 
attempts have been made for constructing a gauge theory of gravitation.
But they are complicated sure enough and not generally acceptable.
Since none of the solutions of the problems of determination of the gauge
group of gravitation and Lagrangian of gravitational field proposed
up to the present seems to be wholly convincing,
one might gain insight into some of unknown features of phenomena of
local internal symmetries and gravity by investigating a new approaches in 
hope of resolving such issues..
Effecting a reconciliation it is our purpose to explore a 
number of fascinating features of {\em generating the gauge 
group of gravitation by hidden local internal symmetry}. In line with this 
we feel it of the most importance to generalize standard 
gauge principle of local internal symmetries. While, a second trend
emerged as a formalism of the reflection of physical fields
and their dynamics from Minkowski space to Riemannian. There is an attempt 
to supply some of the answers in a concise form to the crucial problems of
gravitation and to trace some of the major currents of thoughts under a
novel view-point. Exploiting the whole advantages of the field theory 
in terms of flat space, a particular emphasis will be placed just on 
the formalism of the reflection.
The energy~-momentum conservation laws of 
gravitational interacting fields are well~-defined by exploiting  
whole advantages of auxiliary {\em shadow fields} on flat 
{\em shadow space}.The developed mechanism enables one to infer 
Einstein~'s equation of gravitation, but only with strong difference from
Einstein~'s theory at the vital point of well~-defined energy~-momentum
tensor of gravitational field and conservation laws.
The curvature as well as general distortion of geometry at hidden 
group  $U^{loc}(1)$ are considered. 
This is not a final report on a closed subject, but it is hoped that
suggested theory will serve as useful introduction and that it will 
thereby add the knowledge on the role of local internal symmetries 
in physical concept of curved geometry. \\
In standard picture, suppose a massless gauge field 
${\bf B}_{l}(x_{f})=T^{a}B_{l}^{a}(x_{f})$
with the values in Lie algebra of group G is a local form of expression of 
connection in principle bundle with a structure 
group G. Collection of
matter fields are defined as the sections of vector bundles 
associated with G by reflection $\Phi_{f}:M^{4}\rightarrow E$ that
$p\,\Phi_{f}(x_{f})=x_{f}$, where 
$x_{f} \in M^{4}$ is a space-time point of Minkowski flat space 
specified by index $({}_{f})$. The $\Phi_{f}$ is a column vector denoting 
particular component of field taking values in standard fiber $F_{x_{f}}$
upon $x_{f}$ : $p^{-1}(U^{(f)})=U^{(f)}\otimes F_{x_{f}}$, where 
$U^{(f)}$ is a region of base of principle bundle 
upon which an expansion into direct product $p^{-1}(U^{(f)})=
U^{(f)}\otimes G$ is defined. The various suffixes of 
$\Phi_{f}$ are left implicit.  The fiber is Hilbert vector space 
on which a linear representation $U_{f}(x_{f})=
\exp \left( -T^{a}\theta_{f}^{a}(x_{f}) \right )$ of group G with 
structure constants $C^{abc}$ is given. This space regarded as Lie algebra 
of group G upon which Lie algebra acts according to 
law of adjoint representation: ${\bf B}\, \leftrightarrow \, ad \, 
{\bf B}\: \, \Phi_{f} \, \rightarrow [{\bf B}\, ,\Phi_{f}]$.
To facilitate writing,  
we shall consider, in the main, the most important fields of spin 
$0,\displaystyle\frac{1}{2}, 1$. 
But developed method may be readily extended to the field
of arbitrary spin $s$, since latter will be treated as a system of 2$s$
fermions of half~-integral spin.
In order to generalize standard gauge principle below we proceed 
with preliminary discussion.\\
a. As a starting point we shall assume that under gauge field ${\bf B}_{f}$
the basis vectors $e^{l}_{f}$ transformed into four vector
fields 
\begin{equation}
\label {R21}
\hat{e}^{\mu}(\kappa {\bf B}_{f})=
\hat{D}^{\mu}_{l}(\kappa {\bf B}_{f})e^{l}_{f},
\end{equation}
where $\hat{D}^{\mu}_{l}$ are real~-valued matrix~-functions of 
${\bf B}_{f}$, $\kappa$ is universal coupling constant by which 
gravitational constant will be expressed (see eq.(4.7)). 
The double occurrence of dummy indices, as usual, will be taken to 
denote a summation extended over their all values. Explicit form of 
$\hat{D}^{\mu}_{l}$ will not concern us here, which must be defined 
under concrete physical considerations [1-5] (sec. 5,6). However, 
two of the most common restrictions will be placed upon these functions.
As far as all known interactions are studied on the base of commutative
geometry, then, first of all, the functions $\hat{D}^{\mu}_{l}$ will be 
diagonalized at given
$(\mu, l)$. As a real~-valued function of Hermitian matrix ${\bf B}_{f}$,
each of them is Hermitian too and may be diagonalized by proper unitary 
matrix. Hence
$D^{\mu}_{l}(\kappa {\bf B}_{f})=diag(\lambda^{\mu}_{l1},
\lambda^{\mu}_{l2},\lambda^{\mu}_{l3},\lambda^{\mu}_{l4}),$
provided with eigen~-values $\lambda^{\mu}_{li}$ as the roots of polynomial
characteristic equation
\begin{equation}
\label {R22}
C(\lambda^{\mu}_{l})=det(\lambda^{\mu}_{l}I-\hat{D}^{\mu}_{l})=0.
\end{equation}
Thus
\begin{equation}
\label {R23}
e^{\mu}(\kappa {\bf B}_{f})=
D^{\mu}_{l}(\kappa {\bf B}_{f})e^{l}_{f}
\end{equation}
lead to commutative geometry. Consequently, if $e_{\nu}(\kappa {\bf B}_{f})$
denotes the inverse matrix~-vector $\|D\| \neq 0$, then
$<e^{\mu},e_{\nu}>=D^{\mu}_{l}D_{\nu}^{l}=\delta_{\nu}^{\mu}, \quad
D^{\mu}_{l}D_{\mu}^{k}=\delta_{l}^{k}.$
A bilinear form on vector fields of sections $\tau$ of tangent bundle of
$R^{4}: \quad 
\hat{g}:\tau \otimes \tau \rightarrow C^{\infty} (R^{4}),$
namely the metric $(g_{\mu \nu})$, is a section of conjugate vector bundle 
$S^{2}\tau^{*}$ 
(symmetric part of tensor degree) with corresponding components in basis
$e_{\mu}(\kappa {\bf B}_{f})$
\begin{equation}
\label {R25}
g_{\mu \nu}=<e_{\mu},e_{\nu}>=<e_{\nu},e_{\mu}>=g_{\nu \mu},\quad
e_{\mu}=g_{\mu \nu}e^{\nu}.
\end{equation}
In holonomic basis
$\hat{g}=g_{\mu \nu}dx^{\mu}\otimes dx^{\nu}.$
One now has to impose a second restriction on the functions $D^{\mu}_{l}$ by
placing stringent condition upon the tensor $\omega^{m}_{l}(x_{f})$ implying
\begin{equation}
\label {R26}
\partial^{f}_{l}D^{\mu}_{k}(\kappa {\bf B}_{f})=
\partial^{f}_{k}\left( \omega^{m}_{l}(x_{f})
D^{\mu}_{m}(\kappa {\bf B}_{f})\right),
\end{equation}
where $\partial^{f}_{l}={\displaystyle \frac{\partial}{\partial x_{f}^{l}}}$,
that the $\omega=\omega^{l}_{l}$ should be Lorentz scalar 
function of the trace of curvature form $\Omega$ of connection 
${\bf B}_{l}$ with the values in Lie algebra of group G. That is
\begin{equation}
\label {R27}
\omega = \omega (tr\, \Omega) = \omega (d\,tr \,{\bf B}), 
\end{equation}
provided
\begin{equation}
\label {R28}
\begin{array}{l}
\Omega = \displaystyle{\sum_{l < k}} {\bf F}_{lk}^{f}\,dx^{l}_{f}
\wedge dx^{k}_{f}, \quad {\bf B}={\bf B}_{l}^{f}dx^{l}_{f},\\
{\bf F}_{lk}^{f}=\partial_{l}^{f}{\bf B}_{k}-\partial_{k}^{f}{\bf B}_{l}
-ig[{\bf B}_{l},{\bf B}_{k}].
\end{array}
\end{equation}
There up on a functional $\omega$ has a null variational derivative
$\displaystyle \frac{\delta \omega (tr \,\Omega )}{\delta {\bf B}_{l}} = 0 $
at local variations of connection ${\bf B}_{l} \rightarrow {\bf B}_{l}+
\delta {\bf B}_{l}$, namely, $\omega$ is invariant under Lorentz($\Lambda$) and
G-gauge  groups
$\omega = inv(\Lambda, G).$\\
b. Constructing a diffeomorphism
$x^{\mu}(x^{l}_{f}):M^{4}\rightarrow R^{4},$
the holonomic functions $x^{\mu}(x^{l}_{f})$ satisfy defining
relation
\begin{equation}
\label {R29}
e_{\mu}\psi^{\mu}_{l}=e^{f}_{l} + \chi^{f}_{l}({\bf B}_{f}),
\end{equation}
where
\begin{equation}
\label {R210}
\chi^{f}_{l}({\bf B}_{f})=e_{\mu}\chi^{\mu}_{l}=
-\frac{1}{2}e_{\mu}\int_{0}^{x_{f}}
(\partial^{f}_{k}D^{\mu}_{l}-\partial^{f}_{l}D^{\mu}_{k})dx^{k}_{f}.
\end{equation}
The following 
notational conventions will be used throughout: $\psi^{\mu}_{l}=
\partial^{f}_{l}x^{\mu}$, where the indices $\mu,\nu,\lambda,\tau,\sigma,
\kappa$ stand for variables in $R^{4}$, when $l,k,m,n,i,j$ refer to 
$M^{4}$. A closer examination of eq.(2.9) shows that 
covector $\chi^{f}_{l}({\bf B}_{f})$   realizes coordinates $x^{\mu}$ 
by providing a criteria of integration
\begin{equation}
\label {R211}
\partial^{f}_{k}\psi^{\mu}_{l}=\partial^{f}_{l}\psi^{\mu}_{k}
\end{equation}
and undegeneration $\| \psi \| \neq 0$ [9,10]. Due to eq.(2.6) and 
eq.(2.9), one has Lorentz scalar gauge
invariant functions
\begin{equation}
\label {R212}
\chi=<e^{l}_{f},\chi^{f}_{l}>=\displaystyle {\frac{\omega}{2}}-2,\quad
S=\displaystyle {\frac{1}{4}}\psi_{\mu}^{l}D^{\mu}_{l}=
\displaystyle {\frac{1}{2}(1 +
\displaystyle {\frac{\omega}{4}}}).
\end{equation}
So, out of a set of arbitrary curvilinear coordinates in $R^{4}$ the
{\em real~-curvilinear} coordinates may be distinguished, which satisfy
eq.(2.8) under all possible Lorentz and gauge transformations. There is
a single~-valued conformity between corresponding tensors with various
suffixes on $R^{4}$ and $M^{4}$. While, each co- or contra~-variant index
transformed incorporating respectively with functions $\psi^{\mu}_{l}$
or $\psi_{\mu}^{l}$. Each transformation of real~-curvilinear coordinates
$x'^{\mu'} \rightarrow x^{\mu}$ was generated by some Lorentz and gauge
transformations
\begin{equation}
\label {R213}
\frac{\partial x'^{\mu'}}{\partial x^{\mu}}=\psi^{\mu'}_{l'}(B'_{f})
\psi_{\mu}^{l}(B_{f})\Lambda^{l'}_{l}.
\end{equation}
There would then exist preferred systems and group of transformations of
real~-curvilinear coordinates in $R^{4}$. The wider group of transformations
of arbitrary curvilinear coordinates in $R^{4}$ would then be of no importance 
for the field dynamics.
If an inverse function $\psi_{\mu}^{l}$ meets condition 
\begin{equation}
\label {R214}
\frac{\partial \psi_{\mu}^{l}}{dx^{\nu}}\neq\Gamma^{\lambda}_{\mu\nu}
\psi_{\lambda}^{l},
\end{equation}
where $\Gamma^{\lambda}_{\mu\nu}$ is the usual Christoffel symbol 
agreed with a 
metric $g_{\mu\nu}$, then a curvature of $R^{4}$ is
not vanished [11, 12].\\
In pursuing the original problem further we are led to the principle point
of drastic change of standard gauge scheme
to assume hereafter that {\em single~-valued, smooth, double-sided 
reflection of fields 
$\Phi_{f}\rightarrow \Phi(\Phi_{f}): F_{x_{f}}\rightarrow F_{x}$, 
namely $\Phi_{f}^{l\cdots m}(x_{f})$, takes place under local 
group G}, where $\Phi_{f}\subset F_{x_{f}}$, $\Phi\subset F_{x}$, $F_{x}$
is the fiber upon $x:p^{-1}(U)=U\otimes F_{x}$, $U$ is the region of base 
$R^{4}$.
The tensor suffixes were only put forth in illustration of
a point at issue. Then
\begin{equation}
\label {R215}
\Phi^{\mu\cdots \delta}(x)=\psi^{\mu}_{l}\cdots 
\psi^{\delta}_{m}R({\bf B}_{f})\Phi_{f}^{l\cdots m}(x_{f})\equiv
{(R_{\psi})}^{\mu \cdots \delta}_{l\cdots m}\Phi_{f}^{l\cdots m}(x_{f}),
\end{equation}
where $R({\bf B}_{f})$ is a reflection matrix.
The idea of {\em general gauge principle} may be framed into requirement
of {\em invariance of physical system of fields $\Phi(x)$ under the finite 
local gauge transformations $U_{R}=R'_{\psi}U_{f}R^{+}_{\psi}$ of the 
Lie group of gravitation $G_{R}$ generated by G, where $R'_{\psi}=R_{\psi}
({\bf B}'_{f})$ if gauge field ${\bf B}_{f}(x_{f})$ was transformed under 
G in standard form}.  While the corresponding 
transformations of fields $\Phi(x)$ and their covariant derivatives 
are written 
\begin{equation}
\label {R216}
\begin{array}{l}
\Phi'(x)=U_{R}(x)\Phi(x),
\\
\left( g^{\mu}(x)\nabla_{\mu}\Phi(x)\right)' =U_{R}(x) \left( 
g^{\mu}(x)\nabla_{\mu}\Phi(x)\right).
\end{array}
\end{equation}
The solution of eq.(2.15) may be readily obtained as the 
reflection of covariant derivatives 
\begin{equation}
\label {R217}
g^{\nu}(x)\nabla_{\nu}\Phi^{\mu\cdots \delta}(x)=S(B_{f})
\psi^{\mu}_{l}\cdots \psi^{\delta}_{m}R({\bf B}_{f})\gamma^{k}D_{k}
\Phi_{f}^{l\cdots m}(x_{f}),
\end{equation}
where $D_{l}=\partial_{l}^{f}- ig {\bf B}_{l}(x_{f})$ , 
$S(B_{f})$ is gauge invariant Lorentz scalar, $\nabla_{\mu}$ is 
covariant derivative in $R^{4}$ :
$\nabla_{\mu}=\partial_{\mu}+\Gamma_{\mu}$, provided with connection [11]
$\Gamma_{\mu}(x)=\displaystyle\frac{1}{2}
\Sigma^{\alpha\beta}V^{\nu}_{\alpha}(x)
\partial_{\mu}V_{\beta\nu}(x),$
the $\Sigma^{\alpha\beta}$ are generators of Lorentz group, 
$V^{\mu}_{\alpha}(x)$ are the components of affine tetrad vectors 
$e^{\alpha}$ in used coordinate net $x^{\mu}$: 
$V^{\mu}_{\alpha}(x)=<e^{\mu},e_{\alpha}>$; one has $g^{\mu}(x)\Rightarrow 
e^{\mu}(x)$ and $\gamma^{l}\Rightarrow e^{l}_{f}$ 
for fields $(s=0,1)$; but $g^{\mu}(x)=
V^{\mu}_{\alpha}(x)\gamma^{\alpha}$ for spinor field $(s=\displaystyle
\frac{1}{2})$, where $\gamma^{\alpha}$ are Dirac's matrices.
Since the fields
$\Phi_{f}(x_{f})$ no longer hold, the reflected ones $\Phi(x)$ will be
regarded as the real physical fields. But a conformity eq.(2.14) and eq(2.16)
enables $\Phi_{f}(x_{f})$ to serve as an auxiliary 
{\em shadow fields} on {\em shadow flat space $M^{4}$}. These notions
arise basically  from the most important fact that a 
Lagrangian $L(x)$ of fields $\Phi(x)$
may be obtained under the reflection from a Lagrangian $L_{f}(x_{f})$ of 
corresponding shadow fields and vice versa. Certainly, the $L(x)$ is
also an invariant under the wider group of arbitrary curvilinear 
transformations $x \rightarrow x'$ in $R^{4}$
\begin{equation}
\label {R218}
\left. J_{\psi} L(x)\right|_{inv(G_{R}; \, x\rightarrow x')}=
\left. L_{f}(x_{f})\right|_{inv(\Lambda; \, G)},
\end{equation}
where $J_{\psi}= \| \psi\|\,\surd =\left(1+2\|<e^{f}_{l},\chi^{f}_{k}>\|
+ \| <\chi^{f}_{l},\chi^{f}_{k}>\|\right)^{1/2}$ (see eq.(4.1)).
While, {\em the internal gauge symmetry $G$ remained hidden symmetry,
since it screened by gauge group of gravitation $G_{R}$}.
At last, one notes that the tetrad $e^{\alpha}$ and basis $e^{l}_{f}$
vectors meet a condition
\begin{equation}
\label {R219}
\begin{array}{l}
\rho^{\alpha}_{l}(x,x_{f})=<e^{\alpha},e_{l}^{f}>=V^{\alpha}_{\mu}(x)
D^{\mu}_{l}(\kappa {\bf B}_{f}), \\
<e^{\alpha}_{f},e^{f}_{\beta}>=<e^{\alpha},e_{\beta}>=
\delta^{\alpha}_{\beta}.
\end{array}
\end{equation}
So, Minkowskian metric is written
$\eta^{\alpha\beta}=diag(1,-1,-1,-1)= 
<e^{\alpha}_{f},e_{f}^{\beta}>=<e^{\alpha},e^{\beta}>$.
\section {Reflection Matrix}
\label {ref}
One interesting offshoot of general gauge principle is a formalism of
reflection.
A straightforward calculation for fields $s=0,1$ 
gives the explicit unitary reflection matrix
\begin{equation}
\label {R31}
R(x,x_{f})=R_{f}(x_{f})R_{g}(x)=\exp\left( 
-i\Theta_{f}(x_{f})-
\Theta_{g}(x)\right),
\end{equation}
provided
\begin{equation}
\label {R32}
\Theta_{f}(x_{f})=
g\int_{0}^{x_{f}}{\bf B}_{l}(x_{f})dx_{f}^{l},
\quad \Theta_{g}(x)=
\int_{0}^{x}\left[ R^{+}_{f} \Gamma_{\mu} R_{f}+
\psi^{-1}\partial_{\mu}\psi\right] dx^{\mu},
\quad \psi \equiv \left( \psi^{\mu}_{l}\right),
\end{equation}
where $\Theta_{g}=0$ for scalar field and $\Theta_{g}+\Theta_{g}^{+}=0$
for vector field, because of $\Gamma_{\mu}+\Gamma_{\mu}^{+}=0$
and $\partial_{\mu}(\psi^{-1}\psi)=0$.
The function $S(B_{f})$ has a form eq.(2.11), since
$S(B_{f})=\displaystyle \frac{1}{4}R^{+}(\psi^{l}_{\mu}D_{l}^{\mu})R
=\displaystyle \frac{1}{4}\psi^{l}_{\mu}D_{l}^{\mu}.$
The infinitesimal gauge transformation 
$U_{f}\approx 1-iT^{a}\theta^{a}_{f}(x_{f})$ yields
\begin{equation}
\label {R33}
U_{R}=\exp \left( igC^{abc}T^{a}\int_{0}^{x_{f}}\theta^{b}_{f}B^{c}_{l}
dx^{l}_{f} + \Theta'_{g}-\Theta_{g}\right),
\end{equation}
where the infinitesimal transformation ${B'}^{a}_{l}=B^{a}_{l} + 
C^{abc}\theta^{b}_{f}B^{c}_{l} - \displaystyle\frac{1}{g}\partial^{f}_{l}
\theta^{a}_{f}$ was used.
For example, a Lagrangian of {\em isospinor-scalar} shadow field 
$\varphi_{f}$ is in the form
\begin{equation}
\label {R34}
\begin{array}{l}
L_{f}(x_{f})=J_{\psi}L(x)=J_{\psi}\left[ (e^{\mu}\nabla_{\mu}\varphi)^{+}
(e^{\mu}\nabla_{\mu}\varphi)- m^{2}\varphi^{+}\varphi \right]=\\
=J_{\psi}\left[ S(B_{f})^{2}(D_{l}\varphi_{f})^{+}(D_{l}\varphi_{f})-
m^{2}\varphi_{f}^{+}\varphi_{f}\right].
\end{array}
\end{equation}
A Lagrangian of {\em isospinor-vector} Maxwell~'s shadow field arises in a 
straightforward manner
\begin{equation}
\label {R35}
L_{f}(x_{f})=J_{\psi}L(x)=-\frac{1}{4}J_{\psi}F_{\mu\nu}^{+}F^{\mu\nu}=-
\frac{1}{4}J_{\psi}S(B_{f})^{2}(F^{(f)}_{\mu\nu})^{+}F^{\mu\nu}_{(f)},
\end{equation}
provided
\begin{equation}
\label {R36}
F_{\mu\nu}=\nabla_{\mu}A_{\nu}-\nabla_{\nu}A_{\mu},\quad
F^{(f)}_{\mu\nu}=(\psi^{k}_{\nu}D^{f}_{\mu}-\psi^{k}_{\mu}D^{f}_{\nu})
A^{f}_{k}, \quad D^{f}_{\mu}=D^{l}_{\mu}D_{l},
\end{equation}
with an additional gauge violating term
\begin{equation}
\label {R37}
\begin{array}{l}
L^{f}_{G}(x_{f})=J_{\psi}L_{G}(x)=-\displaystyle \frac{1}{2}\zeta^{-1}_{0}
J_{\psi}(\nabla_{\mu}A^{\mu})^{+}
\nabla_{\nu}A^{\nu}=\\
=-\displaystyle \frac{1}{2}\zeta^{-1}_{0}J_{\psi}S(B_{f})^{2}
(\lambda^{l}_{m}D_{l}A_{f}^{m})^{+}\lambda^{k}_{n}D_{k}A_{f}^{n},
\quad \lambda^{l}_{m}=\displaystyle \frac{1}{2}(\delta^{l}_{m} +
\omega^{l}_{m}),
\end{array}
\end{equation}
where $\zeta^{-1}_{0}$ is gauge fixation parameter. Finally, a Lagrangian
of {\em isospinor-ghost} fields is written
\begin{equation}
\label {R38}
L^{f}_{gh}(x_{f})=J_{\psi}L_{gh}(x)=J_{\psi}
<(e^{\mu}\partial_{\mu}C)^{+},e^{\nu}\partial_{\nu}C>=
J_{\psi}S(B_{f})^{2}(D_{l}C_{f})^{+}D_{l}C_{f}.
\end{equation}
Continuing along this line we come to a discussion of the reflection
of spinor field $\Psi_{f}(x_{f}) (s=\displaystyle\frac{1}{2})$ 
as a solution of eq.(2.15)
\begin{equation}
\label {R39}
\begin{array}{l}
\Psi(x)=R({\bf B}_{f})\Psi_{f}(x_{f}), \quad
\bar{\Psi}(x)=\bar{\Psi}_{f}(x_{f})\widetilde{R}^{+}({\bf B}_{f}),\\
g^{\mu}(x)\nabla_{\mu}\Psi(x)=S(B_{f})R({\bf B}_{f})
\gamma^{l}D_{l}\Psi_{f}(x_{f}),\\
\left( \nabla_{\mu}\bar{\Psi}(x)\right)g^{\mu}(x)=S(B_{f})
\left( D_{l}\bar{\Psi}_{f}(x_{f})\right)\gamma^{l}
\widetilde{R}^{+}({\bf B}_{f}),
\end{array}
\end{equation}
where $\widetilde{R}=\gamma^{0}R\gamma^{0},\quad 
\Sigma^{\alpha\beta}=
\displaystyle\frac{1}{4}[\gamma^{\alpha},\gamma^{\beta}], \quad 
\Gamma_{\mu}(x)=\displaystyle\frac{1}{4}
\Delta_{\mu,\alpha\beta}(x)\gamma^{\alpha}\gamma^{\beta}$, 
$\Delta_{\mu,\alpha\beta}(x)$ are Ricci rotation coefficients. 
The reflection matrix $R$ is 
in the form eq.(3.1), provided we make change
\begin{equation}
\label {R310}
\Theta_{g}(x)=\frac{1}{2}\int_{0}^{x} R^{+}_{f}\left\{ 
g^{\mu}\Gamma_{\mu}R_{f}, g_{\nu}dx^{\nu}\right\},
\end{equation}
the $\{,\}$ is an anticommutator. A calculation now gives
\begin{equation}
\label {R311}
S(B_{f})=\frac{1}{8K}\psi^{l}_{\mu}\left\{
\widetilde{R}^{+}g^{\mu}R,
\gamma_{l}\right\}=inv,
\end{equation}
where
\begin{equation}
\label {R312}
\begin{array}{l}
K =\widetilde{R}^{+}R=\widetilde{R}^{+}_{g}R_{g}=\\
=\exp\left( -\displaystyle \frac{1}{2}\int_{0}^{x}\left( \left\{ R^{+}_{f}
\widetilde{\Gamma}^{+}_{\mu}g^{\mu},g_{\nu}dx^{\nu}\right\}R_{f}+
R^{+}_{f}\left\{ g^{\mu}\Gamma_{\mu}R_{f},
g_{\nu}dx^{\nu}\right\}\right)\right), 
\quad \widetilde{\Gamma}^{+}_{\mu}=
\gamma^{0}\Gamma^{+}_{\mu}\gamma^{0}.
\end{array}
\end{equation}
Taking into account that $[R_{f}, g_{\nu}]=0$ and substituting [13]
\begin{equation}
\label {R313}
\widetilde{\Gamma}^{+}_{\mu}g^{\nu} + g^{\nu}\Gamma_{\mu}=
-\nabla_{\mu}g^{\nu}=0
\end{equation}
into eq.(3.12), we get
\begin{equation}
\label {R314}
K =1.
\end{equation}
Since 
\begin{equation}
\label {R315}
\widetilde{U}_{R}^{+}U_{R}=\widetilde{R}U^{+}_{f}
\widetilde{R'}^{+}R'U_{f}R^{+}=
\widetilde{R}U^{+}_{f}
\widetilde{R}^{+}RU_{f}R^{+},
\end{equation}
where $\widetilde{R'}^{+}R'=\widetilde{R}^{+}R=1$, 
then
\begin{equation}
\label {R316}
\widetilde{U}_{R}^{+}U_{R}=\gamma^{0}U_{R}^{+}\gamma^{0}U_{R}=1.
\end{equation}
A Lagrangian of {\em isospinor-spinor} shadow field may be written
\begin{equation}
\label {R317}
\begin{array}{l}
L_{f}(x_{f})=J_{\psi} L(x)=J_{\psi}\left\{ 
\displaystyle {\frac{i}{2}}\left[ 
\bar{\Psi}(x)g^{\mu}(x)\nabla_{\mu}\Psi(x)-
(\nabla_{\mu}\bar{\Psi}(x))g^{\mu}(x)\Psi(x)\right]-
m\bar{\Psi}(x)\Psi(x)\right\}=\\
=J_{\psi}\left\{S(B_{f})
\displaystyle {\frac{i}{2}}\left[ 
\bar{\Psi}_{f}\gamma^{l}D_{l}\Psi_{f}-
(D_{l}\bar{\Psi}_{f})\gamma^{l}\Psi_{f}\right]-
m\bar{\Psi}_{f}\Psi_{f}\right\}.
\end{array}
\end{equation}
In special case , namely the curvature tensor 
$R^{\lambda}_{\mu\nu\tau}=0$, eq.(2.8) may be satisfied 
globally in $M^{4}$ by putting
\begin{equation}
\label {R318}
\psi^{\mu}_{l}=D^{\mu}_{l}=V^{\mu}_{l}=\frac{\partial x^{\mu}}
{\partial \xi^{l}}, \quad \|D\|\neq 0,\quad \chi^{f}_{l}=0,
\end{equation}
where $\xi^{l}$ are inertial coordinates.
So, $S = J_{\psi}=1$, which means that one simply has constructed 
local G-gauge theory in $M^{4}$ both in curvilinear as well as inertial
coordinates.
\section{Action Principle}
\label{Act}
Field equations may be inferred from an invariant action
\begin{equation}
\label {R41}
S=S_{B_{f}} + S_{\Phi}=\int L_{B_{f}}(x_{f})d^{4}x_{f} +
\int \surd L_{\Phi}(x) d^{4}x.
\end{equation}
A Lagrangian $L_{B_{f}}(x_{f})$ of gauge field ${\bf B}_{l}(x_{f})$
defined on $M^{4}$ is invariant under Lorentz as well as G~-gauge groups. 
But a Lagrangian of the rest of fields $\Phi(x)$ defined on $R^{4}$
is invariant under the gauge group of gravitation $G_{R}$. Consequently, 
the whole action eq.(4.1) is G~-gauge invariant, since $G_{R}$ generated by 
G. Field equations followed at once in terms of Euler-Lagrang variations 
respectively in $M^{4}$ and $R^{4}$  
\begin{equation}
\label {R42}
\begin{array}{l}
\displaystyle \frac{\delta^{f} L_{B_{f}}}{\delta^{f}B^{a}_{l}}=
J_{a}^{l}=-\displaystyle \frac{\delta^{f} L_{\Phi_{f}}}
{\delta^{f} B^{a}_{l}}=-
\displaystyle \frac{\partial g^{\mu\nu}}{\partial B^{a}_{l}}
\displaystyle \frac{\delta (\surd L_{\Phi})}{\delta g^{\mu\nu}}=
\\
=-\displaystyle \frac{1}{2}
\displaystyle \frac{\partial g^{\mu\nu}}{\partial B^{a}_{l}}
\displaystyle \frac{\surd}{\|D\|}
\displaystyle \frac{D_{k\mu}\delta (\surd L_{\Phi})}
{\delta D^{\nu}_{k}}=-\displaystyle \frac{\surd}{2}
\displaystyle \frac{\partial g^{\mu\nu}}{\partial B^{a}_{l}}T_{\mu\nu},\\
\\
\displaystyle \frac{\delta L_{\Phi}}{\delta\Phi}=0,\quad
\displaystyle \frac{\delta L_{\Phi}}{\delta\bar\Phi}=0,
\end{array}
\end{equation}
where $T_{\mu\nu}$ is the energy~-momentum tensor of fields $\Phi (x)$.
Making use of Lagrangian of corresponding shadow fields $\Phi_{f}(x_{f})$,
in generalized sense, one may readily define the energy~-momentum
conservation laws and also exploit whole advantages of field theory in terms
of flat space in order to settle or mitigate the difficulties whenever they
arise including the quantization of gravitation, which will not concern us 
here. Meanwhile, of course, one is free to 
carry out an inverse reflection to $R^{4}$ whenever it will be needed.
To render our discussion here more transparent, below we clarify a relation
between gravitational and coupling constants. So, we consider the theory at 
the limit of Newton~'s non~-relativistic law of gravitation in the case
of static weak gravitational field given by Poisson equation [11]
\begin{equation}
\label {R43}
\nabla^{2}g_{00}=8\pi G\, T_{00}.
\end{equation}
In linear approximation
\begin{equation}
\label {R44}
g_{00}(\kappa {\bf B}_{f})\approx 1 + 2\kappa B^{a}_{k}(x_{f})
\theta_{a}^{k},
\quad \theta_{a}^{k}=\left( \frac{\partial D^{0}_{0} (\kappa {\bf B}_{f})}
{\partial \kappa B^{a}_{k}}\right)_{0} = \mbox{const},
\end{equation}
then
\begin{equation}
\label {R45}
\theta_{a}^{k}\nabla^{2}\kappa B^{a}_{k}=4\pi G\,T_{00}.
\end{equation}
Since, the eq.(4.2) must match onto eq.(4.5) at considered limit, then
the right~-hand sides of both equations should be in the same form
\begin{equation}
\label {R46}
\kappa \theta_{a}^{k}
\displaystyle \frac{\delta^{f} L_{B_{f}}}{\delta^{f}B_{a}^{k}}=
\kappa \theta_{a}^{k}
J^{a}_{k}\approx -\displaystyle \frac{\kappa^{2}}{2}
\displaystyle \frac{\theta_{a}^{k}\partial g^{00}}
{\partial (\kappa B_{a}^{k})}T_{00}=
\kappa^{2}(\theta_{a}^{k}\theta^{a}_{k})T_{00}.
\end{equation}
With this final detail carried for one gets
\begin{equation}
\label {R47}
G=\frac{\kappa^{2}}{4\pi \theta_{c}^{2}}, \quad
\theta_{c}^{2}\equiv (\theta_{a}^{k}\theta^{a}_{k}).
\end{equation}
At this point we emphasize that Weinberg~'s argumentation [14],
namely, a prediction of attraction between particle and antiparticle, 
and repulsion between the same kind of particles, which is valid for 
pure vector theory, no longer holds in suggested theory of gravitation.
Although the ${\bf B}_{l}(x_{f})$ is vector gauge field,
the eq.(4.7) just furnished the prove that only gravitational attraction
is existed.
One final observation is worth recording. A fascinating opportunity
has turned out in the case if one utilizes reflected Lagrangian of 
Einstein~'s gravitational field
\begin{equation}
\label {R48}
L_{B_{f}}(x_{f})=J_{\psi}L_{E}(x)=\frac{J_{\psi}\,R}{16\pi G},
\end{equation}
where $R$ is a scalar curvature. Hence, one readily gets the field equation
\begin{equation}
\label {R49}
\left( R^{\mu}_{\lambda} - \frac{1}{2}\delta ^{\mu}_{\lambda} R\right)
D_{\mu}^{l}\frac{\partial D_{l}^{\lambda}}{\partial B^{a}_{k}}=
-8\pi G U_{a}^{k}(\kappa {\bf B}_{f})=-8\pi G T_{\lambda\nu}
D^{\nu l}\frac{\partial D_{l}^{\lambda}}{\partial B^{a}_{k}},
\end{equation}
which obviously leads to Einsten~'s equation.
Of course in these circumstances it is straightforward
to choose $g^{\mu\nu}$ as the characteristic of gravitational field
without referring to gauge field $B^{a}_{l}(x_{f})$.
However, in this case the energy~-momentum tensor of gravitational 
field is well~-defined. At this vital point suggested theory strongly 
differs from Einstein's classical theory. 
In line with this, one should use the real~-curvilinear coordinates,
in which the gravitational field was well~-defined in the sense that
it cannot be destroyed globally by coordinate transformations.
While, taking into account general rules eq.(2.12), the energy~-momentum 
tensor of gravitational field may be readily obtained 
by expressing the energy~-momentum tensor of vector gauge field 
$B^{a}_{l}(x_{f})$ in terms of metric tensor and its derivatives
\begin{equation}
\label {R410}
\begin{array}{l}
T_{\mu\nu}=g_{\mu\lambda}T^{\lambda}_{\nu}=g_{\mu\lambda}\psi^{\lambda}_{k}
\psi^{i}_{\nu}T^{\,k}_{(f)i}=\\
\\
=g_{\mu\lambda}\psi^{\lambda}_{k}\psi^{i}_{\nu}
\left( {\displaystyle \frac{\partial B^{a}_{l}}{\partial g^{\mu'\nu'}}
\psi^{\sigma}_{i}\partial _{\sigma}g^{\mu'\nu'}
\displaystyle \frac{\partial \left(\partial_{\tau}
g^{\lambda'\tau'}\right)}{\partial\left(\partial^{f}_{k}\,B^{a}_{l}
\right)}
\displaystyle \frac{\partial (J_{\psi} L_{E})}{\partial\left(  
\partial_{\tau} g^{\lambda'\tau'}\right)}
-\delta^{k}_{i}J_{\psi} L_{E} } \right) .
\end{array}
\end{equation}
At last, we should note that eq.(4.8) is not the simplest one among gauge
invariant Lagrangians. Moreover, it must be the same in all cases including
eq.(3.18) too. But in last case it contravenes the standard
gauge theory. There is no need to contemplate such a drastic revision of 
physics. So, for our part we prefer G~-gauge invariant Lagrangian 
in terms of $M^{4}$
\begin{equation}
\label {R411}
L_{B_{f}}(x_{f})=-\frac{1}{4}<{\bf F}^{f}_{lk}(x_{f}),
{\bf F}^{lk}_{f}(x_{f})>_{K}, 
\end{equation}
where ${\bf F}_{lk}^{f}(x_{f})$ is in the form eq.(2.7), 
$<,>_{K}$ is the Killing undegenerate form on the Lie algebra of 
group G for adjoint representation. Certainly, an explicit form of functions
$D^{\mu}_{l}(\kappa {\bf a}_{f})$ should be defined.
\section{Gravitation at $G=U^{loc}(1)$}
The gravitational interaction with hidden Abelian local group
$G=U^{loc}(1)=SO^{loc}(2)$ and one-dimensional trivial algebra
$\hat{g}=R^{1}$ was considered in [1-5], wherein the explicit form of 
transformation function $D^{\mu}_{l}$ is defined by making use of principle
bundle $p:E\rightarrow G(2.3)$. It is worthwhile to consider the major points 
of it anew in concise form and make it complete by calculations,
which will be adjusted to fit the outlined here theory. We start with
a very brief recapitulation of structure of flat manifold
$G(2.3)={}^{*}R^{2}\otimes R^{3}=R^{3}_{+}\oplus R^{3}_{-}$ provided with
the basis vectors $e^{0}_{(\lambda \alpha)}=O_{\lambda}\otimes 
\sigma_{\alpha}$, where $<O_{\lambda},O_{\tau}>={}^{*}\delta_{\lambda\tau}=
1-\delta_{\lambda\tau}$, $<\sigma_{\alpha},\sigma_{\beta}>=\delta_{\alpha
\beta}$ $(\lambda,\tau=\pm : \alpha, \beta=1,2,3)$, $\delta$ is Kronecker 
symbol. A bilinear form on vector fields of sections $\tau$ of tangent bundle 
of $G(2.3)$:  $\hat{g}^{0}:\tau\otimes\tau \rightarrow C^{\infty}(G(2.3))$,
namely, the metric $(g^{0}_{(\lambda\alpha)(\tau\beta)})$ is a section of 
conjugate vector bundle with components $\hat{g^{0}}
(e^{0}_{(\lambda \alpha)}, e^{0}_{(\tau\beta)})$ in basis 
$(e^{0}_{(\lambda \alpha)})$. The $G(2.3)$ decomposes into three-dimensional
ordinary $(R_{f}^{3})$ and time $(T_{f}^{3})$ flat spaces $G(2.3)=
R_{f}^{3}\oplus T_{f}^{3}$ with signatures $sgn(R_{f}^{3})=(+++)$ and
$sgn(T_{f}^{3})=(---)$. Since all directions in $T_{f}^{3}$ are
equivalent, then by notion {\em time} one implies the projection of
time-coordinate on fixed arbitrary universal direction in $T_{f}^{3}$.
By this reduction $T_{f}^{3}\rightarrow T_{f}^{1}$ the transition
$G(2.3)\rightarrow M^{4}=R_{f}^{3}\oplus T_{f}^{1}$ may be performed
whenever it will be needed. 
Under massless gauge field $a_{(\lambda\alpha)}(\eta_{f})$ associating with
$U^{loc}(1)$ the basis $e^{0}_{(\lambda \alpha)}$ transformed at a point
$\eta_{f}\in G(2.3)$ according to eq.(2.3)
\begin{equation}
\label{R51}
e_{(\lambda \alpha)}=D^{(\tau\beta)}_{(\lambda \alpha)}e^{0}_{(\tau\beta)}.
\end{equation}
While, the matrix $D$ is in the form $D=C\otimes R$, where the distortion
transformations $O_{(\lambda\alpha)}=C^{\tau}_{(\lambda\alpha)}
O_{\tau}$ and
$\sigma_{(\lambda\alpha)}=R^{\beta}_{(\lambda\alpha)}
O_{\beta}$ are defined. Thereby $C^{\tau}_{(\lambda\alpha)}=
\delta^{\tau}_{\lambda} + \kappa a_{(\lambda\alpha)}{}^{*}
\delta^{\tau}_{\lambda}$,
but $R$ is a matrix of the group $SO(3)$ of all ordinary rotations of 
the planes, each of which involves two arbitrary basis vectors 
of $R^{3}_{\lambda}$, around  the orthogonal axes. The angles of permissible 
rotations will be determined throughwith a special constraint imposed
upon distortion transformations, namely, a sum of distortions
of corresponding basis vectors $O_{\lambda}$ and 
$\sigma_{\beta}$ has to be zero at given $\lambda$:
\begin{equation}
\label{R52}
<O_{(\lambda\alpha)},O_{\tau}>_{\tau \neq \lambda}+\frac{1}{2}
\varepsilon_{\alpha\beta\gamma}\frac{<\sigma_{(\lambda\beta)},\sigma_{\gamma}>}
{<\sigma_{(\lambda\beta)},\sigma_{\beta}>}=0,
\end{equation}
where $\varepsilon_{\alpha\beta\gamma}$ is an antisymmetric unit tensor.
There up on $\tan\theta_{(\lambda\alpha)}=-\kappa a_{(\lambda\alpha)}$,
where $\theta_{(\lambda\alpha)}$ is the particular rotation around the axis
$\sigma_{\alpha}$ of $R^{3}_{\lambda}$. Inasmuch as the $R$
should be independent of sequence of rotation axes, then  it implies
the mean value $R=\displaystyle \frac{1}{6} \sum_{i \neq j \neq k}
R^{(ijk)}$, where $R^{(ijk)}$ the matrix of rotations carried out 
in sequence $(ijk)$ $(i,j,k=1,2,3)$. As it was seen at the outset {\em the
field $a_{(\lambda\alpha)}$ was generated by the distortion of basis
pseudo-vector $O_{\lambda}$}, when the distortion of $\sigma_{\alpha}$
has followed from eq(5.2).\\
Certainly, the whole theory outlined in sections (1-4) will then hold
provided we simply replace each single index $\mu$ of variables by the pair
$(\lambda\alpha)$ and so on. Following to standard rules, next we construct
the diffeomorphism $\eta^{(\lambda\alpha)}(\eta^{(\tau\beta)}_{f}):
G(2.3) \rightarrow G(23)$ and introduce the action eq.(4.1) for the fields.
In the sequel, a transition from six-dimensional curved manifold $G(23)$
to four-dimensional Riemannian geometry $R^{4}$ is straightforward by
making use of reduction of three time-components $e_{0\alpha}=
\displaystyle \frac{1}{\sqrt{2}}(e_{(+\alpha)}+e_{(-\alpha)})$ of basis 
six-vectors $e_{(\lambda \alpha)}$ to single $e_{0}$ in fixed
universal direction. Actually, since Lagrangian of fields on $R^{4}$
is a function of scalars, namely, $A_{(\lambda \alpha)}B^{(\lambda \alpha)}=
A_{0 \alpha}B^{0 \alpha}+A_{\alpha}B^{\alpha}$, so taking into account that
$A_{0 \alpha}B^{0 \alpha}=A_{0 \alpha}<e^{0\alpha},e^{0\beta}>B_{0 \beta}
=A_{0}<e^{0},e^{0}>B_{0}=A_{0}B^{0}$, one readily may perform a
required transition.
The gravitation field equation is written
\begin{equation}
\label{R53}
\partial_{f}^{(\tau \beta)}\partial^{f}_{(\tau \beta)}
a^{(\lambda\alpha)} -(1-\zeta^{-1}_{0})
\partial_{f}^{(\lambda \alpha)}\partial^{f}_{(\tau \beta)}a^{(\tau \beta)}
=-\frac{1}{2}\sqrt{-g}\frac{\partial 
g^{(\tau \beta)(\mu \gamma)}}{\partial a_{(\lambda\alpha)}}
T_{(\tau \beta)(\mu \gamma)}.
\end{equation}
To render our discussion more transparent, below we consider in detail 
a solution of spherical-symmetric static gravitational field
$a_{(+\alpha)}=a_{(-\alpha)}=\displaystyle \frac{1}{\sqrt{2}}a_{0\alpha}
(r_{f})$. So, $\theta_{(+\alpha)}=\theta_{(-\alpha)} =-\arctan 
(\displaystyle \frac{\kappa}{\sqrt{2}}a_{0\alpha})$ and
$\sigma_{(+\alpha)}=\sigma_{(-\alpha)}$. It is convenient to make use of
spherical coordinates $\sigma_{(+1)}=\sigma_{r}$, $\sigma_{(+2)}=
\sigma_{\theta}$, $\sigma_{(+3)}=\sigma_{\varphi}$. The transition
$G(2.3)\rightarrow M^{4}$ is performed by choosing in $T_{f}^{3}$ the
universal direction along radius-vector: $x_{f}^{0r}=t_{f}$, 
$x_{f}^{0\theta}=x_{f}^{0\varphi}=0$. Then, from eq.(5.1) one gets
\begin{equation}
\label{R54}
e_{0}=D^{0}_{0}e^{0}_{0},\quad
e_{r}=D^{r}_{r}e^{0}_{r},\quad
e_{\theta}=e^{0}_{\theta},\quad
e_{\varphi}=e^{0}_{\varphi},
\end{equation}
provided $D=C\otimes I$ with components 
$D^{0}_{0}=1+\displaystyle \frac{\kappa}{\sqrt{2}}a_{0}$,
$D^{r}_{r}=1-\displaystyle \frac{\kappa}{\sqrt{2}}a_{0}$,
$a_{0}\equiv a_{0 r}$, where
$e_{0}^{0}=\xi_{0}\otimes \sigma_{r},\quad
e^{0}_{r}=\xi\otimes \sigma_{r},\quad
e^{0}_{\theta}=\xi\otimes \sigma_{\theta},\quad
e^{0}_{\varphi}=\xi\otimes \sigma_{\varphi}$, 
$\xi_{0}=\displaystyle \frac{1}{\sqrt{2}}(O_{+}+O_{-})$
and $\xi=\displaystyle \frac{1}{\sqrt{2}}(O_{+}-O_{-})$.
The coordinates $x^{\mu}(t,r,\theta,\varphi)$ implying 
$x^{\mu}(x^{l}_{f}):M^{4}\rightarrow R^{4}$ exist in the whole region 
$p^{-1}(U)\in R^{4}$
\begin{equation}
\label{R55}
\frac{\partial x^{\mu}}{\partial x^{l}_{f}}=\psi^{\mu}_{l}=\frac{1}{2}
(D^{\mu}_{l}+\omega^{m}_{l}D^{\mu}_{m}),
\end{equation}
where according to eq.(5.4) one has $x^{0r}=t$, $x^{0\theta}=x^{0\varphi}
=0$. A straightforward calculation gives non-vanishing components
\begin{equation}
\label{R56}
\begin{array}{l}
\psi^{0}_{0}=\displaystyle \frac{1}{2}D^{0}_{0},\quad
\psi^{0}_{1}=\displaystyle \frac{1}{2}t_{f}\partial^{f}_{r}D^{0}_{0},\quad
\psi^{1}_{1}=D^{r}_{r},\quad
\psi^{2}_{2}=\psi^{3}_{3}=0,\\
\omega^{1}_{1}=\omega^{2}_{2}=\omega^{3}_{3}=1, \quad
\omega^{0}_{1}=t_{f}\partial^{f}_{r}D^{0}_{0}.
\end{array}
\end{equation}
Although $\partial_{r}\psi^{0}_{0}=\Gamma^{0}_{01}\psi^{0}_{0}$ and
$\partial_{r}\psi^{1}_{1}=\Gamma^{1}_{11}\psi^{1}_{1}$, but
\begin{equation}
\label{R57}
\partial_{t}\psi^{0}_{1}=\psi^{0}_{0}\partial_{r}^{f}D^{0}_{0}\neq
\Gamma^{0}_{10}\psi^{0}_{0}=
\psi^{0}_{0}\partial_{r}\ln D^{0}_{0},
\end{equation}
where according to eq.(5.4), the non-vanishing components of Christoffel 
symbol are written $\Gamma^{0}_{01}=\displaystyle \frac{1}{2}
g^{00}\partial_{r} g_{00}$,
$\Gamma^{1}_{00}=-\displaystyle \frac{1}{2}
g^{11}\partial_{r} g_{00}$,
$\Gamma^{1}_{11}=\displaystyle \frac{1}{2}
g^{11}\partial_{r} g_{11}$. So, the condition eq.(2.13) holds, namely,
the curvature of $R^{4}$ is not zero. The curved space $R^{4}$ has the 
group of motions $SO(3)$ with two- dimensional space-like orbits $S^{2}$
where the standard coordinates are $\theta$ and $\varphi$. The stationary 
subgroup of $SO(3)$ acts isotropically upon the tangent space at the point of 
sphere $S^{2}$ of radius $r$. So, the bundle $p:R^{4} \rightarrow R^{2}$
has the fiber $S^{2}=p^{-1}(x)$, $x\in R^{4}$ with a trivial connection on 
it, where $R^{2}$ is the factor-space $R^{4}/SO(3)$. In outside of the 
distribution of matter with the total mass $M$, the eq.(5.3) in 
Feynman gauge reduced to $\nabla^{2}_{f}a_{0}=0$, which has the solution
$\displaystyle \frac{\kappa}{\sqrt{2}}a_{0}=-
\displaystyle \frac{GM}{r_{f}}=-
\displaystyle \frac{r_{g}}{2r_{f}}$ (see eq.(4.7)).
So, the line element is written
\begin{equation}
\label{R58}
d\,s^{2}=(1-\frac{r_{g}}{2r_{f}})^{2}dt^{2}-
(1+\frac{r_{g}}{2r_{f}})^{2}dr^{2}-r^{2}(\sin^{2}\theta d\,\varphi^{2}+
d\,\theta^{2}),
\end{equation}
provided by eq.(5.5) and eq.(5.6).Finally, for example, the explicit form of unitary matrix $U_{R}$ of
gravitation group $G_{R}$ in the case of scalar field reads
\begin{equation}
\label{R510}
U_{R}=R'U_{f}R^{+}=e^{\displaystyle {-i[t_{f}\partial^{f}_{r}
\theta_{f}(r_{f})+ \theta_{f}(r_{f})]}},
\end{equation}
where
\begin{equation}
\label{R511}
R=R_{f}=e^{-\displaystyle \frac{igr_{g}}{2r_{f}}t_{f}},\quad
g(a'_{0}-a_{0})=\partial^{f}_{r}\theta_{f}(r_{f}).
\end{equation}
A Lagrangian of charged scalar shadow field eq.(3.4) is in the form
\begin{equation}
\label{R512}
L_{f}(x_{f})=\frac{1}{2}\left[ \left( \frac{7}{8}\right )^{2} 
(D_{l}\varphi_{f})^{*}
D_{l}\varphi_{f} - m^{2}\varphi_{f}^{*}\varphi_{f} \right],
\end{equation}
where, according to eq.(2.11) and eq.(5.6), one has $J_{\Psi}=
\displaystyle \frac{1}{2}(1+\|\omega^{m}_{l}\|)=
\displaystyle \frac{1}{2}$, $S=\displaystyle \frac{7}{8}$.
\section{Distortion of Flat Manifold $G(2.2.3)$ at $G=U^{loc}(1)$}
Following to [1-5], the foregoing theory can be readily generalized for the 
distortion of 12-dimensional flat manifold $G(2.2.3)={}^{*}R^{2}
\otimes {}^{*}R^{2} \otimes R^{3}=\G1_{\eta}(2.3)\oplus
\G1_{u}(2.3)=\displaystyle\sum^{2}_{\lambda,\mu=1} 
\oplus R^{3}_{\lambda \mu}=
{\R_{x}}^{3}\oplus 
{\T_{x}}^{3}\oplus
{\R_{u}}^{3}\oplus 
{\T_{u}}^{3}$ with corresponding 
basis vectors $e_{(\lambda,\mu,\alpha)}=O_{\lambda,\mu}\otimes
\sigma_{\alpha} \subset G(2.2.3)$, 
${\e1_{i}}^{0}_{(\lambda\alpha)}={\O1_{i}}_{\lambda}\otimes 
\sigma_{\alpha}
\subset \G1_{i}(2.3)$, where 
${\O1_{i}}_{+}=
\displaystyle \frac{1}{\sqrt{2}}(O_{1,1} +\varepsilon_{i} O_{2,1})$,
${\O1_{i}}_{-}=
\displaystyle \frac{1}{\sqrt{2}}(O_{1,2} +\varepsilon_{i} O_{2,2})$,
$\varepsilon_{\eta}=1$, $\varepsilon_{u}=-1$ and 
$<O_{\lambda,\mu},O_{\tau,\nu}>={}^{*}\delta_{\lambda \tau}
{}^{*}\delta_{\mu \nu}$ $(\lambda,\mu, \tau, \nu=1,2)$,
$<\sigma_{\alpha},\sigma_{\beta}>=\delta_{\alpha \beta}$. There up on
$<{\O1_{i}}_{\lambda},{\O1_{i}}_{\tau}>=
\varepsilon_{i}\delta_{ij}{}^{*}\delta_{\lambda \tau}$.
The massless gauge field of distortion $a_{(\lambda,\mu,\alpha)}(\zeta_{f})$ 
with the values in Lie algebra of $U^{loc}(1)$ is a local form of
expression of connection in principle bundle $p:E\rightarrow G(2.2.3)$
with a structure group $U^{loc}(1)$. Collection of matter fields
$\Phi_{f}(\zeta_{f})$ are the sections of vector bundles associated with 
$U^{loc}(1)$ by reflection $\Phi_{f}:G(2.2.3)\rightarrow E$ 
that $p \Phi_{f}(\zeta_{f})=\zeta_{f}$, where the coordinates $
\zeta^{(\lambda,\mu,\alpha)}$ exist in the whole region $p^{-1}(U^{(f)})\in
G(2.2.3)$. Outlined theory will then hold, while each pair
of indices $(\lambda \alpha)$ will be replaced by the 
$(\lambda,\mu, \alpha)$. So
the basis $e_{(\tau,\nu,\beta)}$ transformed
\begin{equation}
\label{R61}
e_{(\lambda\mu\alpha)}=D_{(\lambda\mu\alpha)}^{(\tau,\nu,\beta)}
e_{(\tau,\nu,\beta)},
\end{equation}
provided $D=C\otimes R$, where $O_{(\lambda\mu\alpha)}=
C_{(\lambda\mu\alpha)}^{\tau,\nu} O_{\tau,\nu}$,
$\sigma_{(\lambda\mu\alpha)}=
R_{(\lambda\mu\alpha)}^{\beta} \sigma_{\beta}$. The matrices $C$
generate the group of distortion transformations of bi-pseudo-vectors
$O_{\tau,\nu}$: $C_{(\lambda\mu\alpha)}^{\tau,\nu} =
\delta^{\tau}_{\lambda}\delta^{\nu}_{\mu} +\kappa a_{(\lambda,\mu,\alpha)}
{}^{*}\delta^{\tau}_{\lambda}{}^{*}\delta^{\nu}_{\mu}$, but the 
matrices $R$ are the elements of the group $SO(3)_{\lambda \mu}$ of
ordinary rotations of the planes of corresponding basis vectors of
$R^{3}_{\lambda\mu}$. A special constraint eq.(5.2) holds for basis
vectors of each $R^{3}_{\lambda\mu}$. Thus, the gauge field
$a_{(\lambda,\mu,\alpha)}$ was generated by the distortion 
of bi-pseudo-vectors $O_{\tau,\nu}$.  While the rotation transformations 
$R$ follow due to eq.(5.2) and $R$ implies, as usual, the mean value
with respect to sequence of rotation axes. The angles of permissible rotations 
are $\tan \theta_{(\lambda,\mu,\alpha)}=-\kappa a_{(\lambda,\mu,\alpha)}
(\zeta_{f})$. The action eq.(4.1) now reads
\begin{equation}
\label{R62}
S=S_{a_{f}} + S_{\Phi}=\int L_{a_{f}}(\zeta_{f})\,d\zeta^{(1,1,1)}\wedge
\cdots \wedge d\zeta^{(2,2,3)}+
\int \sqrt{g}L_{\Phi}(\zeta)\,d\zeta^{(111)}\wedge
\cdots \wedge d\zeta^{(223)},
\end{equation}
where $g$ is the determinant of metric tensor on $G(223)$, while the
$\zeta^{(\lambda\mu\alpha)}(\zeta^{(\tau,\nu,\beta)}):G(2.2.3)
\rightarrow G(223)$ was constructed according to eq.(2.8)-eq.(2.10),
provided 
\begin{equation}
\label{R63}
\frac{\partial \zeta^{(\lambda\mu\alpha)}}{\partial \zeta^{(\tau,\nu,\beta)}}
=\psi^{(\lambda\mu\alpha)}_{(\tau,\nu,\beta)}=\frac{1}{2} \left(
D^{(\lambda\mu\alpha)}_{(\tau,\nu,\beta)}+
\omega^{(\rho,\omega,\gamma)}_{(\tau,\nu,\beta)}
D^{(\lambda\mu\alpha)}_{(\rho,\omega,\gamma)}
\right).
\end{equation}
Right through the variational calculations one infers the field equations
eq.(4.2), while
\begin{equation}
\label{R64}
\begin{array}{l}
\partial^{(\tau,\nu,\beta)}
\partial_{(\tau,\nu,\beta)}
a^{(\lambda,\mu,\alpha)} - (1-\zeta^{-1}_{0})\partial
^{(\lambda,\mu,\alpha)}\partial_{(\tau,\nu,\beta)}a^{(\tau,\nu,\beta)}=
J^{(\lambda,\mu,\alpha)}=\\
=-\displaystyle \frac{1}{2}\sqrt{g}\displaystyle \frac{g
^{(\tau\nu\beta)(\rho\omega\gamma)}}{\partial a_{(\lambda,\mu,\alpha)}}
T_{(\tau\nu\beta)(\rho\omega\gamma)},
\end{array}
\end{equation}
provided
\begin{equation}
\label{R65}
T_{(\tau\nu\beta)(\rho\omega\gamma)}=\frac{2\delta(\sqrt{g}L_{\Phi})}
{\sqrt{g}\delta g^{(\tau\nu\beta)(\rho\omega\gamma)}}.
\end{equation}
The curvature of manifold $\G1_{i}(2.3) \rightarrow 
\G1_{i}(23)$ (sec. 5), which leads to 
four-dimensional Riemannian
geometry $R^{4}$, is a familiar distortion 
\begin{equation}
\label{R66}
a_{(1,1,\alpha)}=a_{(2,1,\alpha)}\equiv \frac{1}{\sqrt{2}}{\a_{\eta}}
_{(+\alpha)},\quad
a_{(1,2,\alpha)}=a_{(2,2,\alpha)}\equiv \frac{1}{\sqrt{2}}{\a_{\eta}}
_{(-\alpha)},
\end{equation}
when $\sigma_{(11\alpha)}=\sigma_{(21\alpha)}\equiv 
\sigma_{(+\alpha)}$, $\sigma_{(12\alpha)}=\sigma_{(22\alpha)}\equiv 
\sigma_{(-\alpha)}$. Hence
${\e1_{i}}
_{(\lambda\alpha)}={\O1_{i}}_{(\lambda\alpha)}\otimes 
\sigma_{(\lambda\alpha)}$, where 
${\O1_{i}}_{(\lambda\alpha)}={\O1_{i}}_{\lambda} +
\displaystyle \frac{\kappa}{\sqrt{2}}\varepsilon_{i}
{\a_{\eta}}_{(\lambda\alpha)}{}^{*}\delta^{\tau}_{\lambda}
{\O1_{i}}_{\tau}$. 
In the aftermath $G(223) = \G1_{\eta}(23)
\oplus \G1_{u}(23)$. 
The other important case of inner-distortion
\begin{equation}
\label{R67}
a_{(1,1,\alpha)}=-a_{(2,1,\alpha)}\equiv \frac{1}{\sqrt{2}}{\a_{u}}
_{(+\alpha)},\quad
a_{(1,2,\alpha)}=-a_{(2,2,\alpha)}\equiv \frac{1}{\sqrt{2}}{\a_{u}}
_{(-\alpha)}
\end{equation}
leads to $\sigma_{(11\alpha)}=-\sigma_{(21\alpha)}\equiv 
\sigma_{(+\alpha)}$, $\sigma_{(12\alpha)}=-\sigma_{(22\alpha)}\equiv 
\sigma_{(-\alpha)}$. Hence,
${\O1_{i}}_{(\lambda\alpha)}=
{\O1_{i}}_{\lambda} +
\displaystyle \frac{\kappa}{\sqrt{2}}\varepsilon_{i}
{\a_{u}}_{(\lambda\alpha)}{}^{*}\delta^{\tau}_{\lambda}
{\O1_{i}}_{\tau}$, 
where the ${\e1_{i}}
_{(\lambda\alpha)}={\O1_{i}}_{(\lambda\alpha)}\otimes 
\sigma_{(\lambda\alpha)}$ is the basis in inner-distorted manifold
$\G1_{i}(23)$.
\vskip 0.5truecm
\begin {center}{\Large {\bf Part II. Regular Goyaks}}
\end {center}
\section {The Goyaks and Link-Establishing Processes}
\label {goy}
Next we develop the foundations for our viewpoint and
proceed to general definitions and conjectures of the theory of
goyaks directly. 
Henceforth with in last two parts it is convenient to describe the
theory in terms of manifold $G(2.2.3)$. But, surely, one may
readily perform a transition to $M^{4}$(sec.5,6) whenever it will be 
needed .
We choose a simple setting and consider new formations designed to 
endow certain physical properties and satisfying the general rules
stated below.
At this, we may consider it 
briefly hoping to mitigate a shortage of insufficient rigorous treatment
by the further exposition of the theory and make them complete and discussed
in broad sense in due course.
\begin{guess}
The $6$-dimensional basis vectors ${\e1_{i}}^{0}_{(\lambda \alpha)}$
(or co-vectors ${\e1_{i}}_{0}^{(\lambda \alpha)}$)
we explore from a specific novel point of view, as being
the main characteristics of the real existing structures called "goyaks".
Below we distinguish two type of goyaks: $\eta$-type ($i=\eta$) and $u$-type
$(i=u)$, respectively.
\end{guess}
\begin{guess}
The goyaks establish reciprocal "linkage" between themselves. The
links are described by means of "link-function" vectors 
${\ps1_{\eta}} _{(\lambda\alpha)}(\eta,p_{\eta})$ and
${\ps1_{u}}_{(\lambda\alpha)}(u,p_{u})$ (or co-vectors
${\ps1_{\eta}}^{(\lambda\alpha)}(\eta,p_{\eta})$ and
${\ps1_{u}}^{(\lambda\alpha)}(u,p_{u})$):
\begin{equation}
\label {R71}
{\ps1_{\eta} }_{(\pm\alpha)}(\eta,p_{\eta})=\eta_{(\pm\alpha)}
{\ps1_{\eta} }_{\pm}(\eta,p_{\eta}),\quad 
{\ps1_{u}}_{(\pm\alpha)}(u,p_{u})=u_{(\pm\alpha)}
{\ps1_{u}}_{\pm}(u,p_{u}),
\end{equation}
where the $6$-vectors of 'link-coordinates" -$\eta,u$, and
"link-momenta"  - $p_{\eta},p_{u}$ respectively are
$\eta= {\e1_{\eta}}^{(\lambda\alpha)}_{0}\eta_{(\lambda\alpha)}, \qquad
p_{\eta}= {\e1_{\eta}}^{(\lambda\alpha)}_{0}
{\p1_{\eta}}_{(\lambda\alpha)}, 
\quad
u= {\e1_{u}}^{(\lambda\alpha)}_{0}u_{(\lambda\alpha)}, \qquad
p_{u}= {\e1_{u}}^{(\lambda\alpha)}_{0}
{\p1_{u}}_{(\lambda\alpha)}$. 
In words, the probability of finding the goyak in the state with fixed
link-coordinate and momentum is determined by the square of its state 
wave function ${\ps1_{\eta} }_{\pm}(\eta,p_{\eta}),$ or 
${\ps1_{u}}_{\pm}(u,p_{u})$. This provides a simple intuitive meaning
of state link-functions.
\end{guess}
\begin{guess}
Characterizing by the basis vectors  ${\e1_{i}}^{0}_{(\lambda \alpha)}$
and link-functions eq.(7.1), the goyaks
are called "regular" ones.
The "distorted" goyaks are described in terms of distorted basis vectors
${\e1_{i}}_{(\lambda\alpha)}(\theta)={\O1_{i}}_{(\lambda\alpha)}(\theta)
\otimes \sigma_{(\lambda\alpha)}(\theta)$ (see Part III)
and distorted link-functions ${\ps1_{i}}_{(\lambda\alpha)}(\theta)$, 
where $\theta(\eta,u)=\{ {\ttt_{i}}_{(\lambda\alpha)} \}$ are the angles
of distortion transformations.
\end{guess}
\begin{guess}
The $\eta$-type goyak may accept the linkage only from $u$-type goyak,
which is described by the link-function ${\ps1_{\eta} }_{(\lambda\alpha)}
(\eta,p_{\eta})$ ($\eta$-channel). 
The $u$-type goyak, in its turn, may accept the linkage only
from $\eta$-type goyak, to which the link-function
${\ps1_{u}}_{(\lambda\alpha)}(u,p_{u})$ (u-channel) is corresponded. 
It is assumed that the link-establishing processes between the
goyaks of the same type are absent.
\end{guess}
\begin{guess}
The simplest system made of two goyaks of different types becomes
stable only due to the reciprocal link-establishing processes in the case
of stable linkage. The latter is valid if only the subsidiary condition 
of link-stability holds
\begin{equation}
\label {R72}
\left|\p1_{\eta}\right|={({\p1_{\eta}}^{(\lambda\alpha)},
{\p1_{\eta}}_{(\lambda\alpha)})}^{1/2}=
\left|\p1_{u}\right|={({\p1_{u}}^{(\lambda\alpha)},
{\p1_{u}}_{(\lambda\alpha)})}^{1/2}.
\end{equation}
Otherwise they are unstable.
\end{guess}
\begin{guess}
There is not any restriction on the number of goyaks of both types
getting into the link-establishing processes simultaneously. In order to 
be a stable system the link-stability condition must be held for each 
linkage separately.
\end{guess}
\begin{guess}
The persistent processes of creation and annihilation of goyaks occur
in the different states $s, s',s'',...$ It is assumed that the "creation"
of the goyak in given state $(s)$ is stimulated by its transition to this
state from the other states $(s',s'',...)$, while the "annihilation"
means a vice versa.
Along many states $(s,s',s'',...)$ there is a lowest one ($s_{0}$),
in which all goyaks are regular. In other higher states
the goyaks are distorted. Thus, meanwhile the transition from the
state ($s_{0}$) to any other state $(s)$ and vice versa the goyak
undergoes to distortion transformation.
\end{guess}
\begin{guess}
Hereinafter we will be interested only by a special stable system of
regular goyaks, which is formed in lowest state ($s_{0}$) and made of 
only one goyak of $\eta$-type and infinite number of $u$-type goyaks.
They are characterized by the identical basis structure vectors.
The  $\eta$-type goyak is called "fundamental", but the u-type goyaks -
"ordinary" ones.
\end{guess}
\begin{guess}
The dimension of the basis elements may be
directly regarded as the dimension of associated goyak. In general,
it can be an arbitrary number. But in outlined theory a reduction
of the goyak's dimension just to six ($\lambda=\pm;\quad\alpha=1,2,3$) 
is backed up by the arguments are briefly stated
as follows: It is taken for granted
that the infinite sequence of random transitions along many configurations
of goyaks ($\lambda=1,...;\quad\alpha=1,...$) in an arbitrary given state
are the random vector variables with the differential distribution function
$\varphi(\nu)={\nu}^{n}e^{-\pi{\nu}^{2}}$,
provided by the frequency of the transitions $\nu$.\\
At $n\gg1$ expected mean frequency of transition from the state $n$
is given by
\begin{equation}
\label {R73}
m(n)=\FFr{\IIn_{0}^{\infty}{\nu}^{n}\exp^{(-\pi\nu^{2)}}d\nu}
{\IIn_{0}^{\infty}\exp^{(-\pi\nu^{2)}}d\nu}=\frac{\Gamma
\left( \FFr{n+1}{2} \right)}{2\pi^{(n+1)/2}}.
\end{equation}
The most probable number of dimension of goyak is achieved at the minimum
of the function $m(n)$. The inverse value of $m(n)$ is isomorphic to the value
function of $(n+1)$-dimensional hypersphere of unit radius. This function
is unimodal for the positive values of $(n+1)$ and it is of indefinite sign
for negative values of $(n+1)$.
The maximum extension volume of the formation is achieved at $n=\pm6$
[15].
Therefore the maximum probability corresponds to six-dimensional
goyaks. That is why only six-dimensional goyaks ($\lambda=\pm;
\quad \alpha=1,2,3$) would be considered below.
\end{guess}
\section {Operator Manifold $\hat{G}(2.2.3)$}
\label {Oper}
The processes of creation and annihilation of regular goyaks in the lowest 
state ($s_{0}$)may be described by the formalism similar to 
secondary quantization,
but in the same time it will be the appropriate expansion over the 
geometric structures such as goyaks. We first deal
with a substitution of the basis elements of goyak's structure by the 
corresponding operators of creation and annihilation of the regular
goyaks acting in the configuration space of occupation numbers.
Instead of pseudo-vectors we introduce the following operators supplied 
by additional index ($r$) referring to the quantum numbers of corresponding 
state
\begin{equation}
\label {R81}
\begin{array}{ll}
\hat{O}^{r}_{1}=O^{r}_{1}\alpha_{1},\quad
\hat{O}^{r}_{2}=O^{r}_{2}{\alpha}_{2},\quad
\hat{O}_{r}^{\lambda}={}^{*}\delta^{\lambda\mu}\hat{O}^{r}_{\mu}=
{(\hat{O}^{r}_{\lambda})}^{+},\\
\{ \hat{O}^{r}_{\lambda},\hat{O}^{r'}_{\tau} \}=
\delta_{rr'}{}^{*}\delta_{\lambda\tau}I_{2}, 
\quad
<{O}^{r}_{\lambda},{O}^{r'}_{\tau}>= \delta_{rr'}{}^{*}\delta_{\lambda\tau},
\qquad I_{2}=\left( \begin{array}{cc}
1 \quad 0 \\
0 \quad 1
\end{array}
\right).
\end{array}
\end{equation}
The matrices ${\alpha}_{\lambda}$ satisfy the conditions
\begin{equation}
\label {R82}
\begin{array}{lr}
{\alpha}^{\lambda}={}^{*}\delta^{\lambda\mu}
{\alpha}_{\mu}={({\alpha}_{\lambda})}^{+},\qquad
\{ {\alpha}_{\lambda},{\alpha}_{\tau} \}={}^{*}\delta_
{\lambda\tau}I_{2}.
\end{array}
\end{equation}
For example, they can be in the form
${\alpha}_{1}=\left( \begin{array}{cc}
0 \quad 1 \\
0 \quad 0
\end{array}
\right), \quad
{\alpha}_{2}=\left( \begin{array}{cc}
0 \quad 0 \\
1 \quad 0
\end{array}
\right).$
This forms the starting point for quantization.
Creation operator $\hat{O}^{r}_{1}$ generates one-occupied state
$\mid 1>_{(0)}\equiv\mid 0,\ldots,1,\ldots>_{(0)}$ and the basis vector
$O^{r}_{1}$ with the quantum number $r$,
right through acting on non-occupied vacuum state
$\mid 0>_{(0)}\equiv  \mid 0,0,\ldots>_{(0)}$:
\begin{equation}
\label{R83}
\hat{O}^{r}_{1}\mid 0>_{(0)}={O}^{r}_{1}\mid 1>_{(0)}.
\end{equation}
Accordingly, the action of annihilation operator $\hat{O}^{r}_{2}$
on one-occupied state yields the vacuum state and the basis vector
$O^{r}_{2}$
\begin{equation}
\label{R84}
\hat{O}^{r}_{2}\mid 1>_{(0)}={O}^{r}_{2}\mid 0>_{(0)}.
\end{equation}
So define 
$\hat{O}^{r}_{1}\mid 1>_{(0)}=0,\quad \hat{O}^{r}_{2}\mid 0>_{(0)}=$0.
The matrix realization of the states $\mid 0>_{0}$ and $\mid 1>_{0}$,
for instance, may be as follows:
$\mid 0>_{0}\equiv\chi_{1}=\left( \begin{array}{c}
0 \\
1
\end{array}
\right), \quad
\mid 1>_{0}\equiv\chi_{2}=\left( \begin{array}{c}
1  \\
0
\end{array}
\right).$
The operator of occupation number is
$\hat{N}^{(0)}_{r}=\hat{O}^{r}_{1}\hat{O}^{r}_{2}$,
with the expectation values implying Pauli's exclusion principle
${}_{(0)}<0\mid\hat{N}^{(0)}_{r}\mid 0>_{(0)}=0, \qquad 
{}_{(0)}<1\mid\hat{N}^{(0)}_{r}\mid 1>_{(0)}=1$.
The vacuum state reads
$\chi_{0}\equiv\mid 0>_{(0)}=\displaystyle\prod_{r=1}^{N}(\chi_{1})_{r}$.
With this final detail cared for one-occupied state takes the form
$\chi_{r'}\equiv\mid 1>_{(0)}=(\chi_{2})_{r'}\displaystyle\prod_{r\neq r'}
(\chi_{1})_{r}$.
Continuing along this line, instead of ordinary vectors we introduce
the operators
$\hat{\sigma}^{r}_{\alpha}\equiv\delta_{\alpha\beta\gamma}
\sigma^{r}_{\beta}\widetilde{\sigma}_{\gamma}$,
where $\widetilde{\sigma}_{\gamma}$ are Pauli's matrices, and
\begin{equation}
\label{R85}
<\sigma_{\alpha}^{r},\sigma_{\beta}^{r'}>=\delta_{rr'}\delta_{\alpha\beta},
\quad
\hat{\sigma}^{\alpha}_{r}=\delta^{\alpha\beta}\hat{\sigma}^{r}_{\beta}=
{(\hat{\sigma}_{\alpha}^{r})}^{+}=\hat{\sigma}_{\alpha}^{r},
\quad
\{\hat{\sigma}_{\alpha}^{r},\hat{\sigma}_{\beta}^{r'}\}=2
\delta_{rr'}\delta_{\alpha\beta}I_{2}.
\end{equation}
For the vacuum state $\mid 0>_{(\sigma)}\equiv{\varphi}_{1(\alpha)}$
and one-occupied state $\mid 1_{(\alpha)}>_{(\sigma)}
\equiv{\varphi}_{2(\alpha)}$
we make use of matrix realization
${\varphi}_{1(\alpha)}\equiv\chi_{1}, \quad
{\varphi}_{2(1)}=\left( \begin{array}{c}
1 \\
0
\end{array}
\right), \quad
{\varphi}_{2(2)}=\left( \begin{array}{c}
-i \\
\hspace{0.25cm} 0
\end{array}
\right), \quad
{\varphi}_{2(3)}=\left( \begin{array}{c}
\hspace{0.25cm} 0 \\
-1
\end{array}
\right).$
Then
\begin{equation}
\label{R86}
{\hat{\sigma}}_{\alpha}^{r}\varphi_{1(\alpha)}=\sigma_
{\alpha}^{r}\varphi_{2(\alpha)}=(\sigma_{\alpha}^{r}\widetilde{\sigma}_
{\alpha})\varphi_{1(\alpha)},
\quad
{\hat{\sigma}}_{\alpha}^{r}\varphi_{2(\alpha)}=\sigma_
{\alpha}^{r}\varphi_{1(\alpha)}=(\sigma_{\alpha}^{r}\widetilde{\sigma}_
{\alpha})\varphi_{2(\alpha)}.
\end{equation}
Hence, the single eigen-value
$(\sigma_{\alpha}^{r}\widetilde{\sigma}_{\alpha})$
has associated with it quite different $\varphi_{\lambda(\alpha)}$.
The eigen-value is degenerated with 
degeneracy degree equal 2. Due to it, along many quantum numbers $r$
there is also the quantum number of the spin $\vec{\sigma}$ with the values
$\sigma_{3}=\FFr{1}{2}s\quad (s=\pm1)$.
This rule for spin quantum number is not without an important reason. 
The argument for this conclusion
is compulsory suggested by the properties of operators 
$\hat{\sigma}^{r}_{\alpha}$.
As we will see later on, this consequently {gives rise to spin of
particle}.\\
One-occupied state reads
$\varphi_{r'(\alpha)}={(\varphi_{2(\alpha)})}_{r'}\displaystyle
\prod_{r\neq r'}{(\chi_{1})}_{r}$.
Next we introduce the operators
\begin{equation}
\label{R87}
{\hat{\gamma}}^{r}_{(\lambda,\mu,\alpha)}\equiv{\hat{O}}^{r_{1}}_{\lambda}
\otimes{\hat{O}}^{r_{2}}_{\mu}\otimes{\hat{\sigma}}^{r_{3}}_{\alpha},
\quad
{\hat{\gamma}}_{r}^{(\lambda,\mu,\alpha)}\equiv{\hat{O}}_{r_{1}}^{\lambda}
\otimes{\hat{O}}_{r_{2}}^{\mu}\otimes{\hat{\sigma}}_{r_{3}}^{\alpha}=
{}^{*}\delta^{\lambda\tau}{}^{*}\delta^{\mu\nu}\delta^{\alpha\beta} 
{\hat{\gamma}}^{r}_{(\tau,\nu,\beta)},
\end{equation}
and also the state vector
\begin{equation}
\label{R88}
\chi_{\lambda,\mu,\tau(\alpha)}\equiv\mid\lambda,\mu,\tau(\alpha)>=
\chi_{\lambda}\otimes\chi_{\mu}\otimes\varphi_{\tau(\alpha)}, 
\end{equation}
where $\lambda,\mu,\tau,\nu=
1,2;\quad \alpha,\beta=1,2,3$ and $r\equiv (r_{1},r_{2},r_{3})$.
Hence
${\hat{\gamma}}^{r}_{(\lambda,\mu,\alpha)}\chi_{\tau,\nu,\delta(\beta)}=
({\hat{O}}^{r_{1}}_{\lambda}\chi_{\tau})\otimes
({\hat{O}}^{r_{2}}_{\mu}\chi_{\nu})\otimes({\hat{\sigma}}^{r_{3}}_{\alpha}
\varphi_{\delta(\beta)})$.
Omitting the two-valuedness of state vector we apply
$\mid\lambda,\tau,\delta(\beta)>\equiv\mid\lambda,\tau>$,
and the same time remember that always the summation 
must be extended over the double degeneracy of the spin states $(s=\pm 1)$.\\
With this final detail cared for one infers the explicit forms of corresponding
matrix elements:
\begin{equation}
\label{R89}
\begin{array}{l}
<2,2\mid{\hat{\gamma}}^{r}_{(1,1,\alpha)}\mid 1,1>=e^{r}_{(1,1,\alpha)},\quad
<1,1\mid{\hat{\gamma}}_{r}^{(1,1,\alpha)}\mid 2,2>=e_{r}^{(1,1,\alpha)},\\
<2,1\mid{\hat{\gamma}}^{r}_{(1,2,\alpha)}\mid 1,2>=e^{r}_{(1,2,\alpha)},\quad
<1,2\mid{\hat{\gamma}}_{r}^{(1,2,\alpha)}\mid 2,1>=e_{r}^{(1,2,\alpha)},\\
<1,2\mid{\hat{\gamma}}^{r}_{(2,1,\alpha)}\mid 2,1>=e^{r}_{(2,1,\alpha)},\quad
<2,1\mid{\hat{\gamma}}_{r}^{(2,1,\alpha)}\mid 1,2>=e_{r}^{(2,1,\alpha)},\\
<1,1\mid{\hat{\gamma}}^{r}_{(2,2,\alpha)}\mid 2,2>=e^{r}_{(2,2,\alpha)},\quad
<2,2\mid{\hat{\gamma}}_{r}^{(2,2,\alpha)}\mid 1,1>=e_{r}^{(2,2,\alpha)}.
\end{array}
\end{equation}
The operators of occupation numbers are
\begin{equation}
\label{R810}
\begin{array}{l}
{\hN_{1}}^{rr'}_{\alpha\beta}=
{\hat{\gamma}}^{r}_{(1,1,\alpha)}{\hat{\gamma}}^{r'}_{(2,2,\beta)}=
{\hN_{1}}_{rr'}^{\alpha\beta}=
{\hat{\gamma}}_{r}^{(2,2,\alpha)}{\hat{\gamma}}_{r'}^{(1,1,\beta)}, \\
{\hN_{2}}^{rr'}_{\alpha\beta}=
{\hat{\gamma}}^{r}_{(2,1,\alpha)}{\hat{\gamma}}^{r'}_{(1,2,\beta)}=
{\hN_{2}}_{rr'}^{\alpha\beta}=
{\hat{\gamma}}_{r}^{(1,2,\alpha)}{\hat{\gamma}}_{r'}^{(2,1,\beta)},
\end{array}
\end{equation}
with the expectation values implying Pauli's exclusion principle
\begin{equation}
\label{R811}
\begin{array}{ll}
<2,2\mid{\hN_{1}}_{rr'}^{\alpha\beta}\mid 2,2>=\delta_{rr'}
\delta_{\alpha\beta},
\quad
<1,2\mid{\hN_{2}}_{rr'}^{\alpha\beta}\mid 1,2>=\delta_{rr'}\delta_
{\alpha\beta},\\
<1,1\mid{\hN_{1}}_{rr'}^{\alpha\beta}\mid 1,1>=0,\qquad
<2,1\mid{\hN_{2}}_{rr'}^{\alpha\beta}\mid 2,1>=0.
\end{array}
\end{equation}
The set of operators $\{{\hat{\gamma}}^{r}_{(\lambda,\mu,\alpha)}\}$
eq.(8.7) is the basis in operator manifold
$\hat{G}(2.2.3)={}^{*}\hat{R}^{22}\otimes\hat{R}^{3}$.
Here ${}^{*}\hat{R}^{22}$ is the $2\times2$-dimensional linear 
bi-pseudo operator-space, with the set of the linear unit operator 
bi-pseudo vectors 
$\{ {\hat{O}}^{r_{1}r_{2}}_{\lambda,\mu} \equiv
{\hat{O}}^{r_{1}}_{\lambda}\otimes{\hat{O}}^{r_{2}}_{\mu}\}$, and
$\hat{R}^{3}$ is the three-dimensional real linear operator-space with
the basis consisted of the ordinary unit operator-vectors  
$\{ {\hat{\sigma}}^{r}_{\alpha}\}$. 
By means of mathematical apparatus, which is in close analogy to
differential geometry [9],
a general description of operator manifold $\hat{G}(2.2.3)$ may be
presented. 
Bilinear form on operator-vectors
$\hat{\Phi}(\zeta)={\hat{\gamma}}_{r}^{(\lambda,\mu,\alpha)}
\Phi^{r}_{(\lambda,\mu,\alpha)}(\zeta)\in\hat{G}(2.2.3)$
reads in component form 
$\hat{g}={\hat{g}}_{rr'}^{(\lambda,\mu,\alpha)(\tau,\nu,\beta)}d
\Phi^{r}_{(\lambda,\mu,\alpha)}(\zeta)\otimes d
\Phi^{r'}_{(\tau,\nu,\beta)}(\zeta)$.
Analogical form on operator-co-vectors
$\bar{\hat{\Phi}}(\zeta)={\hat{\gamma}}^{r}_{(\lambda,\mu,\alpha)}
\Phi_{r}^{(\lambda,\mu,\alpha)}(\zeta)\in\hat{G}(2.2.3)$,
can be written, where
$\Phi_{r}^{(\lambda,\mu,\alpha)}(\zeta)=
{\bar{\Phi}}^{r}_{(\lambda,\mu,\alpha)}(\zeta)$.
One easily gets
\begin{equation}
\label{R812}
\begin{array}{l}
<1,1\mid\hat{\Phi}(\zeta)\bar{\hat{\Phi}}(\zeta)\mid 1,1>=
\Phi^{r}_{(1,1,\alpha)}(\zeta)\Phi_{r}^{(1,1,\alpha)}(\zeta),\\
<1,2\mid\hat{\Phi}(\zeta)\bar{\hat{\Phi}}(\zeta)\mid 1,2>=
\Phi^{r}_{(1,2,\alpha)}(\zeta)\Phi_{r}^{(1,2,\alpha)}(\zeta),\\
<2,1\mid\hat{\Phi}(\zeta)\bar{\hat{\Phi}}(\zeta)\mid 2,1>=
\Phi^{r}_{(2,1,\alpha)}(\zeta)\Phi_{r}^{(2,1,\alpha)}(\zeta),\\
<2,2\mid\hat{\Phi}(\zeta)\bar{\hat{\Phi}}(\zeta)\mid 2,2>=
\Phi^{r}_{(2,2,\alpha)}(\zeta)\Phi_{r}^{(2,2,\alpha)}(\zeta).
\end{array}
\end{equation}
Introducing the new state vectors
\begin{equation}
\label{R813}
\begin{array}{l}
\chi^{0}(\nu_{1},\nu_{2},\nu_{3},\nu_{4})=
\mid 1,1>^{\nu_{1}}\cdot\mid 1,2>^{\nu_{2}}\cdot
\mid 2,1>^{\nu_{3}}\cdot\mid 2,2>^{\nu_{4}},\\
\\
\nu_{i}= \left\{ \begin{array}{ll}
                   1   & \mbox{if $\nu=\nu_{i}$}\quad  \mbox{for some $i$,} \\
                   0   & \mbox{otherwise},
                   \end{array}
\right. \\
\\
\mid\chi_{-}(\lambda)>=\left\{ \begin{array}{ll}
                   \chi^{0}(1,0,0,0)   & \lambda=1, \\
                   \chi^{0}(0,0,1,0)   & \lambda=2,
                   \end{array}
\right. \quad
\mid\chi_{+}(\lambda)>=\left\{ \begin{array}{ll}
                   \chi^{0}(0,0,0,1)   & \lambda=1, \\
                   \chi^{0}(0,1,0,0)   & \lambda=2,
                   \end{array}
\right.
\end{array}
\end{equation}
provided
$$
\begin{array}{ll}
<\lambda,\mu,\mid\tau,\nu>=\delta_{\lambda\tau}\delta_{\mu\nu}, \quad
<\chi_{\pm}\mid A\mid \chi_{\mp}>=
\S_{\lambda}<\chi_{\pm}(\lambda)\mid A\mid \chi_{\mp}(\lambda)>,\\
<\chi_{\pm}(\lambda)\mid\chi_{\pm}(\mu)>=\delta_{\lambda\mu}, \quad
<\chi_{\pm}\mid A\mid \chi_{\pm}>=
\S_{\lambda}<\chi_{\pm}(\lambda)\mid A\mid \chi_{\pm}(\lambda)>,\\
<\chi_{\pm}(\lambda)\mid\chi_{\mp}(\mu)>=0,
\end{array}
$$
we get the matrix elements as follows:
\begin{equation}
\label{R814}
\begin{array}{l}
<\chi_{-}\mid\hat{\Phi}(\zeta)\bar{\hat{\Phi}}(\zeta)\mid\chi_{-}>
\equiv\Phi^{2}_{-}(\zeta)=\quad
\Phi^{r}_{(1,1,\alpha)}(\zeta)\Phi_{r}^{(1,1,\alpha)}(\zeta)+
\Phi^{r}_{(2,1,\alpha)}(\zeta)\Phi_{r}^{(2,1,\alpha)}(\zeta),\\
<\chi_{+}\mid\hat{\Phi}(\zeta)\bar{\hat{\Phi}}(\zeta)\mid\chi_{+}>
\equiv\Phi^{2}_{+}(\zeta)=
\Phi^{r}_{(2,2,\alpha)}(\zeta)\Phi_{r}^{(2,2,\alpha)}(\zeta)+
\Phi^{r}_{(1,2,\alpha)}(\zeta)\Phi_{r}^{(1,2,\alpha)}(\zeta).
\end{array}
\end{equation}
The operator manifold decomposes
$\hat{G}(2.2.3)=\hG_{\eta}(2.3)\oplus\hG_{u}(2.3)$,
where $\hG_{i}(2.3)\quad(i=\eta,u)$ is the six-dimensional operator
manifold with the basis $\{ {\hgam_{i}}^{r}_{(\lambda\alpha)} \}\quad
(\lambda=\pm;\quad\alpha=1,2,3;\quad i=\eta,u).$
The latter reads in component form
${\hgam_{i}}^{r}_{(+\alpha)}=\FFr{1}{\sqrt{2}}
({\hgam_{i}}^{r}_{(1,1\alpha)}+\varepsilon_{i}
{\hgam_{i}}^{r}_{(2,1\alpha)}),
\quad
{\hgam_{i}}^{r}_{(-\alpha)}=\FFr{1}{\sqrt{2}}
({\hgam_{i}}^{r}_{(1,2\alpha)}+\varepsilon_{i}{\hgam_{i}}^{r}
_{(2,2\alpha)}).$
The expansions of operator-vectors $\hps_{i}\in\hG_{i}(2.3)$ and 
operator-co-vectors  $\bar{\hps_{i}}\in\hG_{i}(2.3)$ are written 
$\hps_{i}={\hgam_{i}}_{r}^{(\lambda\alpha)}{\ps1_{i}}^{r}_{(\lambda\alpha)},
\qquad\bar{\hps_{i}}=
{\hgam_{i}}^{r}_{(\lambda\alpha)}{\ps1_{i}}_{r}^{(\lambda\alpha)},$
where the components ${\ps1_{\eta}}^{r}_{(\lambda\alpha)}(\eta)$ and
${\ps1_{u}}^{r}_{(\lambda\alpha)}(u)$ are the link-functions of goyaks
of $\eta$-type and u-type, respectively$:
{\ps1_{i}}_{r}^{(\lambda\alpha)}=
{\bar{\ps1_{i}}}^{r}_{(\lambda\alpha)}$.
The operator of occupation number of $i$-type goyak takes the form
${\hN_{i}}_{rr'}^{\alpha\beta}=\varepsilon_{i}{\hgam_{i}}_{r}^{(-\alpha)}
{\hgam_{i}}_{r'}^{(+\beta)}$,
with corresponding expectation values
\begin{equation}
\label {R815}
<\chi_{-}\mid{\hN_{i}}_{rr'}^{\alpha\beta}\mid\chi_{-}>=0, 
\quad
<\chi_{+}\mid{\hN_{i}}_{rr'}^{\alpha\beta}\mid\chi_{+}>
=\varepsilon_{i}<{\e1_{i}}_{r}^{(-\alpha)},{\e1_{i}}_{r'}^{(+\beta)}>=
\delta_{rr'}\delta_{\alpha\beta}.
\end{equation}
Taking into account two-valuedness of degenerate spin states
it follows that the goyaks are the fermions with
the half-integral spins. That is, the link-functions ${\ps1_{\eta}}_
{\lambda}(\eta)$ and ${\ps1_{u}}_{\lambda}(u)$ can be regarded as the
Fermi fields of $\eta$- and $u$-type goyaks.
Explicitly the matrix elements read
\begin{equation}
\label {R816}
\begin{array}{l}
\Phi^{2}_{-}(\zeta)=<\chi_{-}\mid\hps_{\eta}(\eta)\bhp_{\eta}(\eta)
+ \hps_{u}(u)\bhp_{u}(u)\mid\chi_{-}>=
\varepsilon_{i}{\hps_{i}}_
{(+\alpha)}{\hps_{i}}^{(+\alpha)}=\\
={\hps_{\eta}}_{(+\alpha)}(\eta)
{\bhp_{\eta}}_{(+\alpha)}(\eta)
-{\hps_{u}}_{(+\alpha)}(u)
{\bhp_{u}}_{(+\alpha)}(u),\\
\Phi^{2}_{+}(\zeta)=<\chi_{+}\mid\hps_{\eta}(\eta)\bhp_{\eta}(\eta)
+ \hps_{u}(u)\bhp_{u}(u)\mid\chi_{+}>=
\varepsilon_{i}{\hps_{i}}_
{(-\alpha)}{\hps_{i}}^{(-\alpha)}=\\
={\hps_{\eta}}_{(-\alpha)}(\eta)
{\bhp_{\eta}}_{(-\alpha)}(\eta)
-{\hps_{u}}_{(-\alpha)}(u)
{\bhp_{u}}_{(-\alpha)}(u).  
\end{array}
\end{equation}
\section {Quantum Field Aspect of Operator Manifold $\hat{G}(2.2.3)$} 
\label {quant}
The nature of operator manifold $\hat{G}(2.2.3)$ provides its elements 
with both field and geometric aspects. The field aspect will be the subject
for discussion in this section. Our notation will be that of the textbook
by [16].
As far as goyak field may be regarded as a fermion field, its description
either regular or distorted was provided by the theory, which is in 
close analogy to Dirac's conventional wave-mechanical theory of fermions
with spin $\vec{\FFr{1}{2}}$ treated in terms of manifold 
$G(2.2.3)$ [1-5]. The final formulation of quantum theory 
is equivalent to configuration space wave mechanics with antisymmetric state
functions.\\
We remind that it is considered only the special system of regular goyaks, 
which is made of fundamental goyak of $\eta$-type and infinite number of 
$u$-type ordinary goyaks. They are characterized by the identical
structure vectors.
To become stable the goyaks in this system have
established the stable linkage. The link-stability requirement
holds for each ordinary goyak
\begin{equation}
\label{R91}
p^{2}=p^{2}_{\eta}-p^{2}_{u}={\p1_{\eta}}^{(\lambda\alpha)}
{\p1_{\eta}}_{(\lambda\alpha)}-{\p1_{u}}^{(\lambda\alpha)}
{\p1_{u}}_{(\lambda\alpha)}=0.
\end{equation}
The regular goyak field may be considered as
bi-spinor field
$\Psi(\zeta)$ defined on manifold $G(2.2.3)=\G1_{\eta}(2.3)
\oplus\G1_{u}(2.3)$:  $\Psi(\zeta)=\ps1_{\eta}(\eta)\ps1_{u}(u)$ [1-5],
where the $\ps1_{i}$ is a bi-spinor associated 
with the manifold $\G1_{i}(2.3).$
Lagrangian of free field is written down
\begin{equation}
\label{R92}
L=\frac{1}{2} \{ \widetilde{\bar{\Psi}}(\zeta)\gamma^{(\lambda,\mu,\alpha)}\
\partial_{(\lambda,\mu,\alpha)}\widetilde{\Psi}(\zeta)-
\partial_{(\lambda,\mu,\alpha)}\tbp(\zeta)\gamma^{(\lambda,\mu,\alpha)}
\widetilde{\Psi}(\zeta) \},
\end{equation}
provided
\begin{equation}
\label{R93}
\begin{array}{l}
\widetilde{\Psi}(\zeta)=e\otimes\Psi(\zeta)=
\left( \begin{array}{cc}
1\quad 1 \\
1\quad 1 
\end{array}   \right)  
\otimes\Psi(\zeta), 
\quad
\widetilde{\bar{\Psi}}(\zeta)=\bar{\Psi}(\zeta)\otimes e,\quad
\bar{\Psi}(\zeta)=\Psi^{+}(\zeta)\gamma^{0}, \\
\gamma^{(\lambda,\mu,\alpha)}={\widetilde{O}}^{\lambda}\otimes
{\widetilde{O}}^{\mu}\otimes{\widetilde{\sigma}}^{\alpha}, \quad
\partial_{(\lambda,\mu,\alpha)}=\partial/\partial\zeta^
{(\lambda,\mu,\alpha)}, 
\quad
{\widetilde{O}}^{\lambda}=
{}^{*}\delta^{\lambda\mu}{\widetilde{O}}_{\mu}=
{({\widetilde{O}}_{\lambda})}^{+},\\
{\widetilde{O}}_{1}=\FFr{1}{\sqrt{2}}\left( \begin{array}{cc}
1 \qquad 1 \\
\hspace{-0.2cm}-1 \hspace{-0.25cm} \quad -1
\end{array}   \right), \quad
{\widetilde{O}}_{2}=\FFr{1}{\sqrt{2}}\left( \begin{array}{cc}
1 \quad -1 \\
1 \quad -1
\end{array}   \right),
\end{array}
\end{equation}
and $\gamma^{0},\quad\gamma^{\alpha}$ are Dirac's matrices
in standard representation.
Field equations follow at once right through the variational
principle of least action
\begin{equation}
\label{R94}
\begin{array}{l}
\hat{p}\widetilde{\Psi}(\zeta)=i(\widetilde{\gamma}\partial)
\widetilde{\Psi}(\zeta)=i\gamma^{(\lambda,\mu,\alpha)}\partial_
{(\lambda,\mu,\alpha)}\widetilde{\Psi}(\zeta)=0,\\
\widetilde{\bar{\Psi}}(\zeta){\hat{p}}^{+}=
-i\widetilde{\bar{\Psi}}(\zeta){(\widetilde{\gamma}\partial)}^{+}
=-i\partial_{(\lambda,\mu,\alpha)}\widetilde{\bar{\Psi}}(\zeta)
\gamma_{(\lambda,\mu,\alpha)}=0,
\end{array}
\end{equation}
where ${(\gamma^{(\lambda,\mu,\alpha)})}^{+}=
{}^{*}\delta^{\lambda\tau}{}^{*}\delta^{\mu\nu}\delta^{\alpha\beta}
\gamma^{(\tau,\nu,\beta)}=\gamma_{(\lambda,\mu,\alpha)}$.
Explicitly the eq.(9.4) reads
\begin{equation}
\label{R95}
(\hp1_{\eta}-m)\ps1_{\eta}(\eta)=0,\quad       
\bar{\ps1_{\eta}}(\eta)(\hp1_{\eta}-m)=0,\quad 
(\hp1_{u}-m)\ps1_{u}(u)=0, \quad \bar{\ps1_{u}}(u)(\hp1_{u}-m)=0,
\end{equation}
where the function of mass at rest of bi-spinor fields $\ps1_{\eta}$ and
$\ps1_{u}$ is introduced \\
$m\Psi(\zeta)\equiv\hp1_{u}\Psi(\zeta)= \hp1_{\eta}\Psi(\zeta)$ and 
\begin{equation}
\label{R96}
\begin{array}{ll}
{\gam_{\eta}}^{(\lambda\alpha)}={\widetilde{\O1_{\eta}}}^{\lambda}\otimes
{\widetilde{\sigma}}^{\alpha}=\xi_{0}\otimes\gamma^{(\lambda\alpha)}=
\xi_{0}\otimes{\widetilde{O}}^{\lambda}\otimes
{\widetilde{\sigma}}^{\alpha},\\
{\gam_{u}}^{(\lambda\alpha)}={\widetilde{\O1_{u}}}^{\lambda}\otimes
{\widetilde{\sigma}}^{\alpha}=\xi\otimes\gamma^{(\lambda\alpha)}=
\xi\otimes{\widetilde{O}}^{\lambda}\otimes
{\widetilde{\sigma}}^{\alpha},\\
\xi_{0}=\left( \begin{array}{lr}
1\qquad 0 \\
0 \quad \hspace{-.05cm}-1
\end{array}   \right)
\quad
\xi=\left( \begin{array}{lr}
0\quad 1 \\
\hspace{-.25cm}-1 \quad 0
\end{array}   \right),
\quad
{\xi_{0}}^{2}=-\xi^{2}=1,\quad \{\xi_{0},\xi\}=0,\\
{({\gam_{i}}^{(\lambda\alpha)})}^{+}=\varepsilon_{i}
{}^{*}\delta^{\lambda\tau}\delta^{\alpha\beta}{\gam_{i}}^{(\tau\beta)}=
\varepsilon_{i}{\gam_{i}}_{(\lambda\alpha)},\quad 
\hpr_{\eta}=\gamma^{(\lambda\alpha)}{\pr_{\eta}}_{(\lambda\alpha)},\\
{(\gamma^{(\lambda\alpha)})}^{+}=
{}^{*}\delta^{\lambda\tau}\delta^{\alpha\beta}\gamma^{(\tau\beta)},
\quad
\hpr_{u}=\gamma^{(\lambda\alpha)}{\pr_{u}}_{(\lambda\alpha)},\quad
\hp1_{\eta}=i\hpr_{\eta}, \\
\hp1_{u}=i\hpr_{u},\quad
{\pr_{\eta}}_{(\lambda\alpha)}=\partial/\partial\eta^{(\lambda\alpha)},
\quad{\pr_{u}}_{(\lambda\alpha)}=\partial/\partial u^{(\lambda\alpha)}.
\end{array}
\end{equation}
The state of free goyak of $i$-type with definite values of link-momentum
$\p1_{i}$ and spin projection $s$ is described by means of plane waves,
respectively (in units $\hbar=1,\quad c=1$):
\begin{equation}   
\label{R97}
\begin{array}{l}
{\ps1_{\eta}}_{p_{\eta}}(\eta)={\left({\FFr{m}{E_{\eta}}}
\right)} ^{1/2}
\u1_{\eta}(p_{\eta},s)e^{-ip_{\eta}\eta}, 
\quad
{\ps1_{u}}_{p_{u}}(u)={\left( {\FFr{m}{E_{u}}}
\right)} ^{1/2}
\u1_{u}(p_{u},s)e^{-ip_{u}u}.
\end{array}
\end{equation}
It is denoted
$E_{i}\equiv{\p1_{i}}_{0}={({\p1_{i}}_{0\alpha},{\p1_{i}}_{0\alpha})}^{1/2},
\quad
{\p1_{i}}_{0\alpha}=\FFr{1}{\sqrt{2}}({\p1_{i}}_{(+\alpha)}+
{\p1_{i}}_{(-\alpha)})$,
and
\begin{equation}
\label{R98}
(\hp1_{i}-m)\u1_{i}(p_{i},s)=0,
\quad
\bar{\u1_{i}}(p_{i},s)(\hp1_{i}-m)=0.
\end{equation}
The amplitudes of waves $\u1_{i}(p_{i},s)$ are normalized bi-spinors.
It is necessary to consider also the solutions of negative
frequencies
\begin{equation}   
\label{R99}
{\ps1_{\eta}}_{-p_{\eta}}(\eta)={\left({\FFr{m}{E_{\eta}}}
\right) }^{1/2}
\n1_{\eta}(p_{\eta},s)e^{ip_{\eta}\eta}, 
\quad
{\ps1_{u}}_{-p_{u}}(u)={\left( {\FFr{m}{E_{u}}}
\right) }^{1/2}
\n1_{u}(p_{u},s)e^{ip_{u}u},
\end{equation}
and
\begin{equation}   
\label{R910}
(\hp1_{i}+m)\n1_{i}(p_{i},s)=0,
\quad
\bar{\n1_{i}}(p_{i},s)(\hp1_{i}+m)=0,
\end{equation}
where
${\p1_{i}}_{\alpha}=\FFr{1}{\sqrt{2}}({\p1_{i}}_{(+\alpha)}-
{\p1_{i}}_{(-\alpha)}),
\quad
p^{2}_{\eta}=E^{2}_{\eta}-{\vec{p}}^{2}_{\eta}=p^{2}_{u}=
E^{2}_{u}-{\vec{p}}^{2}_{u}=m^{2}$.
For the spinors the useful relations of orthogonality and
completeness hold.
Turning to secondary quantization of goyak's field 
we make use of localized wave packets constructed by
means of superposition of plane wave solutions furnished by creation
and annihilation operators in agreement with Pauli's principle
\begin{equation}
\label{R911}
\hps_{\eta}(\eta)= \S_{\pm s}\int\frac{d^{3}p_{\eta}}{{(2\pi)}^{3/2}}
\hps_{\eta}(p_{\eta},s,\eta),
\quad
\hps_{u}(u)= \S_{\pm s}\int\frac{d^{3}p_{u}}{{(2\pi)}^{3/2}}
\hps_{u}(p_{u},s,u),
\end{equation}
where, as usual, it is denoted
\begin{equation}   
\label{R912}
\begin{array}{ll}
\hps_{\eta}(p_{\eta},s,\eta)={\hgam_{\eta}}^{(\lambda\alpha)}(p_{\eta},s)
{\ps1_{\eta}}_{(\lambda\alpha)}(p_{\eta},s,\eta),
\quad
\bhp_{\eta}(p_{\eta},s,\eta)={\hgam_{\eta}}_{(\lambda\alpha)}(p_{\eta},s)
{\ps1_{\eta}}^{(\lambda\alpha)}(p_{\eta},s,\eta),\\
\hps_{u}(p_{u},s,u)={\hgam_{u}}^{(\lambda\alpha)}(p_{u},s)
{\ps1_{u}}_{(\lambda\alpha)}(p_{u},s,u),
\quad
\bhp_{u}(p_{u},s,u)={\hgam_{u}}_{(\lambda\alpha)}(p_{u},s)
{\ps1_{u}}^{(\lambda\alpha)}(p_{u},s,u).
\end{array}
\end{equation}
One has
${\ps1_{\eta}}_{(\pm\alpha)}=\eta_{(\pm\alpha)}{\ps1_{\eta}}_{\pm},\quad
{\ps1_{u}}_{(\pm\alpha)}=u_{(\pm\alpha)}{\ps1_{u}}_{\pm}, 
\quad
{\ps1_{\eta}}^{(\pm\alpha)}=\eta^{(\pm\alpha)}{\ps1_{\eta}}^{\pm},\quad
{\ps1_{u}}^{(\pm\alpha)}=u^{(\pm\alpha)}{\ps1_{u}}^{\pm}, $
provided
${\ps1_{i}}_{+}\equiv{\ps1_{i}}_{p_{i}},\quad{\ps1_{i}}_{-}\equiv
{\ps1_{i}}_{-p_{i}},\quad{\ps1_{i}}^{\lambda}={\bar{\ps1_{i}}}_{\lambda},
\quad{\ps1_{i}}^{(\lambda\alpha)}={\bar{\ps1_{i}}}_{(\lambda\alpha)}$.
A closer examination of the properties of the matrix elements of 
the anticommutators of expansion coefficients shows that
\begin{equation}   
\label{R913}
\begin{array}{l}
<\chi_{-}\mid \{ {\hgam_{i}}^{(+\alpha)}(p_{i},s),
{\hgam_{j}}_{(+\beta)}(p'_{j},s')\}\mid\chi_{-}>=\\
=<{\he_{i}}^{(+\alpha)}(p_{i},s),{\he_{j}}_{(+\beta)}(p'_{j},s')>=
\varepsilon_{i}\delta_{ij}\delta_{ss'}\delta_{\alpha\beta}\delta^{(3)}
({\vec{p}}_{i}- {\vec{p'}}_{i}),\\
<\chi_{+}\mid \{ {\hgam_{j}}^{(-\beta)}(p'_{j},s'),
{\hgam_{i}}_{(-\alpha)}(p_{i},s)\}\mid\chi_{+}>=\\
=<{\he_{j}}^{(-\beta)}(p'_{j},s'),{\he_{i}}_{(-\alpha)}(p_{i},s)>=
\varepsilon_{i}\delta_{ij}\delta_{ss'}\delta_{\alpha\beta}\delta^{(3)}
({\vec{p}}_{i}- {\vec{p'}}_{i}).
\end{array}
\end{equation}
We can also consider the analogical wave packets of operator-vector
fields of operator manifold $\hat{G}(2.2.3)$.
Explicitly the matrix element of anticommutator reads
\begin{equation}
\label{R914}
\begin{array}{l}
<\chi_{\pm}\mid\{ {\hat{\gamma}}^{(\lambda,\mu,\alpha)}(p,s),
{\hat{\gamma}}_{(\tau,\nu,\beta)}(p',s')\}\mid\chi_{\pm}>=\\
=<e^{(\lambda,\mu,\alpha)}(p,s),e_{(\tau,\nu,\beta)}(p',s')>=
\delta_{ss'}\delta^{\lambda}_{\tau}\delta^{\nu}_{\mu}\delta^{\alpha}_{\beta}
\delta^{(3)}(\vec{p}-\vec{p'}).
\end{array}
\end{equation}
The following relations along many others are established:
\begin{equation}
\label{R915}
\begin{array}{c}
\S_{\lambda=\pm}<\chi_{\lambda}\mid\hps_{\eta}(\eta)\bhp_{\eta}
({\eta}')\mid
\chi_{\lambda}>_{0}=-i(\eta{\eta}')_{0}\G1_{\eta}(\eta-{\eta}'),\\
\S_{\lambda=\pm}<\chi_{\lambda}\mid\hps_{u}(u)\bhp_{u}
(u')\mid
\chi_{\lambda}>_{0}=i(uu')_{0}\G1_{u}(u-u'),\\
\S_{\lambda=\pm}<\chi_{\lambda}\mid\hat{\Phi}(\zeta)
\bar{\hat{\Phi}}({\zeta}')\mid
\chi_{\lambda}>_{0}=-i(\zeta\zeta')_{0}\G1_{\zeta}(\zeta-{\zeta}')=\\
=-i\left[ (\eta{\eta}')_{0}\G1_{\eta}(\eta-{\eta}')-(uu')_{0}\G1_{u}(u-u')
\right],
\end{array}
\end{equation}
where the subscript $(_{0})$ $(<>_{0}, ()_{0})$ specifies the cone
$(xy)_{0}=x_{(+\alpha)}y^{(+\alpha)}=x_{(-\alpha)}y^{(-\alpha)}$.
The Green's functions are used
\begin{equation}
\label{R916}
\G1_{\eta}(\eta-{\eta}')=-(i\hpr_{\eta}+m)\Dlt_{\eta}(\eta-{\eta}'),
\quad
\G1_{u}(u-u')=-(i\hpr_{u}+m)\Dlt_{u}(u-u'),
\end{equation}
where the  
$\Dlt_{\eta}(\eta-{\eta}')$ and $\Dlt_{u}(u-u')$
are invariant singular functions. Thus
\begin{equation}
\label{R917}
\begin{array}{l}
\S_{\lambda=\pm}<\chi_{\lambda}\mid\hat{\Phi}(\zeta)
\bar{\hat{\Phi}}(\zeta)\mid
\chi_{\lambda}>= 
\S_{\lambda=\pm}<\chi_{\lambda}\mid
\bar{\hat{\Phi}}(\zeta)\hat{\Phi}(\zeta)\mid\chi_{\lambda}>= \\
-i\Lm_{\zeta\rightarrow\zeta'}(\zeta\zeta')\G1_{\zeta}(\zeta-\zeta')=
-i\left[ \Lm_{\eta\rightarrow\eta'}(\eta\eta')\G1_{\eta}(\eta-\eta')-
\Lm_{u\rightarrow u'}(uu')\G1_{u}(u-u')\right].
\end{array}
\end{equation}
We may consider so-called "causal" Green's functions ${\G1_{\eta}}_{F},
{\G1_{u}}_{F}$ and ${\G1_{\zeta}}_{F}$ for $\eta-,u-$ and $\zeta$-type
goyak fields, respectively. Usually they appeared in quantum field theory
as an expression of "causality", in place of the retarded functions and 
characterized by the boundary condition that only positive frequency
occur for $\eta_{0}>0\quad(u_{0}>0)$, only negative for
$\eta_{0}<0\quad(u_{0}<0)$:
\begin{equation}
\label{R918}
(\zeta\zeta')_{0}{\G1_{\zeta}}_{F}(\zeta-\zeta')=
(\eta\eta')_{0}{\G1_{\eta}}_{F}(\eta-\eta')- (uu')_{0}{\G1_{u}}_{F}(u-u').
\end{equation}
\subsection {Realization of The Manifold $G(2.2.3)$}
Realization of the manifold $G(2.2.3)$ is expressed in the milder 
restriction imposed on the matrix element
$
\S_{\lambda=\pm}<\chi_{\lambda}\mid\hat{\Phi}(\zeta)
\bar{\hat{\Phi}}(\zeta)\mid
\chi_{\lambda}>, 
$
which is as a geometric object required to be finite. The latter reads
\begin{equation}
\label{R919}
\S_{\lambda=\pm}<\chi_{\lambda}\mid\hat{\Phi}(\zeta)
\bar{\hat{\Phi}}(\zeta)\mid
\chi_{\lambda}>=
\S_{\lambda=\pm}<\chi_{\lambda}\mid
\bar{\hat{\Phi}}(\zeta)\hat{\Phi}(\zeta)\mid
\chi_{\lambda}>=\zeta^{2}{\G1_{\zeta}}_{F}(0) < \infty.
\end{equation}
This requirement should begin to manifest its virtue in the special
case, when due to eq.(9.1) the following relations hold for each
ordinary goyak:
\begin{equation}
\label{R920}
\begin{array}{l}
{\G1_{\zeta}}_{F}(0)={\G1_{\eta}}_{F}(0)=\IIn\FFr{d^{4}p_{\eta}}{(2\pi)^{4}}
\FFr{p_{\eta}+m}{p^{2}+i\varepsilon}={\G1_{u}}_{F}(0)=
\IIn\FFr{d^{4}p_{u}}{(2\pi)^{4}}
\FFr{p_{u}+m}{p^{2}+i\varepsilon}= \\
= \Lm_{u\rightarrow u'}\left[ -i\S_{{\vec{p}}_{u}}{\ps1_{u}}_{{p}_{u}}(u)
{\bp_{u}}_{{p}_{u}}(u')\theta (u_{0}-u'_{0})+
i\S_{{\vec{p}}_{u}}{\bp_{u}}_{{-p}_{u}}(u'){\ps1_{u}}_{{-p}_{u}}(u)
\theta (u'_{0}-u_{0}) \right].
\end{array}
\end{equation}
Then satisfying the condition eq.(9.19), the length of each vector
$\zeta=e^{(\lambda,\mu,\alpha)}\zeta_{(\lambda,\mu,\alpha)}\in G(2.2.3)$
(see eq.(9.18)) compulsory should be equaled zero
\begin{equation}
\label{R921}
\zeta^{2}=\eta^{2}-u^{2}=0.
\end{equation}
Thus, it brings us to the conclusion: {\em the main requirement eq.(9.19)
provided by eq.(9.20) yields the realization of the flat manifold $G(2.2.3)$},
which subsequently leads to Minkowski flat space $M^{4}$ (Part I).
\begin{equation}
\label{R922}
G(2.2.3)=\G1_{\eta}(2.3)\oplus\G1_{u}(2.3).
\end{equation}
The condition eq.(9.21) provides the invariance of line element in
$\G1_{i}(2.3)$, as well as in $M^{4}$. 
In the case of distorted manifold $G(223)$ (sec. 6) [1-5], 
instead of eq.(9.21) one obviously has
\begin{equation}
\label{R923}
d\,\zeta^{2}=d\,\eta^{2}-d\,u^{2}=0,\quad 
d\,\eta^{2}\left.\right|_{6\rightarrow 4}=d\,s^{2}=g_{\mu\nu}
d\,x^{\mu}d\,x^{\nu}=inv=d\,u^{2}.
\end{equation}
Thus, {\em the principle of Relativity comes into being with the geometry.}
We have reached this important conclusion, which furnishes the proof of
our main idea that {\em the geometry is derivative}.
\section {A System of Identical Goyaks} 
\label {ident}
Next we discuss briefly the quantum theory situation, which corresponds 
to simultaneously presence of many identical goyak fields. Certainly, 
if there are $n$ identical goyaks with link-coordinates
$\zeta_{1},\zeta_{2},...,\zeta_{n}$ the antisymmetrical state function
$\Psi$ will be a function of all of them and presents the system of
$n$ fermions $\Psi(\zeta_{1},\zeta_{2},...,\zeta_{n})$, which implies
the Fermi-Dirac statistics.
To describe the $n$-goyak fermion system by means of quantum field theory, 
it will be advantageous to make use of convenient method of constructing 
the state vector of physical system by proceeding from the vacuum state
as a very point of origin.
One ought to modify the operators eq.(8.7) in order to provide
an anticommutation is being valid in both cases acting on the same as well as
different states:
\begin{equation}
\label{R101}
{\hat{\gamma}}_{(\lambda,\mu,\alpha)}^{r}\Rightarrow
{\hat{\gamma}}_{(\lambda,\mu,\alpha)}^{r}\eta_{r}^{\lambda\mu},
\quad
{\hat{\gamma}}^{(\lambda,\mu,\alpha)}_{r}\Rightarrow
\eta_{r}^{\lambda\mu}{\hat{\gamma}}^{(\lambda,\mu,\alpha)}_{r}=
{\hat{\gamma}}^{(\lambda,\mu,\alpha)}_{r}\eta_{r}^{\lambda\mu},
\quad
{(\eta_{r}^{\lambda\mu})}^{+}=\eta_{r}^{\lambda\mu},
\end{equation}
for fixed $\lambda,\mu,\alpha$,
where $\eta_{r}^{\lambda\mu}$ is a diagonal operator in the space 
of occupation numbers.\\
Therewith, at $r_{i}<r_{j}$ one gets
\begin{equation}
\label{R102}
{\hat{\gamma}}_{(\lambda,\mu,\alpha)}^{r_{i}}\eta_{r_{j}}^{\lambda\mu}=
-\eta_{r_{j}}^{\lambda\mu}{\hat{\gamma}}_{(\lambda,\mu,\alpha)}^{r_{i}},
\quad
{\hat{\gamma}}_{(\lambda,\mu,\alpha)}^{r_{j}}\eta_{r_{i}}^{\lambda\mu}=
\eta_{r_{i}}^{\lambda\mu}{\hat{\gamma}}_{(\lambda,\mu,\alpha)}^{r_{j}}.
\end{equation}
The operators of corresponding occupation numbers, for fixed 
$\lambda,\mu,\alpha$, are
\begin{equation}
\label{R103}
{\hat{N}}_{r}^{\lambda\mu}={\hat{\gamma}}_{(\lambda,\mu,\alpha)}^{r}
{\hat{\gamma}}^{(\lambda,\mu,\alpha)}_{r}={\hat{N}}_{r}^{\lambda}\otimes
{\hat{N}}_{r}^{\mu},
\end{equation}
where we make use of 
${\hat{N}}^{1}_{r}=\hat{O}^{r}_{1}\hat{O}_{r}^{1}=(\alpha_{1}\alpha_{2})_{r},$
${\hat{N}}^{2}_{r}=
\hat{O}^{r}_{2}\hat{O}_{r}^{2}=(\alpha_{2}\alpha_{1})_{r}.$
As far as diagonal operators $(1-2{\hat{N}}_{r}^{\lambda\mu})$. 
anticommute with the
${\hat{\gamma}}_{(\lambda,\mu,\alpha)}^{r}$, then\\
$\eta_{r_{i}}^{\lambda\mu}=\displaystyle\prod_{r =1}^{r_{i}-1}(1-
2{\hat{N}}_{r}^{\lambda\mu})$,
provided
\begin{equation}
\label{R104}
\eta_{r_{i}}^{11}\Psi(n_{1},\ldots,n_{N};0;0;0)=\pd_{r =1}^{r_{i}-1}
(-1)^{n_{r}}\Psi(n_{1},\ldots,n_{N};0;0;0),
\end{equation}
and so on.
Here the occupation numbers $n_{r}(m_{r},q_{r},t_{r})$ are introduced, 
which refer to the $r$-th states corresponding to operators
${\hat{\gamma}}_{(1,1,\alpha)}^{r}({\hat{\gamma}}_{(1,2,\alpha)}^{r},
{\hat{\gamma}}_{(2,1,\alpha)}^{r},{\hat{\gamma}}_{(2,2,\alpha)}^{r})$
either empty ($n_{r},\ldots,t_{r}=0$) or occupied
($n_{r},\ldots,t_{r}=1$).
To save writing we abbreviate the modified operators by the
same symbols.
The creation operator ${\hat{\gamma}}_{(\lambda,\mu,\alpha)}^{r_{i}}$ by
acting on free state $\mid 0>_{r_{i}}$ yields the one-occupied state
$\mid 1>_{r_{i}}$ with the phase $+$ or $-$ depending of parity of the number
of goyaks in the states $ r < r_{i}$.
Modified operators satisfy the same anticommutation relations
of the operators eq.(8.7). 
It is convenient to make use of notation
${\hat{\gamma}}^{(\lambda,\mu,\alpha)}_{r}\equiv
{e}^{(\lambda,\mu,\alpha)}_{r}{\hat{b}}^{\lambda\mu}_{(r\alpha)},
\quad
{\hat{\gamma}}_{(\lambda,\mu,\alpha)}^{r}\equiv
{e}_{(\lambda,\mu,\alpha)}^{r}{\hat{b}}_{\lambda\mu}^{(r\alpha)}$,
and abbreviate the pair of indices $(r\alpha)$ by the single symbol $r$. 
Then, non vanishing general anticommutation
relations hold
\begin{equation}
\label{R105}
\begin{array}{l}
<1,1\mid\{ {\hat{b}}^{11}_{r},{\hat{b}}_{11}^{r'}
\}\mid 1,1>=
<2,2\mid\{ {\hat{b}}^{22}_{r},{\hat{b}}_{22}^{r'}
\}\mid 2,2>=\\
=<1,2\mid\{ {\hat{b}}^{12}_{r},{\hat{b}}_{12}^{r'}
\}\mid 1,2>=
<2,1\mid\{ {\hat{b}}^{21}_{r},{\hat{b}}_{(21}^{r'}
\}\mid 2,1>=\delta^{r'}_{r}.
\end{array}
\end{equation}
The operator of occupation number in terms of new operators is
${\hat{N}}^{\lambda\mu}_{r}={\hat{b}}^{r}_{\lambda\mu}
{\hat{b}}_{r}^{\lambda\mu}$.
The vacuum state reads eq.(8.13), with the normalization requirement
\begin{equation}
\label{R106}
<\chi^{0}(\nu'_{1},\nu'_{2},\nu'_{3},\nu'_{4})\mid
\chi^{0}(\nu_{1},\nu_{2},\nu_{3},\nu_{4})>=\prod_{i=1}^{4}
\delta_{\nu_{i}\nu'_{i}}.
\end{equation}
The state vectors
\begin{equation}
\label{R107}
\begin{array}{l}
\chi({\{n_{r}\}}^{N}_{1};{\{m_{r}\}}^{M}_{1};
{\{q_{r}\}}^{Q}_{1};{\{t_{r}\}}^{T}_{1};
{\{\nu_{r}\}}^{4}_{1})=
{(\hat{b}^{N}_{11})}^{n_{\scriptscriptstyle N}}\cdots
{(\hat{b}^{1}_{11})}^{n_{1}}\cdot\\
\cdot {(\hat{b}^{M}_{12})}^{m_{\scriptscriptstyle M}}\cdots
{(\hat{b}^{1}_{12})}^{m_{1}}\cdot
{(\hat{b}^{Q}_{21})}^{q_{\scriptscriptstyle Q}}\cdots
{(\hat{b}^{1}_{21})}^{q_{1}}\cdot
\cdot{(\hat{b}^{T}_{22})}^{t_{\scriptscriptstyle T}}\cdots
{(\hat{b}^{1}_{22})}^{t_{1}}
\chi^{0}(\nu_{1},\nu_{2},\nu_{3},\nu_{4}),
\end{array}
\end{equation}
where ${\{n_{r}\}}^{N}_{1}=n_{1},\ldots,n_{N}$ and so on, are the 
eigen-functions
of modified operators. They form a whole set of orthogonal vectors
\begin{equation}
\label{R108}
\begin{array}{l}
<\chi({\{n'_{r}\}}^{N}_{1};{\{m'_{r}\}}^{M}_{1};
{\{q'_{r}\}}^{Q}_{1};{\{t'_{r}\}}^{T}_{1};
{\{\nu'_{r}\}}^{4}_{1})\mid
\chi({\{n_{r}\}}^{N}_{1};{\{m_{r}\}}^{M}_{1};
{\{q_{r}\}}^{Q}_{1};{\{t_{r}\}}^{T}_{1};
{\{\nu_{r}\}}^{4}_{1})>=\\
=\displaystyle \prod_{r=1}^{N}\delta_{n_{r}n'_{r}}\cdot
\displaystyle \prod_{r=1}^{M}\delta_{m_{r}m'_{r}}\cdot
\displaystyle \prod_{r=1}^{Q}\delta_{q_{r}q'_{r}}\cdot
\displaystyle \prod_{r=1}^{T}\delta_{t_{r}t'_{r}}\cdot
\displaystyle \prod_{r=1}^{4}\delta_{\nu_{r}\nu'_{r}}.
\end{array}
\end{equation}
Considering an arbitrary superposition
\begin{equation}
\label{R109}
\chi=
\S_{\begin{array}{l}
{\scriptstyle n_{1},...,n_{N}}=0 \\
{\scriptstyle m_{1},...,m_{M}=0} \\
{\scriptstyle q_{1},...,q_{Q}=0}  \\
{\scriptstyle t_{1},...,t_{T}=0}
\end{array}}^{1}
c'({\{n_{r}\}}^{N}_{1};{\{m_{r}\}}^{M}_{1};
{\{q_{r}\}}^{Q}_{1};{\{t_{r}\}}^{T}_{1})
\chi({\{n_{r}\}}^{N}_{1};{\{m_{r}\}}^{M}_{1};
{\{q_{r}\}}^{Q}_{1};{\{t_{r}\}}^{T}_{1};
{\{\nu_{r}\}}^{4}_{1}),
\end{equation}
the coefficients $c'$ of expansion are the corresponding amplitudes
of probabilities:
\begin{equation}
\label{R1010}
<\chi\mid\chi>=
\S_{\begin{array}{l}
{\scriptstyle n_{1},...,n_{N}=0} \\
{\scriptstyle m_{1},...,m_{M}=0} \\
{\scriptstyle q_{1},...,q_{Q}=0}  \\
{\scriptstyle t_{1},...,t_{T}=0}
\end{array}}^{1}
{\left| c'({\{n_{r}\}}^{N}_{1};{\{m_{r}\}}^{M}_{1};
{\{q_{r}\}}^{Q}_{1};{\{t_{r}\}}^{T}_{1}) \right|}^{2}.
\end{equation}
The non-vanishing matrix elements of
operators $\hat{b}^{11}_{r_{k}}$ and $\hat{b}_{11}^{r_{k}}$ read
\begin{equation}
\label{R1011}
\begin{array}{l}
<\chi({\{n'_{r}\}}^{N}_{1};0;0;0;1,0,0,0)\left|   \right.
\hat{b}^{11}_{r_{k}}
\chi({\{n_{r}\}}^{N}_{1};0;0;0;1,0,0,0)>=\\
=<1,1\mid \hat{b}^{11}_{r'_{1}}\cdots\hat{b}^{11}_{r'_{n}}\cdot
\hat{b}^{11}_{r_{k}}\cdot
\hat{b}_{11}^{r_{n}}\cdots\hat{b}_{11}^{r_{1}}\mid 1,1>=\\
= \left\{ \begin{array}{ll}
(-1)^{n-k}  & \mbox{if $n_{r}=n'_{r}$ for $r\neq r_{k}$ and 
$n'_{r_{k}}=0;n_{r_{k}}=1$}, \\
0           & \mbox{otherwise},
\end{array}  \right.   \\
<\chi({\{n'_{r}\}}^{N}_{1};0;0;0;1,0,0,0)\left|   \right.
\hat{b}_{11}^{r_{k}}
\chi({\{n_{r}\}}^{N}_{1};0;0;0;1,0,0,0)>=\\
=<1,1\mid \hat{b}^{11}_{r'_{1}}\cdots\hat{b}^{11}_{r'_{n}}\cdot
\hat{b}_{11}^{r_{k}}\cdot
\hat{b}_{11}^{r_{n}}\cdots\hat{b}_{11}^{r_{1}}\mid 1,1>=\\
= \left\{ \begin{array}{ll}
(-1)^{n'-k'}  & \mbox{if $n_{r}=n'_{r}$ for $r\neq r_{k}$ and 
$n_{r_{k}}=0;n'_{r_{k}}=1$}, \\
0           & \mbox{otherwise},
\end{array} \right.
\end{array}
\end{equation}
where one denotes
$n=\S_{r=1}^{N}n_{r}, \quad n'=\S_{r=1}^{N}n'_{r}$,
the $r_{k}$ and $r'_{k}$ are $k$-th and $k'$-th terms of regulated sets of
$\{r_{1},\ldots ,r_{n}\} \quad (r_{1}<r_{2}<\cdots <r_{n})$ and
$\{r'_{1},\ldots ,r'_{n}\} \quad (r'_{1}<r'_{2}<\cdots <r'_{n})$,
respectively.
Continuing along this line we get a whole set of explicit forms of
matrix elements for the rest of operators
$\hat{b}^{\lambda \mu}_{r_{k}}$ and $\hat{b}_{\lambda \mu}^{r_{k}}$.
There up on 
\begin{equation}
\label{R1012}
\begin{array}{l}
\S_{\{\nu_{r}\}=0}^{1}<\chi^{0}\mid\hat{\Phi}(\zeta)
\mid\chi>=\S_{\{\nu_{r}\}=0}^{1}<\chi^{0}\mid
{\hat{\gamma}}^{(\lambda,\mu,\alpha)}_{r}{\Phi}_{(\lambda,\mu,\alpha)}^{r}
(\zeta)\mid\chi>= \\
=\S_{r=1}^{N}c'_{n_{r}}e^{(1,1,\alpha)}_{n_{r}}
{\Phi}_{(1,1,\alpha)}^{n_{r}}+
\S_{r=1}^{M}c'_{m_{r}}e^{(1,2,\alpha)}_{m_{r}}
{\Phi}_{(1,2,\alpha)}^{m_{r}}+ 
\S_{r=1}^{Q}c'_{q_{r}}e^{(2,1,\alpha)}_{q_{r}}
{\Phi}_{(2,1,\alpha)}^{q_{r}}+
\S_{r=1}^{T}c'_{t_{r}}e^{(2,2,\alpha)}_{t_{r}}
{\Phi}_{(2,2,\alpha)}^{t_{r}},
\end{array}
\end{equation}
provided
\begin{equation}
\label{R1013}
\begin{array}{ll}
c'_{n_{r}}\equiv\delta_{1n_{r}}c'(0,\ldots, n_{r},\ldots,0;0;0;0),\quad
c'_{m_{r}}\equiv\delta_{1m_{r}}c'(0;0,\ldots m_{r},\ldots,0;0;0),\\
c'_{q_{r}}\equiv\delta_{1q_{r}}c'(0;0;0,\ldots q_{r},\ldots,0;0),\quad
c'_{t_{r}}\equiv\delta_{1t_{r}}c'(0;0;0;0,\ldots t_{r},\ldots,0).
\end{array}
\end{equation}
Hereinafter we change the notation to
\begin{equation}
\label{R1014}
\begin{array}{l}
\bar{c}(r^{11})=c'_{n_{r}},\quad \bar{c}(r^{21})=c'_{q_{r}},
\quad N_{11}=N, \quad N_{21}=Q, \\
\bar{c}(r^{12})=c'_{m_{r}}, \quad \bar{c}(r^{22})=c'_{t_{r}},
\quad N_{12}=M, \quad N_{22}=T,
\end{array}
\end{equation}
and make use of
\begin{equation}
\label{R1015}
\begin{array}{ll}
F_{r^{\lambda\mu}}= \S_{\alpha}
e^{(\lambda,\mu,\alpha)}_{r^{\lambda\mu}}
{\Phi}_{(\lambda,\mu,\alpha)}^{r^{\lambda\mu}},
\quad
F^{r^{\lambda\mu}}= \S_{\alpha}
e_{(\lambda,\mu,\alpha)}^{{r}^{\lambda\mu}}
{\Phi}^{(\lambda,\mu,\alpha)}_{r^{\lambda\mu}}=\bar{F}_{r^{\lambda\mu}},\\
\S_{\{\nu_{r}\}=0}^{1}<\chi^{0}\mid\hat{A}\mid\chi>
\equiv<\chi^{0}\parallel\hat{A}\parallel\chi>,
\quad
\S_{\{\nu_{r}\}=0}^{1}<\chi\mid\hat{A}\mid\chi^{0}>
\equiv<\chi\parallel\hat{A}\parallel\chi^{0}>.
\end{array}
\end{equation}
Then, the matrix elements of operator-vector and co-vector fields
take the forms
\begin{equation}
\label{R1016}
\begin{array}{ll}
<\chi^{0}\parallel\hat{\Phi}(\zeta)\parallel\chi>=
\S_{\lambda\mu=1}^{2}\S_{r^{\lambda\mu}=1}^{N_{\lambda\mu}}
\bar{c}(r^{\lambda\mu})F_{r^{\lambda\mu}}(\zeta),\\
<\chi\parallel\bar{\hat{\Phi}}(\zeta)\parallel\chi^{0}>=
\S_{\lambda\mu=1}^{2}\S_{r^{\lambda\mu}=1}^{N_{\lambda\mu}}
{\bar{c}}^{*}(r^{\lambda\mu})F^{r^{\lambda\mu}}(\zeta).
\end{array}
\end{equation}
Explicitly the matrix element of $n$-operator
vector fields, in general, reads
\begin{equation}
\label{R1017}
\begin{array}{l}
\FFr{1}{\sqrt{n!}}<\chi^{0}\parallel\hat{\Phi}({\zeta}_{1})
\cdots \hat{\Phi}({\zeta}_{n})\parallel\chi>=\\
=\FFr{1}{\sqrt{n!}}\left\{\S_{\lambda\mu=1}^{2} \right\}_{1}^{n}
\S_{r_{1}^{\lambda\mu},\ldots r_{n}^{\lambda\mu}=1}^{N_{\lambda\mu}}
\bar{c}(r_{1}^{\lambda\mu},\ldots , r_{n}^{\lambda\mu})
\S_{\sigma\in S_{n}}sgn(\sigma)
F_{r_{1}^{\lambda\mu}} (\zeta_{1}) \cdots
F_{r_{n}^{\lambda\mu}} (\zeta_{n})=\\
=\FFr{1}{\sqrt{n!}}\left\{\S_{\lambda\mu=1}^{2} \right\}_{1}^{n}
\S_{r_{1}^{\lambda\mu},\ldots r_{n}^{\lambda\mu}=1}^{N_{\lambda\mu}}
\bar{c}(r_{1}^{\lambda\mu},\ldots ,r_{n}^{\lambda\mu})
\left\|{ \begin{array}{ll}
F_{r_{1}^{\lambda\mu}}(\zeta_{1})\cdots F_{r_{n}^{\lambda\mu}}(\zeta_{1})\\
\cdots\cdots\cdots\cdots\cdots\cdots \\
\cdots\cdots\cdots\cdots\cdots\cdots \\
F_{r_{1}^{\lambda\mu}}(\zeta_{n})\cdots F_{r_{n}^{\lambda\mu}}(\zeta_{n})
\end{array}} \right\|.
\end{array}
\end{equation}
The summation is extended over all permutations of indices
$(r_{1}^{\lambda\mu},\ldots, r_{n}^{\lambda\mu})$ of the integers
$1,2,\ldots,n,$ whereas the antisymmetrical eigen-functions are sums
of the same terms with alternating signs in dependence of a parity
$sgn(\sigma)$ of transposition; $\left\{\S_{\lambda\mu}^{2} \right\}_{1}^{n}
\equiv \S_{\lambda_{1}\mu_{1}}^{2}\ldots\S_{\lambda_{n}\mu_{n}}^{2}$
and $r^{\lambda\mu}_{i}\equiv r^{\lambda_{i}\mu_{i}}$.
Here we make use of
\begin{equation}
\label{R1018}
\begin{array}{l}
\bar{c}(r^{11}_{1},\ldots,r^{11}_{n})=c'(n_{1},\ldots,n_{N};0;0;0),
\quad
\bar{c}(r^{12}_{1},\ldots,r^{12}_{n})=c'(0;m_{1},\ldots,m_{M};0;0),\\
\bar{c}(r^{21}_{1},\ldots,r^{21}_{n})=c'(0;0;q_{1},\ldots,q_{Q};0),
\quad
\bar{c}(r^{22}_{1},\ldots,r^{22}_{n})=c'(0;0;0;t_{1},\ldots,t_{T}).
\end{array}
\end{equation}
Analogical expression can be obtained for co-vector fields.
Following to this procedure one also gets
\begin{equation}
\label{R1019}
\begin{array}{l}
\S_{\lambda =1}^{2}
<\chi({\{n'_{r}\}}^{N}_{1};0;0;0;\lambda;0)\left| \right.
{\hb_{\eta}}^{+}_{r_{k}}
\chi({\{n_{r}\}}^{N}_{1};0;0;0;\lambda;0)>=\\
= \left\{ \begin{array}{ll}
(-1)^{n-k}  & \mbox{if $n_{r}=n'_{r}$ for $r\neq r_{k}$ and 
$n'_{r_{k}}=0;n_{r_{k}}=1,$} \\
0           & \mbox{otherwise},
\end{array}  \right. \\
\S_{\lambda =1}^{2}
<\chi({\{n'_{r}\}}^{N}_{1};0;0;0;\lambda;0)\left| \right.
{\hb_{\eta}}_{+}^{r_{k}}
\chi({\{n_{r}\}}^{N}_{1};0;0;0;\lambda;0)>=\\
= \left\{ \begin{array}{ll}
(-1)^{n'-k'}  & \mbox{if $n_{r}=n'_{r}$ for $r\neq r_{k}$ and 
$n_{r_{k}}=0;n'_{r_{k}}=1,$} \\
0           & \mbox{otherwise}
\end{array}  \right.
\end{array}
\end{equation}
and so on.
Thus
\begin{equation}
\label{R1020}
\begin{array}{l}
<\chi^{0}\parallel\hps_{\eta}(\eta)\bhp_{\eta}(\eta')\parallel\chi^{0}>=
\S_{\lambda\tau=1}^{2}\S_{r^{\lambda}=1}^{N^{1}_{\lambda}}
\S_{r^{\tau}=1}^{N^{1}_{\tau}}
\bar{c}(r^{\lambda})\bar{c}^{*}(r^{\tau}){\F_{\eta}}_{r^{\lambda}}(\eta),
{\bar{\F_{\eta}}}_{r^{\tau}}(\eta'), \\
<\chi^{0}\parallel\hps_{u}(u)\bhp_{u}(u')\parallel\chi^{0}>=
\S_{\lambda\tau=1}^{2}\S_{r^{\lambda}=1}^{N^{2}_{\lambda}}
\S_{r^{\tau}=1}^{N^{2}_{\tau}}
\bar{c}(r^{\lambda})\bar{c}^{*}(r^{\tau}){\F_{u}}_{r^{\lambda}}(u),
{\bar{\F_{u}}}_{r^{\tau}}(u').
\end{array}
\end{equation}
The expression of generalized Green's function
$\widetilde{\G1_{\eta}}(\eta-\eta')$ may be written 
\begin{equation}
\label{R1021}
\widetilde{\G1_{\eta}}(\eta-\eta')
=\S_{r^{\lambda}=1}^{N^{1}_{\lambda}}
{\mid\bar{c}(r^{\lambda})\mid}^{2}
{\G1_{\eta}}^{r^{\lambda}}_{\lambda}(\eta-\eta'),
\quad
{\G1_{\eta}}^{r^{\lambda}}_{\lambda}(\eta-\eta')=-iT(
{\ps1_{\eta}}_{\lambda}^{{r}^{\lambda}}(\eta)
{\bar{\ps1_{\eta}}}_{\lambda}^{{r}^{\lambda}}(\eta')),
\end{equation}
with incorporating of orthogonality relation
$<{\e1_{\eta}}^{(\lambda\alpha)}_{r^{\lambda}},
{\e1_{\eta}}_{(\tau\beta)}^{r^{\tau}}>=\delta^{\lambda}_{\tau}
\delta^{\alpha}_{\beta}\delta^{r^{\tau}}_{r^{\lambda}}$.
It is easy to get
\begin{equation}
\label{R1022}
\widetilde{\G1_{u}}(u-u')
=\S_{r^{\lambda}=1}^{N^{2}_{\lambda}}
{\mid\bar{c}(r^{\lambda})\mid}^{2}
{\G1_{u}}^{r^{\lambda}}_{\lambda}(u-u'), 
\quad
\widetilde{\G1_{\zeta}}(\zeta-\zeta')
=\S_{r^{\lambda\mu}=1}^{N_{\lambda\mu}}
{\mid\bar{c}(r^{\lambda\mu})\mid}^{2}
{\G1_{\zeta}}^{r^{\lambda\mu}}_{\lambda\mu}(\zeta-\zeta'),
\end{equation}
provided
\begin{equation}
\label{R1023}
{\G1_{u}}^{r^{\lambda}}_{\lambda}(u-u')=-iT(
{\ps1_{u}}_{\lambda}^{{r}^{\lambda}}(u)
{\bar{\ps1_{u}}}_{\lambda}^{{r}^{\lambda}}(u')), 
\quad
{\G1_{\zeta}}^{r^{\lambda\mu}}_{\lambda\mu}(\zeta-\zeta')=-iT(
{\Phi_{\lambda\mu}}^{r^{\lambda\mu}}(\zeta)
{\bar{\Phi}}_{\lambda\mu}^{r^{\lambda\mu}}(\zeta')),
\end{equation}
and
$<{\e1_{u}}^{(\lambda\alpha)}_{r^{\lambda}},
{\e1_{u}}_{(\tau\beta)}^{r^{\tau}}>=-\delta^{\lambda}_{\tau}
\delta^{\alpha}_{\beta}\delta^{r^{\tau}}_{r^{\lambda}}. 
\quad
<{\e1_{\zeta}}^{(\lambda,\mu,\alpha)}_{r^{\lambda\mu}},
{\e1_{\zeta}}_{(\tau,\nu,\beta)}^{r^{\tau\nu}}>=\delta^{\lambda}_{\tau}
\delta^{\mu}_{\nu}
\delta^{\alpha}_{\beta}\delta^{r^{\tau\nu}}_{r^{\lambda\mu}}$.
The generalized causal Green's functions 
associate with the fields defined on the wave manifolds
$\widetilde{G}(2.2.3)=\widetilde{\G1_{\eta}}(2.3)\oplus
\widetilde{\G1_{u}}(2.3)$,
in which the bases are the sets of vectors
$\{{\e1_{\zeta}}_{(\lambda,\mu,\alpha)}^{r^{\lambda\mu}}\},
\{{\e1_{\eta}}_{(\lambda\alpha)}^{r^{\lambda}}\}$ and
$\{{\e1_{u}}_{(\lambda\alpha)}^{r^{\lambda}}\},$ respectively.
\section {Geometric Aspect of Operator Manifold}
\label {diff}
The other interesting offshoot of this generalization is a geometric aspect.
Here we shall briefly continue the differential geometric treatment of the 
operator manifold $\hat{G}(2.2.3)$ 
and perform a passage from the operator manifold to wave manifold 
$\widetilde{G}(2.2.3)$. 
Let consider, for example, the operator-tensor of 
$\widehat{(n,0)}$-type, which is
$\hat{\Phi}(\zeta_{1})\otimes\cdots\otimes\hat{\Phi}(\zeta_{n})$,
where $\otimes $ stands for tensor product. 
We readily obtain the matrix element of 
operator-tensor of $\widehat{(n,0)}$-type as follows:
\begin{equation}
\label{R111}
\begin{array}{l}
\FFr{1}{\sqrt{n!}}<\chi^{0}\parallel\hat{\Phi}({\zeta}_{1})\otimes
\cdots \otimes\hat{\Phi}({\zeta}_{n})\parallel\chi>=\\
=\left\{\S_{\lambda\mu=1}^{2}\right\}_{1}^{n}
\S_{r_{1}^{\lambda\mu},\ldots ,r_{n}^{\lambda\mu}=1}^{N_{\lambda\mu}}
\bar{c}(r_{1}^{\lambda\mu},\ldots ,r_{n}^{\lambda\mu})
\S_{\sigma\in S_{n}}sgn(\sigma)
F_{r_{1}^{\lambda\mu}}(\zeta_{1})\otimes\cdots\otimes
F_{r_{n}^{\lambda\mu}}(\zeta_{n})= \\
=\left\{\S_{\lambda\mu=1}^{2}\right\}_{1}^{n}
\S_{r_{1}^{\lambda\mu},\ldots ,r_{n}^{\lambda\mu}=1}^{N_{\lambda\mu}}
\bar{c}(r_{1}^{\lambda\mu},\ldots ,r_{n}^{\lambda\mu})
F_{r_{1}^{\lambda\mu}}(\zeta_{1})\wedge\cdots\wedge
F_{r_{n}^{\lambda\mu}}(\zeta_{n}),
\end{array}
\end{equation}
where $\wedge$ stands for exterior product. 
It is a straightforward to get also the explicit form of the 
operator-tensor  $\widehat{(0,n)}$.
The matrix element eq.(11.1) is the geometric objects belonging 
to manifold $\widetilde{G}(2.2.3)$, which expose an antisymmetric part of 
tensor degree. 
We have reached an important conclusion, that {\em by constructing matrix 
elements of operator-tensors of $\hat{G}(2.2.3)$ one produces the 
external products on wave manifold $\widetilde{G}(2.2.3)$}. 
There up on, {\em the matrix elements of symmetric 
operator-tensor identically equal zero}. 
To facilitate writing,
we abbreviate the set of indices $(\lambda_{i},\mu_{i},\alpha_{i})$
by the single symbol $i$. 
The antisymmetric tensor $T_{i_{1}\ldots i_{k}}$ may be defined by single 
number $T_{1\ldots k}$. It is convenient to express it in terms of 
differential operator-forms on $\hat{G}(2.2.3)$. 
As the basis in the space of such tensor the operator elements
$d\Phi^{i_{1}}\wedge\cdots\wedge d\Phi^{i_{k}}=
\S_{\sigma\in S_{k}}sgn(\sigma)
{\hat{\gamma}}^{\sigma(i_{1}}\otimes\cdots\otimes
{\hat{\gamma}}^{i_{k})}$
are considered, where the summation is extended over all permutations
$(i_{1},\ldots,i_{k}).$ 
Corresponding to  $T_{i_{1}\ldots i_{k}}$ differential operator-form
$\hat{\omega}$ reads
$\hat{\omega}=\S_{i_{1}<\ldots<i_{k}}T_{i_{1}\ldots i_{k}}
d\Phi^{i_{1}}\wedge\cdots\wedge d\Phi^{i_{k}},$
the matrix element of which gives rise to
\begin{equation}
\label{R112}
\begin{array}{l}
<\chi^{0}\parallel \hat{\omega} \parallel\chi>= \\
=\S_{i_{1}<\ldots<i_{k}}
\left\{\S_{\lambda,\mu=1}^{2}\right\}_{1}^{k}
\S_{r_{1}^{\lambda\mu},\ldots ,r_{k}^{\lambda\mu}=1}^{N_{\lambda\mu}}
\bar{c}(r_{1}^{\lambda\mu},\ldots ,r_{k}^{\lambda\mu})
{T(r_{1}^{\lambda\mu},\ldots ,r_{k}^{\lambda\mu})}_{i_{1}\ldots i_{k}}
d\Phi^{i_{1}}_{r_{1}^{\lambda\mu}}\wedge\cdots\wedge 
d\Phi^{i_{k}}_{r_{k}^{\lambda\mu}}.
\end{array}
\end{equation}
So define the external differential by operator-form $d\hat{\omega}$
of $(k+1)$ degree
\begin{equation}
\label{R113}
d\hat{\omega}=\S_{\begin{array}{l}
{\scriptstyle i_{0}} \\
{\scriptstyle i_{1}<\ldots<i_{k}}
\end{array}}
\FFr{\partial T_{i_{1}\ldots i_{k}}}{\partial\Phi^{i_{0}}}d\Phi^{i_{0}}\wedge
d\Phi^{i_{1}}\wedge\cdots\wedge d\Phi^{i_{k}}
=\S_{i_{1}<\ldots<i_{k}}(dT_{i_{1}\ldots i_{k}})\wedge
d\Phi^{i_{1}}\wedge\cdots\wedge d\Phi^{i_{k}},
\end{equation}
then
\begin{equation}
\label{R114}
\begin{array}{l}
<\chi^{0}\parallel d\hat{\omega} \parallel\chi>= \\
=\S_{i_{1}<\ldots<i_{k}}
\left\{\S_{\lambda,\mu=1}^{2}\right\}_{1}^{k}
\S_{r_{1}^{\lambda\mu},\ldots ,r_{k}^{\lambda\mu}=1}^{N_{\lambda\mu}}
\bar{c}(r_{1}^{\lambda\mu},\ldots r_{k}^{\lambda\mu})
{(dT(r_{1}^{\lambda\mu},\ldots ,r_{k}^{\lambda\mu})}_{i_{1}\ldots i_{k}})
\wedge 
d\Phi^{i_{1}}_{r_{1}^{\lambda\mu}}\wedge\cdots\wedge 
{d\Phi}^{i_{k}}_{r_{k}^{\lambda\mu}}.
\end{array}
\end{equation}
This is the external differential of the corresponding form on wave manifold 
$\widetilde{G}(2.2.3) $.
Continuing along this line we may employ the analog of integration of the 
forms.
{\em The matrix element 
of any geometric object of operator 
manifold $\hat{G}(2.2.3)$ yields corresponding geometric object of wave 
manifold $\widetilde{G}(2.2.3)$}. Thus, {\em all geometric objects 
belonging to the latter can be constructed by means of matrix elements of 
corresponding geometric objects of the operator manifold $\hat{G}(2.2.3)$}.
\section {The Groups $\hat{F}$ and  $\widetilde{F}$}
\label {group}
Next we deal with the treatment of principles of the theory of the groups
$\hat{F},\widetilde{F}$ and their representations. They refer to continuous 
and discrete transformations of symmetries of operator $\hat{G}(2.2.3)$ and 
wave $\widetilde{G}(2.2.3)$ manifolds, respectively. 
The action of linear operator $\hat{F}$ upon any operator-vector of
$\hat{G}(2.2.3)$ is completely defined by its action upon the operator basis
$\{\hat{\gamma}\}$. 
We consider the homogeneous 
$\hat{F}$ transformations of basis vectors eq.(8.7)
\begin{equation}
\label{121}
{\hat{\gamma'}}^{(\lambda,\mu,\alpha)}=
{\hat{F}}^{(\lambda,\mu,\alpha)}_{(\tau,\nu,\beta)}
{\hat{\gamma}}^{(\tau,\nu,\beta)}, \quad
{\hat{\gamma'}}_{(\lambda,\mu,\alpha)}=
{\hat{\gamma}}_{(\tau,\nu,\beta)}
{\hat{F}}_{(\lambda,\mu,\alpha)}^{(\tau,\nu,\beta)}, \quad
(\hat{\gamma'}=\hat{F}\hat{\gamma},\quad \bar{\hat{\gamma'}}
=\bar{\hat{\gamma}}{\bar{\hat{F}}}^{T}),
\end{equation}
provided by ${\hat{F}}^{(\lambda,\mu,\alpha)}_{(\tau,\nu,\beta)}$, which is a
matrix operator with respect to basis $\{\hat{\gamma}\}$, and
$\hat{F}{\hat{F}}^{T}={\hat{F}}^{T}\hat{F}=\hat{\delta}$.
The $\hat{\delta}$ is called the operator Kronecker symbol. 
It is being fashioned after the conventional symbol and best
visualized as
$<\chi^{0}\parallel{\hat{\delta}}^{(\tau,\nu,\beta)}_{(\rho,\omega,\gamma)}
\parallel\chi^{0}>=\delta^{(\tau,\nu,\beta)}
_{(\rho,\omega,\gamma)}=\delta^{\tau}_{\rho}\delta^{\nu}_{\omega}
\delta^{\beta}_{\gamma}$.
The transformations eq.(12.1) leave unchanged the matrix element of the norm of
operator-vector.The unification of homogeneous $\hat{F}$-transformations with
$12$-dimensional operator-transfers has reflected the properties of continuous
symmetries of operator manifold $\hat{G}(2.2.3)$. Meanwhile the matrix
elements of the constant operator $12$-vectors $\hat{a}\in\hat{G}(2.2.3)$
yield the constant $12$-vectors $a\in G(2.2.3)$.\\
The matrix element of eq.(12.1) yields corresponding transformations of
homogeneous wave group $\widetilde{F}$
\begin{equation}
\label{122}
\begin{array}{l}
<\chi^{0}\parallel\hat{\gamma'}\parallel\chi>=
<\chi^{0}\parallel\hat{F}\parallel\chi^{0}>
<\chi^{0}\parallel\hat{\gamma}\parallel\chi>=
\widetilde{F}
<\chi^{0}\parallel\hat{\gamma}\parallel\chi>, \\
<\chi\parallel\bar{\hat{\gamma'}}\parallel\chi^{0}>=
<\chi\parallel\bar{\hat{\gamma}}\parallel\chi^{0}>
<\chi^{0}\parallel{\hat{F}}^{T}\parallel\chi^{0}>=
<\chi\parallel\bar{\hat{\gamma}}\parallel\chi^{0}>
{\widetilde{F}}^{T}.
\end{array}
\end{equation}
It reads in component form
\begin{equation}
\label{123}
\begin{array}{l}
{\widetilde{F}}^{(\lambda,\mu,\alpha)}_{(\tau,\nu,\beta)}=
{\left(\S_{r^{\tau\nu}=1}^{N_{\tau\nu}}\bar{c}(r^{\tau\nu})\right)}^{-1}
\S_{r^{\lambda\mu}=1}^{N_{\lambda\mu}}\bar{c}(r^{\lambda\mu})
K^{(\lambda,\mu,\alpha)}_{(\tau,\nu,\beta)}(r^{\lambda\mu},r^{\tau\nu}), \\
{\widetilde{F}}_{(\lambda,\mu,\alpha)}^{(\tau,\nu,\beta)}=
{\left(\S_{r^{\tau\nu}=1}^{N_{\tau\nu}}{\bar{c}}^{*}(r^{\tau\nu})\right)}^{-1}
\S_{r^{\lambda\mu}=1}^{N_{\lambda\mu}}{\bar{c}^{*}}(r^{\lambda\mu})
K_{(\lambda,\mu,\alpha)}^{(\tau,\nu,\beta)}(r^{\tau\nu},r^{\lambda\mu}), \\
\end{array}
\end{equation}
provided
\begin{equation}
\label{124}
\begin{array}{l}
K^{(\lambda,\mu,\alpha)}_{(\tau,\nu,\beta)}(r^{\lambda\mu},r^{\tau\nu})=
<e'^{(\lambda,\mu,\alpha)}_{r^{\lambda\mu}},
e^{r^{\tau\nu}}_{(\tau,\nu,\beta)}>,\\ 
K_{(\lambda,\mu,\alpha)}^{(\tau,\nu,\beta)}(r^{\tau\nu},r^{\lambda\mu})=
(K^{T})_{(\lambda,\mu,\alpha)}^{(\tau,\nu,\beta)}=
<e'^{(\tau,\nu,\beta)}_{r^{\tau\nu}},
e^{r^{\lambda\mu}}_{(\lambda,\mu,\alpha)}>,
\end{array}
\end{equation}
which are known as the elements of homogeneous $\hat{K}$ group of continuous
symmetries of the manifold $G(2.2.3)$ [1-5]. 
Any homogeneous  $\hat{K}$-transformation
is defined by $12$ independent parameters, and the $12\times 12$-dimensional
matrix  $\hat{K}$ is orthogonal and unimodal.
In general the group $\hat{K}$ is decomposed into $8$ components of
connectedness $K^{(++)}_{+},K^{(++)}_{-},\\K^{(+-)}_{+},K^{(+-)}_{-},
K^{(-+)}_{+},K^{(-+)}_{-},K^{(--)}_{+},K^{(--)}_{-}$.
The sings $(+)$ and $(-)$ in the parentheses refer respectively to
orthochronous and non-orthochronous transformations, but the subscripts
$+$ and $-$ specify the special $(\mid K\mid=1)$ and non-special
$(\mid K\mid=-1)$ transformations. 
According to it, the operator group $\hat{F}$ and wave group $\widetilde{F}$
are decomposed into $8$ components of connectedness, respectively
$\hat{F}^{(++)}_{+},\hat{F}^{(++)}_{-},\hat{F}^{(+-)}_{+},\hat{F}^{(+-)}_{-},
\hat{F}^{(-+)}_{+},\hat{F}^{(-+)}_{-},\hat{F}^{(--)}_{+},\hat{F}^{(--)}_{-}$;\\
$\widetilde{F}^{(++)}_{+},\widetilde{F}^{(++)}_{-},\widetilde{F}^{(+-)}_{+},
\widetilde{F}^{(+-)}_{-},
\widetilde{F}^{(-+)}_{+},\widetilde{F}^{(-+)}_{-},\widetilde{F}^{(--)}_{+},
\widetilde{F}^{(--)}_{-}$. 
In particular, when the matrix K runs upon the homogeneous special group
$\widehat{SO}(6.6)$, hence corresponding matrices  $\hat{F}$ and
$\widetilde{F}$ run upon the homogeneous special groups
${\widehat{SO}(6.6)}_{\hat{F}}$ and ${\widehat{SO}(6.6)}_{\widetilde{F}}$. \\
Alongside with continuous transformations some discrete transformations
can be distinguished: \\
1) The positive co-contra transformation $\hat{P}_{+}$ is given by
\begin{equation}
\label{125}
({\hat{\gamma'}}^{(\lambda,\mu,\alpha)})=
{\hat{P}}_{+}
({\hat{\gamma}}_{(\lambda,\mu,\alpha)})=
({\hat{\gamma}}^{(\lambda,\mu,\alpha)}), 
\quad
({\hat{\gamma'}}_{(\lambda,\mu,\alpha)})=
({\hat{\gamma}}_{(\lambda,\mu,\alpha)})
{\hat{P}}_{+}^{T}=
({\hat{\gamma}}_{(\lambda,\mu,\alpha)}),
\end{equation}
provided ${\hat{P}}_{+}{\hat{P}}_{+}^{T}={\hat{P}}_{+}^{2}=1$.\\
2) The negative co-contra transformation $\hat{P}_{-}$ is defined as
${\hat{P}}_{-}=-{\hat{P}}_{+}$.\\
3) The total co-co or contra-contra transformations $\hat{P}_{+-}$ is in the
form ${\hat{P}}_{+-}={\hat{P}}_{+}{\hat{P}}_{-}=-1$.
The matrix elements yield corresponding diagonal
discrete transformations of wave group $\widetilde{P}$:
\begin{equation}
\label{126}
\begin{array}{ll}
\left({\widetilde{P}}_{\pm}\right)^{(\lambda,\mu,\alpha)}_{(\lambda,\mu,
\alpha)}=
<\chi^{0}\parallel
\left({\hat{P}}_{\pm}\right)^{(\lambda,\mu,\alpha)}_{(\lambda,\mu,\alpha)}
\parallel\chi^{0}>= \quad
{\left(\S_{r^{\lambda\mu}=1}^{N_{\lambda\mu}}{\bar{c}}^{*}(r^{\lambda\mu})
\right)}^{-1}
\S_{r^{\lambda\mu}=1}^{N_{\lambda\mu}}\bar{c}(r^{\lambda\mu})
{(I_{\pm})}^{(\lambda,\mu,\alpha)}_{(\lambda,\mu,\alpha)}(r^{\lambda\mu}),\\
\left({\widetilde{P}}_{\pm}\right)^{T(\lambda,\mu,\alpha)}_{(\lambda,\mu,
\alpha)}=
<\chi^{0}\parallel
\left({\hat{P}}_{\pm}\right)^{T(\lambda,\mu,\alpha)}_{(\lambda,\mu,\alpha)}
\parallel\chi^{0}>= \quad
{\left(\S_{r^{\lambda\mu}=1}^{N_{\lambda\mu}}\bar{c}(r^{\lambda\mu})
\right)}^{-1}
\S_{r^{\lambda\mu}=1}^{N_{\lambda\mu}}{\bar{c}}^{*}(r^{\lambda\mu})
{(I_{\pm}^{*})}^{(\lambda,\mu,\alpha)}_{(\lambda,\mu,\alpha)}
(r^{\lambda\mu}),
\end{array}
\end{equation}
where ${(I_{\pm})}^{(\lambda,\mu,\alpha)}_{(\lambda,\mu,\alpha)}
(r^{\lambda\mu})=
<e^{(\lambda,\mu,\alpha)},e_{(\lambda,\mu,\alpha)}>$
are the elements of the group $I$ of discrete transformations in $G(2.2.3)$.
Therewith $I_{+}$ is the particular special orthochron-orthochronous
$K^{(++)}_{+}$ transformation
\begin{equation}
\label{127}
\begin{array}{l}
{I_{+}}^{T}={I_{+}}^{*}={(I_{+})}^{+}, \quad {(I_{+})}^{2}=1, \quad
\mid I_{+}\mid=1, \\
({I_{+}})^{(2,1,\alpha)}_{(1,1,\alpha)} =
({I_{+}})^{(1,2,\alpha)}_{(2,2,\alpha)}=0, \quad
({I_{+}})^{(1,1,\alpha)}_{(1,1,\alpha)} =
({I_{+}})^{(2,2,\alpha)}_{(2,2,\alpha)}=1.
\end{array}
\end{equation}
Accordingly $I_{-}\in K^{(--)}_{-},\quad
I_{+-}\in K^{(--)}_{-}$.
It follows that
${\hat{P}}_{+}\in {\hat{F}}^{(++)}_{+}, \quad {\hat{P}}_{-}\in 
{\hat{F}}^{(--)}_{-},\quad
{\hat{P}}_{+-}\in {\hat{F}}^{(--)}_{-},
{\widetilde{P}}_{+}\in {\widetilde{F}}^{(++)}_{+}, \quad \
{\widetilde{P}}_{-}\in {\widetilde{F}}^{(--)}_{-},\quad
{\widetilde{P}}_{+-}\in {\widetilde{F}}^{(--)}_{-}$.
Thus the group of discrete transformations $\hat{P}$ and $\widetilde{P}$ are
consisted respectively of identical $\hat{E};\widetilde{E}$ and
${\hat{P}}_{+},{\hat{P}}_{-},{\hat{P}}_{+-}; {\widetilde{P}}_{+,}
{\widetilde{P}}_{-},{\widetilde{P}}_{+-}$ transformations. Each of the groups
$\hat{P}$ and $\widetilde{P}$ have only four single-valued irreducible
representations, which are all one-dimensional. 
The important property of Lie algebra of the groups
${\widehat{SO}(6.6)}_{\hat{F}}$ and ${\widehat{SO}(6.6)}_{\widetilde{F}}$
can be revealed. The infinitesimal generators
of ${\widehat{SO}(6.6)}_{\widetilde{F}}$, which are the vacuum expectation
of corresponding generators of ${\widehat{SO}(6.6)}_{\hat{F}}$,
subsequently may be reduced to independent vectors
satisfying the commutation relations of ordinary three-dimensional
rotation group ${SO(3)}_{\pm}$. 
Actually, in accordance with eq.(12.3), the infinitesimal generators
$({\hh_{i}}_{\alpha},{\HH_{i}}_{\alpha}) \quad(i=\eta,u;\alpha=1,2,3)$
of the group ${\widehat{SO}(6.6)}_{\hat{F}}$ and
$({\th_{i}}_{\alpha},{\TH_{i}}_{\alpha})$
of the group ${\widehat{SO}(6.6)}_{\widetilde{F}}$ take the form
\begin{equation}
\label{128}
\begin{array}{l}
{\th_{i}}_{\alpha}=\e1_{i}\otimes{\widetilde{h}}_{\alpha}=
<\chi^{0}\parallel{\hh_{i}}_{\alpha}\parallel\chi^{0}>=
\e1_{i}\otimes<\chi^{0}\parallel{\hat{h}}_{\alpha}\parallel\chi^{0}>, \\
{\TH_{i}}_{\alpha}=\e1_{i}\otimes{\widetilde{H}}_{\alpha}=
<\chi^{0}\parallel{\HH_{i}}_{\alpha}\parallel\chi^{0}>=
\e1_{i}\otimes<\chi^{0}\parallel{\hat{H}}_{\alpha}\parallel\chi^{0}>,
\end{array}
\end{equation}
at $r^{-}=r^{+}\equiv r$ provided
\begin{equation}
\label{129}
\begin{array}{l}
{\widetilde{h}}_{\alpha}=<\chi^{0}\parallel{\hat{h}}_{\alpha}\parallel
\chi^{0}>=
<\chi^{0}\parallel i{\hat{H}}^{[(-\alpha)(+\alpha)]}\parallel\chi^{0}>= \\
=i{\left(\S_{r=1}^{N}\bar{c}(r)\right)}^{-1}\left(\S_{r=1}^{N}\bar{c}(r)
\right)
H^{[(-\alpha)(+\alpha)]}=iH^{[(-\alpha)(+\alpha)]}=h_{\alpha}, \\
{\widetilde{H}}_{\alpha}=<\chi^{0}\parallel{\hat{H}}_{\alpha}\parallel
\chi^{0}>=
H_{\alpha}, \quad
\e1_{\eta}=\frac{1}{2} \left( \begin{array}{ll}
1  & 1 \\
1  & 1
\end{array}  \right), \quad
\e1_{u}=\frac{1}{2} \left( \begin{array}{ll}
1  & -1 \\
1  & -1
\end{array}    \right).
\end{array}
\end{equation}
Here the  $h_{\alpha}$ and $H_{\alpha} \quad(\alpha=1,2,3)$ are the
generators of the group $\widehat{SO}(3.3)$.
As far as the generators of the group $\widehat{SO}(6.6)$ obey the 
transpositional relations of the angular momenta,
then the generators of the groups
${\widehat{SO}(6.6)}_{\hat{F}}$ and ${\widehat{SO}(6.6)}_{\widetilde{F}}$
have undergone the same reduction. Due to it, the irreducible representation
$\hat{D}(p_{1},q_{1};p_{2},q_{2})$ of the group ${\widehat
{SO}(6.6)}_{\hat{F}}$, which can be treated as the reducible representation
$\hat{D}(p_{1})\otimes\hat{D}(q_{1})\otimes\hat{D}(p_{2})\otimes
\hat{D}(q_{2})$
of ordinary three-dimensional rotation group $SO(3)$, is decomposed into
irreducible representation of the group $SO(3)$.
\vskip 0.5truecm
\begin {center}{\Large {\bf Part III. Distorted Goyaks}}
\end {center}
\section {The Principle of Identity of Regular Goyaks}
\label {prid}
Next we extend
the scopes of treatment by considering the processes involving also
distorted goyaks. 
We start with formulation of {\em the principle of identity of regular
goyaks}, the main idea of which comes to the following:{\em since a 
generation of each regular goyak in the lowest state $(s_{0})$ is the 
result of its transition into that state from the arbitrary states
$(s,s',\ldots)$, where the goyaks are assumed to be distorted,
then all ordinary regular goyaks of $u$-type
are identical structures}.
Realizing a distortion of the basis $\{e_{(\lambda\alpha)}=O_{\lambda}
\otimes {\sigma}_{\alpha} \}$, the set of matrices
$\{D(\theta)=C(b)\otimes R(\theta)\}$ are the elements of local
distortion group (sec. 5,6).
This principle regarding the system of regular goyaks, which
is formed in the lowest state $(s_{0})$, holds only for the ordinary goyaks
of $u$-type. It is due to the fact that in this system there is only one
goyak of $\eta$-type (fundamental) and infinite number of $u$-type goyaks
(ordinary).
There are a number of advantages to the suggestion of the principle
of identity of regular goyaks. It is impressive to see later on
that this principle, underlies the most
important Gauge principle and the concepts of
unitary groups associating with the internal symmetries of elementary
particles.
\section {The Distorted Ordinary Goyak's Function} 
\label {disgoy}
Distorted ordinary goyak is described by means of link-function
\begin{equation}
\label{R141}
{\ps1_{u}}_{(\lambda\alpha)}(\theta_{+})=u_{(\lambda\alpha)}
{\ps1_{u}}_{\lambda}(\theta_{+}),\quad
{\ps1_{u}}^{(\lambda\alpha)}(\theta_{-})=u^{(\lambda\alpha)}
{\ps1_{u}}^{\lambda}(\theta_{-}),
\end{equation}
where $u_{(\lambda\alpha)}$ are distorted link-coordinates. We take for
granted that the distortion transformations right through the angles
$\theta_{+}(\theta_{(+k)},\quad k=1,2,3)$ and
$\theta_{-}(\theta_{(-k)})$ hold for goyaks
$({\ps1_{u}}_{\lambda}(\theta_{+}))$ and anti-goyaks
$({\ps1_{u}}^{\lambda}(\theta_{-}))$, respectively.
The function $\ps1_{u}(\theta)=({\ps1_{u}}_{\lambda}(\theta_{+}),
{\ps1_{u}}^{\lambda}(\theta_{-}))$ as the bi-spinor Fermi field,
can be derived from Dirac's wave equation of distorted goyak
(anti-goyak)(see appendix)
\begin{equation}
\label{R142}
\begin{array}{l}
\left[ i{\g1_{u}}^{(\lambda\alpha)}(\theta)({\pr_{u}}_{(\lambda\alpha)}-
{\Gam_{u}}_{(\lambda\alpha)}(\theta))-m \right] \ps1_{u}(\theta)=0,\\
\bar{\ps1_{u}}(\theta)
\left[ i({\lpr_{u}}_{(\lambda\alpha)}-{\bar{\Gam_{u}}}_{(\lambda\alpha)}
(\theta))
{\g1_{u}}^{(\lambda\alpha)}(\theta)-m \right]=0,
\end{array}
\end{equation}
where
${\g1_{u}}^{(\lambda\alpha)}(\theta)={\V_{u}}^{(\lambda\alpha)}_{(\tau\beta)}
(\theta){\gamma}^{(\tau\beta)}_{0}, 
\quad
{\gamma}^{(\pm\alpha)}_{0}=\FFr{1}{\sqrt{2}}(\gamma^{0}\sigma^{\alpha}
\pm \gamma^{\alpha})$, 
$\gamma^{0},\gamma^{\alpha}$ , as usual, are Dirac's matrices. The quantities
${\Gam_{u}}_{(\lambda\alpha)}(\theta),({\bar{\Gam_{u}}}_{(\lambda\alpha)}
(\theta))$
are expressed by Ricci rotation coefficients.
To save writing, henceforth within this section we abbreviate the
indices $(\lambda\alpha)$ by the single symbol $\mu$, and Latin indices
$(im)\quad(i=\pm, m=1,2,3)$ by $i$. 
Also we denote ${\p1_{u}}_{\mu}\equiv p_{\mu}, \quad
{\Gam_{u}}_{\mu}\equiv \Gamma_{\mu}, \quad
\ps1_{u}\equiv \Psi, \quad
\g1_{u}\equiv \gam $.
Then the equations (14.2) give rise to
\begin{equation}
\label{R143}
\begin{array}{l}
\left[ \gamma^{\mu}(\theta)(\hp1_{\mu}-
i\Gamma_{\mu}(\theta))-m \right] \Psi(\theta)=0,\\
\bar{\Psi}(\theta)
\left[ (\hp1_{\mu}-i{\bar{\Gamma}}_{\mu}(\theta))
\gamma^{\mu}(\theta)-m \right]=0.
\end{array}
\end{equation}
In order to solve eq.(14.3), it is advantageous to transform it
into the following one
\begin{equation}
\label{R144}
\{-\partial^{2}-m^{2}-{(\gamma\Gamma)}^{2}+2(\Gamma\gamma)+
(\gamma\partial)(\gamma\Gamma) \}\Psi=0,
\end{equation}
provided
\begin{equation}
\label{R145}
\begin{array}{l}
g^{\mu\nu}({\hat{p}}_{\mu}-i\Gamma_{\mu})({\hat{p}}_{\nu}-i\Gamma_{\nu})=
-\partial^{2}-\Gamma^{2}+2(\Gamma\partial)+(\partial\Gamma), \\
\FFr{1}{2}\sigma^{\mu\nu}F_{\mu\nu}=(\gamma\partial)(\gamma\Gamma)-
(\partial\Gamma),\quad (\gamma\partial)(\gamma\Gamma)=
\gamma^{\mu}\gamma^{\nu}\partial_{\mu}\Gamma_{\nu}, \\
\FFr{1}{2}\sigma^{\mu\nu}[\Gamma_{\mu},\Gamma_{\nu} ]=
{(\gamma\Gamma)}^{2}- \Gamma^{2}, \quad
\partial^{2}=\partial^{\mu}\partial_{\mu}, \quad
\Gamma^{2}=\Gamma^{\mu}\Gamma_{\mu}, \\
\gamma^{\mu}\gamma^{\nu}=g^{\mu\nu}+\sigma^{\mu\nu},
\quad
2g^{\mu\nu}=\{ \gamma^{\mu},\gamma^{\nu} \},\quad
2\sigma^{\mu\nu}= [\gamma^{\mu},\gamma^{\nu} ],\quad
\F_{\mu\nu}=\partial_{\mu}\Gamma_{\nu}-
\partial_{\nu}\Gamma_{\mu}.
\end{array}
\end{equation}
We are looking for the solution given in the form
$\Psi=e^{-ipu}F(\varphi)$,
where $p_{\mu}$ is a constant six-vector $pu=p_{\mu}u_{\mu}$.
Next we take for granted that the field of distortion is switched on
at $u_{0}=-\infty $ infinitely slowly. Then the function $\Psi$ must match
onto the wave function of ordinary regular goyak. Smoothness of the match
requires that the numbers $p_{\mu}$ become the components of link-momentum
of regular goyak and satisfy the boundary condition
$p_{\mu}p_{\mu}=m^{2}=p^{2}_{\eta}$.
Due to this requirement, we cancel unwanted solutions,
and also clear up the normalization of wave functions 
\begin{equation}
\label{R146}
\int\Psi^{*}_{p'}\Psi_{p}d^{3}u=\int{\bar{\Psi}}_{p'}\gamma^{0}\Psi_{p}
d^{3}u={(2\pi)}^{3}\delta(\vec{p'}-\vec{p}).
\end{equation}
At $\sqrt{-g}\neq1 $ the gradient of the function $\varphi $ takes the form
$\partial_{\mu}\varphi=V^{i}_{\mu}k_{i}, \quad
\partial^{\mu}\varphi=V_{i}^{\mu}k^{i}$,
where $k_{i}$ are arbitrary constant numbers satisfying the condition
$k_{i}k_{i}=0$.
The $V^{i}_{\mu}(V_{i}^{\mu})$ are congruence parameters of curves (Latin
indices refer to tetrad components). Thus
$V_{i}^{\mu}V^{j}_{\mu}=g^{\mu\nu}V_{\nu i}V^{j}_{\mu}=
{(g_{0})}^{j}_{i}={\delta}^{j}_{i}$ and
$\partial^{\mu}\varphi\partial_{\mu}\varphi=
(V_{i}^{\mu}V^{j}_{\mu})k^{i}k_{j}=0$.
A substitution in eq.(14.4) gives rise to $F'=A(\theta)F$,
where $(\cdots)'$ stands for the derivative with respect to $\varphi$, and
\begin{equation}
\label{R147}
A(\theta)=\FFr{2i(p\Gamma) + m^{2}-p^{2}+{(\gamma\Gamma)}^{2}-
(\gamma\partial)(\gamma\Gamma)}{2i(kVp)-(kDV)};
\end{equation}
\begin{equation}
\label{R148}
\begin{array}{ll}
(kVp)=k^{i}V_{i}^{\mu}p_{\mu},\quad
(kDV)=k^{i}D_{\mu}V_{i}^{\mu}, 
\quad
D_{\mu}=\partial_{\mu}-2\Gamma_{\mu}, \\
p^{2}=p^{\mu}p_{\mu}=g^{\mu\nu}(\theta)p_{\mu}p_{\nu}\quad
(kVdu)=k_{i}V^{i}_{\mu}du^{\mu}.
\end{array}
\end{equation}
Here we are interested in the right-handed eigen-vectors $F_{r}\quad
(r=1,2,3,4)$ corresponding to eigen-values $\mu_{r}$ of matrix $A:
AF_{r}=\mu_{r}F_{r}$. They are the roots of polynomial characteristic
equation $c(\mu)=\left\|(\mu I-A)\right\|=0$.
Thus, one gets $F'_{r}=\mu_{r}F_{r}$.
We may think of the function $F$ as being the product
$F=\displaystyle \prod^{4}_{r=1}F_{r}$,
hence
$(\ln F)'=\S^{4}_{r=1}\mu_{r}=trA$,
where the trace of matrix $A$ is denoted by $trA$. 
Then $(\ln F)'=X_{R}(\theta)-iX_{J}(\theta)$,
provided
\begin{equation}
\label{R149}
\begin{array}{l}
X_{R}(\theta)=
trA_{R}(\theta)=tr \left\{ \FFr{-(kDV)\left[ m^{2}-p^{2}+
{(\gamma\Gamma)}^{2}-(\gamma\partial)(\gamma\Gamma)\right]+4(kVp)
(p\Gamma)}{(kDV)^{2}+4(kVp)^{2}} \right\},\\
X_{J}(\theta)=
trA_{J}(\theta)=2tr \left\{ \FFr{(kVp)\left[ m^{2}-p^{2}+
{(\gamma\Gamma)}^{2}-
(\gamma\partial)(\gamma\Gamma)\right]+(kDV)(p\Gamma)}{(kDV)^{2}+
4(kVp)^{2}} \right\}.
\end{array}
\end{equation}
One infers at once that the solution clearly is
\begin{equation}
\label{R1410}
F(\theta)=C{\left( \frac{m}{E_{u}} \right)}^{1/2}u\exp
\{\chi_{R}(\theta)-i\chi_{J}(\theta)\},
\end{equation}
where $C$ is a normalization constant, $u$ is a constant bi-spinor, and
\begin{equation}
\label{R1411}
\chi_{R}(\theta)=\IIn^{u^{\mu}}_{0}(kVdu)X_{R}(\theta), 
\quad
\chi_{J}(\theta)= \IIn^{u^{\mu}}_{0}(kVdu)X_{J}(\theta).
\end{equation}
So define
\begin{equation}
\label{R1412}
\Psi(\theta)\Rightarrow {\ps1_{u}}_{\lambda}(\theta_{+k})=
\left.{\bar{\ps1_{u}}}^{\lambda}(\theta_{-k})
\right|_{\theta_{-k}=\theta_{+k}},
\quad
\bar{\Psi}(\theta)\Rightarrow \bar{{\ps1_{u}}}_{\lambda}(\theta_{+k})=
\left.{\ps1_{u}}^{\lambda}(\theta_{-k})
\right|_{\theta_{-k}=\theta_{+k}},
\end{equation}
then
\begin{equation}
\label{R1413}
{\ps1_{u}}_{\lambda}(\theta_{+k})=F(\theta_{+k})e^{-ipu}=
f_{(+)}(\theta_{+k}){\ps1_{u}}_{\lambda}, 
\quad
{\ps1_{u}}^{\lambda}(\theta_{-k})=
{\ps1_{u}}^{\lambda}f_{(-)}(\theta_{-k}),
\end{equation}
provided by the transformation functions
\begin{equation}
\label{R1414}
\begin{array}{l}
f_{(+)}(\theta_{+k})=Ce^{\chi_{R}(\theta_{+k})-
i\chi_{J}(\theta_{+k})}, \\
f_{(-)}(\theta_{-k})=\left.f_{(+)}^{*}(\theta_{+k}) 
\right|_{\theta_{+k}=\theta_{-k}}=
Ce^{\chi_{R}(\theta_{-k})+i\chi_{J}(\theta_{-k})}.
\end{array}
\end{equation}
The ${\ps1_{u}}_{\lambda}({\ps1_{u}}^{\lambda})$ is the plane wave function 
of regular ordinary goyak.
It was assumed that constant bi-spinor $u$ coincides with the amplitude
of plane wave, just because of requirement of boundary smoothness, that
${\ps1_{u}}^{J}(\theta)$ must match onto the wave function of ordinary
regular goyak at $u_{0}=-\infty $.
The inverse transformations at fixed $(k)$ are given by
\begin{equation}
\label{R1415}
{\ps1_{u}}_{\lambda}=f_{(+)}^{-1}(\theta_{+k})
{\ps1_{u}}_{\lambda}(\theta_{+k}), 
\quad
{\ps1_{u}}^{\lambda}=
{\ps1_{u}}^{\lambda}(\theta_{-k})f_{(-)}^{-1}(\theta_{-k}).
\end{equation}
\section {Quarks and Color Confinement Principle} 
\label {Color}
Until now we agreed to consider only a special stable system of regular
goyaks, which comes into being in the lowest state $(s_{0})$. As it was
seen, it consists of one $\eta$-type goyak and infinite number of $u$-type
goyaks. They realize the geometry $G(2.2.3)$ by satisfying the necessary
requirements. But such kind of realization
is a trivial one, which gives rise to geometry without particles
and interactions.
However, there is still an other a quite different choice of realization
of geometry, which subsequently leads to geometry  $G(2.2.3)$, with
the particles and interactions, by the following recipe: Everything
said in section 7 will then remain valid provided we make the simple
changes. We admit, that the distorted ordinary goyaks
have took participation in the realization of geometry
instead of regular ordinary goyaks.
The ordinary laws regarding these changes apply that we must make
use of localized wave packets of distorted ordinary goyaks
\begin{equation}
\label{R151}
\begin{array}{l}
\hps_{u}(\theta)=\S_{\pm s}\int\frac{d^{3}p_{u}}{{(2\pi)}^{3/2}}
\left( {\hgam_{u}}^{(+\alpha)}_{k}{\ps1_{u}}_{(+\alpha)}(\theta_{+k})+
{\hgam_{u}}^{(-\alpha)}_{k}{\ps1_{u}}_{(-\alpha)}(\theta_{+k})\right), \\
\bar{\hps_{u}}(\theta)=\S_{\pm s}\int\frac{d^{3}p_{u}}{{(2\pi)}^{3/2}}
\left( {\hgam_{u}}_{(+\alpha)}^{k}{\ps1_{u}}^{(+\alpha)}(\theta_{-k})+
{\hgam_{u}}_{(-\alpha)}^{k}{\ps1_{u}}^{(-\alpha)}(\theta_{-k})\right),
\end{array}
\end{equation}
where as usual the summation is extended over all dummy indices, and
\begin{equation}
\label{R152}
\begin{array}{l}
{\ps1_{u}}_{(+\alpha)}(\theta_{+k})=u_{(+\alpha)}
{\ps1_{u}}_{+}(\theta_{+k}), \quad
{\ps1_{u}}_{(-\alpha)}(\theta_{+k})=u_{(-\alpha)}
{\ps1_{u}}_{-}(\theta_{+k}), \\
{\ps1_{u}}^{(+\alpha)}(\theta_{-k})=u^{(+\alpha)}
{\ps1_{u}}^{+}(\theta_{-k}), \quad
{\ps1_{u}}^{(-\alpha)}(\theta_{-k})=u^{(-\alpha)}
{\ps1_{u}}^{-}(\theta_{-k}).
\end{array}
\end{equation}
The matrix elements of anti-commutators
of generalized expansion
coefficients take the form
\begin{equation}
\label{R153}
\begin{array}{l}
<\chi_{-}\mid \{ {\hgam_{u}}^{(+\alpha)}_{k}(p,s),
{\hgam_{u}}_{(+\beta)}^{k'}(p',s') \}\mid\chi_{-}>= \\
=<\chi_{+}\mid \{{\hgam_{u}}_{(+\beta)}^{k'}(p',s'),
{\hgam_{u}}^{(+\alpha)}_{k}(p,s)\}\mid\chi_{+}>=
-\delta_{ss'}\delta_{\alpha\beta}\delta_{kk'}\delta^{3}(\vec{p}-\vec{p'}).
\end{array}
\end{equation}
The expressions of causal Green's functions
of distorted ordinary goyaks and their generalization
can be readily written 
\begin{equation}
\label{R154}
\begin{array}{l}
{\G1_{u}}^{\theta}_{F}(\theta_{+}-\theta_{-})=-i
<\chi_{-}\mid
T{\hps_{u}}(\theta_{+}){\bar{\hps_{u}}}
(\theta_{-})\mid\chi_{-}>_{0}/(u_{+}u_{-})_{0}= \\
=-i\IIn\FFr{d^{3}p_{u}}{{(2\pi)}^{3/2}}
{\ps1_{u}}_{+p}(\theta_{+k})
{\bar{\ps1_{u}}}_{+p}(\theta_{-k})\theta(u^{0}_{+}-u^{0}_{-})+ 
i\IIn\FFr{d^{3}p_{u}}{{(2\pi)}^{3/2}}
{\bar{\ps1_{u}}}_{-p}(\theta_{-k})
{\ps1_{u}}_{-p}(\theta_{+k})
\theta(u^{0}_{-}-u^{0}_{+});
\end{array}
\end{equation}
and
${\TG_{u}}^{\theta}_{F}(\theta_{+}-\theta_{-})
=\S_{r^{\lambda}=1}^{N_{\lambda}}{\mid\bar{c}(r^{\lambda})\mid}^{2}
{\G1_{u}}^{r^{\lambda}}_{\lambda}(\theta_{+}-\theta_{-})$,
provided
${\G1_{u}}^{r^{\lambda}}_{\lambda}(\theta_{+}-\theta_{-})=-iT
({\ps1_{u}}^{r^{\lambda}}_{\lambda}(\theta_{+})
{\bar{\ps1_{u}}}^{r^{\lambda}}_{\lambda}(\theta_{-}))$,
where $\theta_{\pm}(\eta_{\pm},u_{\pm}).$
Geometry realization requirement now must be satisfied for each
ordinary goyak in terms of
\begin{equation}
\label{R155}
{\G1_{u}}^{\theta}_{F}(0)={\G1_{\eta}}_{F}(0) \quad \mbox{or}\quad
{\TG_{u}}^{\theta}_{F}(0)={\TG_{\eta}}_{F}(0).
\end{equation}
They are valid in the case if following relations hold for
each distorted ordinary goyak:
\begin{equation}
\label{R156}
\begin{array}{l}
\S_{k}{\ps1_{u}}_{+p}(\theta_{+k}(\eta ,u))
{\bar{\ps1_{u}}}_{+p}(\theta_{-k}(\eta ,u))= \\
=\S_{k}{\ps1_{u}}'_{+p}(\theta'_{+k}(\eta ,u))
{\bar{\ps1_{u}}}'_{+p}(\theta'_{-k}(\eta ,u))=
\cdots = {\ps1_{\eta}}_{+p}(\eta)\bar{{\ps1_{\eta}}}_{+p}(\eta),
\end{array}
\end{equation}
and
\begin{equation}
\label{R157}
\begin{array}{l}
\S_{k}{\bar{\ps1_{u}}}_{-p}(\theta_{-k}(\eta ,u))
{\ps1_{u}}_{-p}(\theta_{+k}(\eta ,u))= \\
=\S_{k}{\bar{\ps1_{u}}}'_{-p}(\theta'_{-k}(\eta ,u))
{\ps1_{u}}'_{-p}(\theta'_{+k}(\eta ,u))=
\cdots =\bar{{\ps1_{\eta}}}_{-p}(\eta){\ps1_{\eta}}_{-p}(\eta).
\end{array}
\end{equation}
That is, {\em  distorted ordinary goyaks are being met in the special 
permissible
combinations to realize the geometry}. 
One final observation is worth recording.
The following relations hold for the distortion
transformation functions $f_{(\pm)}(\theta_{\pm})$:
\begin{equation}
\label{R158}
\begin{array}{l}
\S_{k}f_{(+)}(\theta_{+k})f_{(-)}(\theta_{-k} )=
\S_{k}f'_{(+)}(\theta'_{+k})f'_{(-)}(\theta'_{-k} )=
\cdots =inv,\\
\S_{k}f_{(-)}(\theta_{-k})f_{(+)}(\theta_{+k} )=
\S_{k}f'_{(-)}(\theta'_{-k})f'_{(+)}(\theta'_{+k} )=
\cdots =inv.
\end{array}
\end{equation}
We have arrived at a quite promising and entirely satisfactory
proposal to answer some of the original questions,
namely those: what is the essential content of our notion
of quarks and quantum number of color? What is the physical origin of them
and especially the Color Confinement principle? To answer these queries below,
in special case of local angles $\theta_{\pm k}(\eta,u)$ with $(k)$ running
from $1$ to $3$, we may think of the function
${\ps1_{u}}_{\lambda}(\theta_{+k}(\eta,u))$ at fixed $(k)$ as being 
$u$-component of bi-spinor field of "quark" ${\hat{q}}_{k}$, and of
${\ps1_{u}}^{\lambda}(\theta_{-k}(\eta,u))$ - an $u$-component of conjugate
bi-spinor field of "anti-quark" ${\bar{\hat{q}}}_{k}$.
Thus, {\em the local rotations through the angles $\theta_{+k}(\eta,u)$ and
$\theta_{-k}(\eta,u)$ yield the quarks and anti-quarks, respectively}.
The index $(k)$ refers to so-called "color" degrees of freedom in the 
case of rotations through the angles  $\theta_{+k}(\eta,u)$ and "anti-color" 
degrees of freedom in the case of $\theta_{-k}(\eta,u)$.
Here we leave the $\eta$-components of quark bi-spinor fields implicit, which
are plane waves.
Surely, one may readily perform a transition to conventional quark fields
$\Psi_{f}(x_{f})$ on $M^{4}$ (sec. 5.6).
Thus,{\em a quark is a fermion with the half-integral spin and certain color
degree of freedom}. As it was seen there are exactly three colors and the
usual nomenclature is: ${\hat{q}}_{1}$=red quark, ${\hat{q}}_{2}$=
green quark, ${\hat{q}}_{3}$ =blue quark.
Due to eq.(15.8), one gets
\begin{equation}
\label{R159}
\begin{array}{l}
\S_{k}{\hat{q}}_{kp}{\bar{\hat{q}}}_{kp} =
\S_{k}{\hat{q'}}_{kp}{\bar{\hat{q'}}}_{kp}=
\cdots=inv={\ps1_{\eta}}_{+p}(\eta){\bar{\ps1_{\eta}}}_{+p}(\eta), \\
\S_{k}{\bar{\hat{q}}}_{-kp}{\hat{q}}_{-kp}=
\S_{k}{\bar{\hat{q'}}}_{-kp}{\hat{q'}}_{-kp}=
\cdots=inv={\bar{\ps1_{\eta}}}_{-p}(\eta){\ps1_{\eta}}_{-p}(\eta),
\end{array}
\end{equation}
which utilize the whole idea of {\em Color(Quark) Confinement principle}.
According to it {\em the quarks act in the special combinations
of "color singlets" to realize the geometry}. \\
Only two color singlets are available (see section 16.3)
\begin{equation}
\label{R1510}
(q\bar{q})=\FFr{1}{\sqrt{3}}\delta_{kk'}{\hat{q}}_{k}{\bar{\hat{q}}}_{k'}=
inv, 
\quad
(qqq)=\FFr{1}{\sqrt{6}}\varepsilon_{klm}{\hat{q}}_{k}{\hat{q}}_{l}
{\hat{q}}_{m}=inv.
\end{equation}
If this picture of the color confinement turns out to reflect
the actual situation in nature, quarks would be fields without free particles
associated to them. Color would be confined, and the spectrum of hadrons
(permissible combinations of quarks) would emerge as the spectrum of the
color singlet states. Unwanted states (since not seen) like quarks $(q)$,
(color triplets) or diquarks $(qq)$ (color sextets) are eliminated by
construction at the very beginning. 
\section {Gauge Principle; Internal Symmetries}
\label {gauge}
As it was seen, the principle of identity holds for ordinary
regular goyaks, according to it each goyak in the lowest state can be
regarded as a result of transition into the state $(s_{0})$ from an
arbitrary state, in which the goyaks assumed to be distorted. 
This is succinctly stated below
\begin{equation}
\label{R161}
\begin{array}{l}
{\ps1_{u}}_{\lambda}=f^{-1}_{(+)}(\theta_{+k})
{\ps1_{u}}_{\lambda}(\theta_{+k})=
f^{-1}_{(+)}(\theta'_{+l})
{\ps1'_{u}}_{\lambda}(\theta'_{+l})=
\cdots, \\
{\ps1_{u}}^{\lambda}=
{\ps1_{u}}^{\lambda}(\theta_{-k})f^{-1}_{(-)}(\theta_{-k})=
{\ps1_{u}}'^{\lambda}(\theta'_{-l})f^{-1}_{(-)}(\theta'_{-l})=
\cdots \\
\end{array}
\end{equation}
Hence the following transformations may be implemented on the
distorted ordinary goyaks: 
\begin{equation}
\label{R162}
\begin{array}{l}
{\ps1_{u}}'_{\lambda}(\theta'_{+l})=
f_{(+)}(\theta'_{+l}){\ps1_{u}}_{\lambda}=
f_{(+)}(\theta'_{+l})f^{-1}_{(+)}(\theta_{+k})
{\ps1_{u}}_{\lambda}(\theta_{+k}),\\
{\ps1_{u}}'^{\lambda}(\theta'_{-l})=
{\ps1_{u}}^{\lambda}f_{(-)}(\theta'_{-l})=
{\ps1_{u}}^{\lambda}(\theta_{-k})f^{-1}_{(-)}(\theta_{-k})
f_{(-)}(\theta'_{-l}).
\end{array}
\end{equation}
The transformations take
the standard form
\begin{equation}
\label{R163}
\begin{array}{l}
{\ps1_{u}}'_{\lambda}(\theta'_{+l})=
f^{(+)}_{lk}{\ps1_{u}}_{\lambda}(\theta_{+k})=
f(\theta'_{+l},\theta_{+k}){\ps1_{u}}_{\lambda}(\theta_{+k}),\\
{\ps1_{u}}'^{\lambda}(\theta'_{-l})=
{\ps1_{u}}^{\lambda}(\theta_{-k})f^{(-)}_{kl}=
\left.{\ps1_{u}}^{\lambda}(\theta_{-k})f^{*}(\theta'_{-l},\theta_{-k})
\right|_{\begin{array}{l}
\theta'_{-l}=\theta'_{+l}\\
\theta_{-k}=\theta_{+k},
\end{array}}
\end{array}
\end{equation}
provided
\begin{equation}
\label{R164}
\begin{array}{l}
f^{(+)}_{lk}=\exp \{ \chi^{R}_{lk}-i\chi^{J}_{lk} \}, 
\quad f^{(-)}_{kl}={(f^{(+)}_{lk})}^{*},\\
\chi^{R}_{lk}=\chi_{R}(\theta'_{+l})-\chi_{R}(\theta_{+k}),
\quad
\chi^{J}_{lk}=\chi_{J}(\theta'_{+l})-\chi_{J}(\theta_{+k}).
\end{array}
\end{equation}
The transformation functions are regarded as the operators in the space 
of internal degrees of freedom labeled by $(\pm k)$, corresponding to 
distortion rotations around the axes $(\pm k)$ by the
angles  $\theta_{\pm k}(\eta,u)$. 
We make proposition that {\em local distortion rotations through the angles
$\theta_{\pm k}(\eta,u)$ with different $(k)$ are incompatible}. \\
It is clear that because of incompatibility they can not be realized
simultaneously. Hence the operators $f^{(\pm)}_{lk}$  with different $(lk)$ 
cannot have simultaneous
eigen-states, i.e. measurements on their observables do not yield definite
values (they "spread"). 
Transformation operators $f^{(\pm)}_{lk}$ obey the commutation relations of 
{\em incompatibility} of distortion rotations
\begin{equation}
\label{R165}
\begin{array}{l}
f^{(+)}_{lk}f^{(+)}_{cd}-f^{(+)}_{ld}f^{(+)}_{ck}=\|f^{(+)}\|
\varepsilon_{lcm}\varepsilon_{kdn}f^{(-)}_{nm},\\
f^{(-)}_{kl}f^{(-)}_{dc}-f^{(-)}_{dl}f^{(-)}_{kc}=\|f^{(-)}\|
\varepsilon_{lcm}\varepsilon_{kdn}f^{(+)}_{mn},
\end{array}
\end{equation}
where $l,k,c,d,m,n=1,2,3$.
As far as distorted ordinary goyaks have took
participation in the realization of geometry $G(2.2.3)$ instead of
regular ones,  
the principle of identity of regular goyaks directly leads to the equivalent
statement, which is known in physics under the name of {\em Gauge Principle:
an action integral of any physical system, defined in flat manifold
$G(2.2.3)$, must be invariant under arbitrary transformations eq.(16.3)}. 
This principle is valid for any physical system, which can be treated as a
definite system of distorted ordinary goyaks with arbitrary internal
degrees of freedom. 
In accordance with eq.(9.2) the total bi-spinor field of
distorted goyak, which realizes the representation
$D(\frac{1}{2},0)\otimes D(0,\frac{1}{2})\otimes
D(\frac{1}{2},0)\otimes D(0,\frac{1}{2})$ of the group $\widehat{SO}(6.6)$,
reads in component form
$\Psi_{k}(\zeta)=\ps1_{\eta}(\eta){\ps1_{u}}_{\lambda}(\theta_{+k}), 
\quad
{\bar{\Psi}}_{k}(\zeta)={\bar{\ps1_{u}}}_{\lambda}(\theta_{+k})
\bar{\ps1_{\eta}}(\eta),$
where $\ps1_{\eta}(\eta)$ is simply plane wave bi-spinor. 
Taking into account eq.(16.3), we get the equivalent transformations 
implemented on the total fields
\begin{equation}
\label{R166}
\Psi'_{l}(\zeta)=
f(\theta'_{+l}(\zeta),\theta_{+k}(\zeta))\Psi_{k}(\zeta),
\quad
{\bar{\Psi'}}_{l}(\zeta)=
{\bar{\Psi}}_{k}(\zeta)f^{*}(\theta'_{-l}(\zeta),\theta_{-k}(\zeta)).
\end{equation}
They can be written succinctly 
\begin{equation}
\label{R167}
\Psi'(\zeta)=U(\theta(\zeta))\Psi(\zeta), 
\quad
\bar{\Psi'}(\zeta)=\bar{\Psi}(\zeta)U^{+}(\theta(\zeta)).
\end{equation}
Matrix notation is employed in eq.(16.7):
$\Psi=\{ \Psi_{k} \}, \quad U(\theta)= \{ f(\theta'_{+l},\theta_{+k}) \} $.
Particularly, in the case of local transformations through the angles
$\theta_{\pm k}(\zeta)$ with the color index $(k)$ running from $1$ to
$3$, the transformations eq.(16.7) yield the following ones implemented
on the quark fields defined on manifold $G(2.2.3)$:
\begin{equation}
\label{R168}
{\hat{q'}}_{l}(\zeta)=
f(\theta'_{+l}(\zeta),\theta_{+k}(\zeta)){\hat{q}}_{k}(\zeta),
\quad
{\bar{\hat{q}}}'_{l}(\zeta)=
{\bar{\hat{q}}}_{k}(\zeta)f^{*}(\theta'_{-l}(\zeta),\theta_{-k}(\zeta)),
\end{equation}
or in matrix notation
\begin{equation}
\label{R169}
\hat{q'}(\zeta)=U(\theta(\zeta))\hat{q}(\zeta), 
\quad
\bar{\hat{q'}}(\zeta)=\bar{\hat{q}}(\zeta)U^{+}(\theta(\zeta)),
\end{equation}
where $\hat{q}(\zeta)=\{ {\hat{q}}_{k}(\zeta) \}=
\{ \ps1_{\eta}(\eta){\hat{q}}_{k} \}$. 
As we will see below, due to the incompatibility commutation relations
(16.5), the transformation matrices $\{ U \}$  generate
the unitary group of internal symmetries $U(1), SU(2),$ 
$SU(3)$. 
In order to utilize a gauge principle in its concrete
expression, below we discuss different possible models.
\subsection {The Local Group $U^{loc}(1)$ of Electromagnetic Interactions}
At the beginning we discuss the most simple case of one-dimensional local
transformations, through the local angles $\theta_{+1}(\zeta)$ and
$\theta_{-1}(\zeta)$
\begin{equation}
\label{R1610}
f^{(+)}=\left( \begin{array}{lll}
f^{(+)}_{11}  & 0  & 0 \\ 
0             & 1  & 0 \\
0             & 0  & 1
\end{array} \right), \quad
f^{(-)}={(f^{(+)})}^{+}.
\end{equation}
The commutation relations (16.5) of incompatibility of distortion rotations
reduced to identity $f^{(+)}_{11}=\|f^{(+)}\|.$
In considered case 
$\chi_{R}(\theta_{+1})=\chi_{R}(\theta_{-1})$,
and
$f^{(+)}_{11}=f_{(+)}(\theta_{+1})f_{(-)}(\theta_{-1})=
f(\theta_{+1},\theta_{-1}).$
Combining eq.(16.4) and eq.(16.9), it holds
\begin{equation}
\label{R1611}
\Psi'(\zeta)=U(\theta)\Psi(\zeta), 
\quad
\bar{\Psi'}(\zeta)=\bar{\Psi}(\zeta)U^{*}(\theta),
\end{equation}
provided by the transformation function
\begin{equation}
\label{R1612}
f^{(+)}_{11}=U(\theta)=f(\theta_{+1}(\zeta),\theta_{-1}(\zeta))=
\exp \{ -i\chi^{(+)}_{J}(\theta_{+1})+i\chi^{(-)}_{J}(\theta_{-1}) \}.
\end{equation}
It may give rise to
$U(\theta)=e^{-i\theta}$,
where
$\theta \equiv \chi^{(+)}_{J}(\theta_{+1})-\chi^{(-)}_{J}(\theta_{-1})$.
The strength of interaction is specified by a single coupling $Q$,
which is called electrical charge. 
A set of transformations generates a commutative
Abelian unitary local group of electromagnetic interactions  $U^{loc}(1)$.
That is, Lie group $G$ is realized as a $G= U^{loc}(1)= SO^{loc}(2)$, with
one-dimensional trivial algebra $\hat{g}_{1}=R^{1}$. The invariance under
the local group $U^{loc}(1)$ leads to electromagnetic field, the massless
quanta of which - photons are {\em electrically neutral}, just because of the
conditions (15.8) and eq.(16.12):
\begin{equation}
\label{R1613}
f(\theta_{+1},\theta_{-1})=f(\theta'_{+1},\theta'_{-1})=
\cdots inv.
\end{equation}
\subsection {Unitary Local Group $SU^{loc}(2)$ of Weak Interactions}
Next we consider a particular case of two-dimensional local transformations
through the angles $\theta_{\pm m}(\zeta)$ around two axes $(m=1,2)$.
The matrix function of transformation is written down
\begin{equation}
\label{R1614}
f^{(+)}=\left( \begin{array}{lll}
f^{(+)}_{11}  & f^{(+)}_{12}  & 0 \\ 
f^{(+)}_{21}    & f^{(+)}_{22}  & 0 \\
0               & 0             & 1
\end{array} \right), \quad
f^{(-)}={(f^{(+)})}^{+}.
\end{equation}
So, the commutation relations (16.5) of incompatibility of distortion 
rotations give rise to non-trivial conditions
\begin{equation}
\label{R1615}
\begin{array}{ll}
f^{(+)}_{11}=\left\| f^{(+)}\right\| {(f^{(+)}_{22})}^{*}, \qquad
f^{(+)}_{21}=-\left\| f^{(+)}\right\| {(f^{(+)}_{12})}^{*},\\
f^{(+)}_{12}=-\left\| f^{(+)}\right\| {(f^{(+)}_{21})}^{*},\quad
f^{(+)}_{22}=\left\| f^{(+)}\right\| {(f^{(+)}_{11})}^{*},
\end{array} 
\end{equation}
Hence $\left\| f^{(+)}\right\|=1.$ One can readily infer the matrix 
$U(\theta)$ of gauge transformations of collection of fundamental fields
\begin{equation}
\label{R1616}
U=e^{-i\vec{T}\vec{\theta}}=
\left( \begin{array}{ll}
f^{(+)}_{11}  & f^{(+)}_{12}  \\ 
f^{(+)}_{21}  & f^{(+)}_{22} 
\end{array} \right),
\end{equation}
where $T_{i} \quad(i=1,2,3)$ are the matrix representation of generators
of the group $SU(2)$:
$T_{i}=\displaystyle \frac{1}{2}\sigma_{i}, 
\quad [T_{i},T_{j}]=i\varepsilon_{ijk} T_{k}$,
$\sigma_{i}$ are Pauli's matrices, 
$\vec{\theta}(\theta_{1},\theta_{2},\theta_{2})$ are the transformation
parameters  of  $SU(2)$, and due to eq.(16.15)
$U^{+}U=I$.
The fundamental fields will come in multiplets, which form a 
basis for representations of the isospin group $SU(2)$
\begin{equation}
\label{R1617}
U(\theta)= \left[ \cos \FFr{\theta}{2}-\FFr{i}{\theta}
\sin\FFr{\theta}{2} \left( \matrix{
\theta_{3}               & \theta_{1}-i\theta_{2}\cr
\theta_{1}+i\theta_{2}   & -\theta_{3}\cr} \right)
\right],
\end{equation}
$(\theta=\mid\vec{\theta}\mid)$. We get easily
\begin{equation}
\label{R1618}
\begin{array}{l}
\FFr{\theta_{1}}{\theta}=\FFr{e^{\chi^{R}_{12}}\sin\chi^{J}_{12}}
{\sqrt{1-e^{2\chi^{R}_{11}}{\cos}^{2}\chi^{J}_{11}}}, 
\quad
\FFr{\theta_{2}}{\theta}=-\FFr{e^{\chi^{R}_{12}}\cos\chi^{J}_{12}}
{\sqrt{1-e^{2\chi^{R}_{11}}{\cos}^{2}\chi^{J}_{11}}},\\
\FFr{\theta_{3}}{\theta}=\FFr{e^{\chi^{R}_{11}}\sin\chi^{J}_{11}}
{\sqrt{1-e^{2\chi^{R}_{11}}{\cos}^{2}\chi^{J}_{11}}}, \quad
\theta = 2\arccos \left(e^{\chi^{R}_{11}}\cos\chi^{J}_{11}\right),
\quad e^{\chi^{R}_{11}}\leq 1. 
\end{array}
\end{equation}
The following relations hold:
\begin{equation}
\label{R1619}
\chi^{R}_{11}=\chi^{R}_{22}, \quad\chi^{R}_{12}=\chi^{R}_{21}
\quad
\chi^{J}_{11}+\chi^{J}_{22}=0, \quad \chi^{J}_{21}+\chi^{J}_{12}=\pi,
\quad
\chi^{R}_{12}=\FFr{1}{2}\ln \left(1-e^{2\chi^{R}_{11}}\right).
\end{equation}
Hence, three functions $\chi^{R}_{11}, \chi^{J}_{11}$ and $\chi^{J}_{12}$
or the angles $\theta'_{+1},\theta_{+1}$ and $\theta_{+2}$ are parameters
of the group $SU^{loc}(2)$ 
\begin{equation}
\label{R1620}
\chi^{R}_{11}=\chi_{R}(\theta'_{+1})-\chi_{R}(\theta_{+1}),
\quad
\chi^{J}_{11}=\chi_{J}(\theta'_{+1})-\chi_{J}(\theta_{+1}),
\quad
\chi^{J}_{12}=\chi_{J}(\theta'_{+1})-\chi_{J}(\theta_{+2}).
\end{equation}
Therefore, the local gauge transformations of collection of fundamental
fields read $\Psi'(\zeta)=U(\theta)\Psi(\zeta)$,
where physical field $\Psi(\zeta)$ is a column vector.
Continuing along this line, the Lagrangian must be invariant under local
gauge transformations, as well by introducing non-Abelian vector
gauge fields of weak interactions.

\subsection {Unitary Local Group $SU^{loc}(3)$ of Strong Interactions}
Finally we turn to the case, when gauge transformations around all three
axes are local. Then, there are nine possible transformations with the
functions $f^{(+)}_{lk}, \quad(l,k=1,2,3)$:
\begin{equation}
\label{R1621}
f^{(+)}=\left( \begin{array}{lll}
f^{(+)}_{11}  & f^{(+)}_{12}  & f^{(+)}_{13} \\ 
f^{(+)}_{21}  & f^{(+)}_{22}  & f^{(+)}_{23} \\
f^{(+)}_{31}  & f^{(+)}_{32}  & f^{(+)}_{33}
\end{array} \right), \quad
f^{(-)}={(f^{(+)})}^{+}.
\end{equation}
Incompatibility commutation relations (16.5) yield the unitary condition
$U^{-1}=U^{+}, \\
f^{(+)}\equiv U$, and also $\left\| U \right\| =1$. Then
$U(\theta)=e^{-\frac{i}{2}\vec{\lambda}\vec{\theta}}$,
where $\FFr{\lambda_{i}}{2} \quad(i=1,\ldots,8)$ are the matrix
representation of generators of the group $SU(3)$:
$\left[ \FFr{\lambda_{i}}{2},\FFr{\lambda_{k}}{2} \right]=
if_{ikl}\FFr{\lambda_{l}}{2}$,
the antisymmetric structure constants are denoted by $f_{ikl}$. 
The transformations are implemented on the column vector fundamental
quark fields
$\hat{q}'=U(\theta)\hat{q}$.
Right through differentiation one infers at once 
$\vec{\lambda}\vec{d\theta}=2iU^{+}dU,$
or
\begin{equation}
\label{R1622}
\vec{\theta}=-\IIn Im\left(tr \left(\vec{\lambda}
\left( f^{(-)}df^{(+)} \right) 
\right)\right),
\end{equation}
provided
$Re\left(tr \left(\vec{\lambda}\left( f^{(-)}df^{(+)}\right) \right)
\right)\equiv 0.$
At the infinitesimal transformations $\theta_{i} \ll 1$
\begin{equation}
\label{R1623}
\left(\matrix{
q'_{1}\cr
q'_{2}\cr
q'_{3}\cr} \right)
=\left\{ 1-\FFr{i}{2} \left( \matrix{
\theta_{3}+\FFr{1}{\sqrt{3}}\theta_{8}  & \theta_{1}-i\theta_{2}  
& \theta_{4}-i\theta_{5}\cr
\theta_{1}+i\theta_{2}    & -\theta_{3}+\FFr{1}{\sqrt{3}}\theta_{8}   
& \theta_{6}-i\theta_{7}\cr
\theta_{4}+i\theta_{5}    & \theta_{6}+i\theta_{7}   
& -\FFr{2}{\sqrt{3}}\theta_{8}
\cr} \right)  \right\}\cdot 
\left( \matrix{
q_{1}\cr
q_{2}\cr
q_{3}\cr} \right),
\end{equation}
we get
\begin{equation}
\label{R1624}
\begin{array}{ll}
\theta_{1}\approx 2e^{\chi^{R}_{12}}\sin\chi^{J}_{12},\quad   
\theta_{3}\approx \sin\chi^{J}_{33}+2\sin\chi^{J}_{11},\quad       
\theta_{5}\approx 2(1- e^{\chi^{R}_{13}}\cos\chi^{J}_{13}),\\
\theta_{2}\approx 2(1- e^{\chi^{R}_{12}}\cos\chi^{J}_{12}),\quad  
\theta_{4}\approx 2e^{\chi^{R}_{13}}\sin\chi^{J}_{13},\quad       
\theta_{6}\approx 2e^{\chi^{R}_{23}}\sin\chi^{J}_{23},\\
\theta_{7}\approx 2(1- e^{\chi^{R}_{23}}\cos\chi^{J}_{23}),\quad
\theta_{8}\approx -\sqrt{3}\sin\chi^{J}_{33},
\end{array}
\end{equation}
provided
\begin{equation}
\label{R1625}
\chi^{R}_{ll}\approx 0,
\quad
\chi^{R}_{lk}\approx \chi^{R}_{kl}, \quad 
\chi^{J}_{lk}\approx \chi^{J}_{kl}, \quad (l \neq k)\quad
\sin\chi^{J}_{11}+\sin\chi^{J}_{22}+\sin\chi^{J}_{33}\approx 0.
\end{equation}
As it was seen in section $15$, the existence of the internal symmetry group
$SU^{loc}_{C}(3)$ allows oneself to introduce a gauge theory in color space,
with the color charges as exactly conserved quantities. The local color
transformations are implemented on the colored quarks right through a
$SU^{loc}_{C}(3)$ rotation matrix U eq.(16.21) in the fundamental 
representation. 
\section {Discussion and Conclusions}
At this point we cut short our exposition of the
theory and reflect upon the results far obtained.
A number of conclusions may be drawn and the main features of suggested
theory  are outlined below.
We have finally arrived at an entirely satisfactory proposal to answer
all original questions posed in introduction. The following scenario is
cleared up, the whole idea of which comes to this: the basic concepts
of geometry, fundamental fields with various quantum numbers, internal 
symmetries and so forth; also the basic principles of Relativity, Quantum, 
Gauge and Color Confinement, are all derivative.
They come into being simultaneously.\\
We have mainly developed the mathematical framework
for our viewpoint. We recognize that this kind of mathematical treatment 
has been necessarily introductory by nature, hence our discussion 
has been rather general and abstract. The numerous issues in suggested 
theory still remain to be resolved. Surely the more realistic complete 
theory is a subject for further research.\\
We start with a quite promising proposal of generating
the group of gravitation by hidden local internal symmetries. Meanwhile, under
the reflection of shadow fields from Minkowski flat space to 
Riemannian, one framed the idea of general gauge principle of local internal
symmetries G into requirement of invariance of physical system of reflected
fields on $R^{4}$ under the Lie group of gravitation $G_{R}$ of local
gauge transformations generated by G. This yields the invariance under
wider group of arbitrary curvilinear coordinate transformations in $R^{4}$.
While the energy~-momentum conservation laws are well~-defined.
The fascinating prospect emerged for resolving or mitigating a shortage of 
controversial problems of gravitation including its quantization by 
exploiting whole advantages of field theory in terms of flat space.
In the aftermath, one may carry out an inverse reflection into $R^{4}$
whenever it will be needed.
It was proved that only gravitational attraction exists. One may easily
infer Einstein's equation of gravitation, but with strong difference 
at the vital point of well~-defined energy~-momentum tensor of gravitational
field and conservation laws. Nevertheless, for our part we prefer 
standard gauge invariant Lagrangian eq.(4.11), while the functions
eq.(2.3) ought to be defined.
We considered the gravitational interaction with
hidden Abelian group $G=U^{loc}(1)$ with the base $G(2.3)$
of principle bundle as well as
the general distortion of manifold $G(2.2.3)$, which yields both the 
curvature and inner-distortion of space and time.
Regarding the persistent processes of creation and 
annihilation of regular goyaks in the
lowest state, we develop the formalism of operator manifold
$\hat{G}(2.2.3)$, which is the mathematical foundation for
our viewpoint in the second and third parts. This is a 
guiding formalism of our approach
and a still wider generalization of familiar methods of 
secondary quantization with appropriate expansion over the 
processes involving geometric objects. This generalization yields
the quantization of geometry, which differs in principle from
all earlier suggested schemes.
One readily found out a contingency arisen at
the very beginning that all states of goyak are degenerate with a degree
of degeneracy equal 2. That is, along many quantum numbers of definite state
of goyak there is half-integral spin number. Thus, the goyaks are turned
out to be fermions with the half-integral spins. As it was seen in later
sections, this gives rise to the spins of particles.
The nature of $\hat{G}(2.2.3)$ provides its elements with both the field 
and geometric aspects.
In this context it comes out that the geometry is
derivative by nature. It may be realized in the special system of regular
goyaks if only some subsidiary condition holds for each participating
ordinary goyak. In the sequel, the manifold $G(2.2.3)$ gives rise to 
the Minkowski flat space $M^{4}$.
By this we have arrived at an answer to the question
of the physical origin of the space-time and the Principle of Relativity. 
We have briefly treated the quantum theory
situation corresponding to simultaneous presence of many identical goyak
fields. 
In last part  we extend the
scopes of treatment by considering the processes involving distorted
goyaks. For the second choice of realization of geometry we have
following recipe: it is admitted that distorted ordinary goyaks
have took participation in the realization of geometry instead of regular
ordinary goyaks. It comes out that distorted goyaks are acting
in special permissible combinations to realize the geometry.
The wave functions and conjugate functions of distorted goyaks
enable one to introduce the bi-spinor (conjugate) field of quarks and
anti-quarks as the operator fields in the
color space of internal degrees of freedom, corresponding to local
distortion transformations around axes $k=1,2,3$.
We have assumed that local distortion rotations around the different axes are
incompatible, i.e. they cannot occur simultaneously.
It has been shown that quarks are being met in some special combinations
to emerge in geometry only in color singlets. That is, they obey exact Color
Confinement principle and the spectrum of hadrons would emerge as the
spectrum of the color singlet states.
As it was seen , the identity principle holds for
ordinary regular goyaks, according to it each goyak in the lowest state
may be regarded as the result of transformation from arbitrary higher state,
in which the goyaks are assumed to be distorted. It is impressive to learn 
that this principle, underlies the most important gauge principle.
A global charge conservation corresponds to global gauge invariance. But
a requirement of expansion of Lie group from global to local symmetry
can be satisfied by introducing  the gauge fields of interactions.
This invariance in its concrete expressions yields the different models
of possible interactions: one-dimensional transformations provide a local
internal symmetry of electromagnetic interactions, with an Abelian group
$U^{loc}(1)$: two-dimensional transformations lead to local internal symmetry
of weak interactions, with the local non-Abelian unitary group $SU^{loc}(2)$;
and finally, three-dimensional transformations yield the strong interactions,
with  the local group $SU^{loc}_{C}(3)$.\\
All these results assure us that the more realistic
final theory of particles and interactions can be found within the
context of the theory of goyaks.
We believe we have made good headway by presenting a reasonable framework
whereby one will be able to verify the basic ideas of suggested theory.
\vskip 0.5truecm
\centerline {\bf\large Acknowledgements}
\vskip 0.1\baselineskip
\noindent
This work was supported in part by the International Atomic Energy Agency
and the United Nations Educational, Scientific and Cultural Organization.
It is a pleasure to thank for their hospitality the International
Centre for Theoretical Physics, Trieste, where part of this work
was done, and its director Professor Abdus Salam.
I express my gratitude to colleagues for fruitful discussions and useful
comments on the various issues treated in this paper, among them
Professor V.A.Ambartsumian, Professor J.Strathdee and also all 
participants of the seminars.
I am greatly indebted to A.M.Vardanian and K.L.Yerknapetian
for their steady support.
\section * {Appendix}
\subsection * {The Wave Equation of Distorted Goyak}
\label {app}
\renewcommand {\theequation}{A.\arabic {equation}}
As a Fermi field defined in distorted manifold $\G1_{u}(23)$, the bi-spinor 
field of distorted ordinary goyak is described by invariant Lagrangian, 
which, according to tetrad formalism [13] 
can be written [1-5]:
\begin{equation}
\label{A1}
\begin{array}{l}
\sqrt{-g}L_{F}=\FFr{\sqrt{-g}}{2} \left\{-i \bp_{u}(\theta)
{\g1_{u}}^{(\lambda\alpha)}({\pr_{u}}_{(\lambda\alpha)}
{\Gam_{u}}_{(\lambda\alpha)})
\ps1_{u}(\theta)+ \right.\\
\\
\left.+i\bp_{u}(\theta)
({\lpr_{u}}_{(\lambda\alpha)}-{\bGam_{u}}_{(\lambda\alpha)})
{\g1_{u}}^{(\lambda\alpha)}\ps1_{u}(\theta)+
2m\bp_{u}(\theta)\ps1_{u}(\theta) \right\},
\end{array}
\end{equation}
where ${\g1_{u}}^{(\lambda\alpha)}(\theta)$ are matrix realization of basis
${\e1_{u}}^{(\lambda\alpha)}(\theta)$:
\begin{equation}
\label{A2}
{\g1_{u}}^{(\lambda\alpha)}(\theta)={\V_{u}}^{(\lambda\alpha)}
_{(\tau\beta)}(\theta)\gamma_{0}^{(\tau\beta)}.
\end{equation}
The matrix functions ${\Gam_{u}}_{(\lambda\alpha)}(\theta)$ and
${\bGam_{u}}_{(\lambda\alpha)}(\theta)$ in terms of Ricci rotation
coefficients read
\begin{equation}
\label{A3}
{\Gam_{u}}_{(\lambda\alpha)}(\theta)=\FFr{1}{4}
\Delta_{(\lambda\alpha)(i,l)(m,p)}\gamma^{(i,l)}_{0}\gamma^{(m,p)}_{0},
\quad
{\bGam_{u}}_{(\lambda\alpha)}(\theta)=\FFr{1}{4}
\Delta_{(\lambda\alpha)(i,l)(m,p)}\gamma^{(m,p)}_{0}\gamma^{(i,l)}_{0},
\end{equation}
provided
$\Delta_{(\lambda\alpha)(i,l)(m,p)}=
\Delta_{(\lambda\alpha)(\tau\beta)(\rho\gamma)}V^{(\tau\beta)}_{(i,l)}
V^{(\rho\gamma)}_{(m,p)}$,
and
\begin{equation}
\label{A4}
\begin{array}{l}
\Delta_{(\lambda\alpha)(\tau\beta)(\rho\gamma)}=\FFr{1}{2}
\left( {\pr_{u}}_{(\tau\beta)}{\g1_{u}}_{(\lambda\alpha)(\rho\gamma)}-
{\pr_{u}}_{(\rho\gamma)}{\g1_{u}}_{(\lambda\alpha)(\tau\beta)}-
{\pr_{u}}_{(\lambda\alpha)}{\g1_{u}}_{(\rho\gamma)(\tau\beta)}\right)+
V_{(\tau\beta)}^{(i,l)}{\pr_{u}}_{(\lambda\alpha)}
V_{(\rho\gamma)(i,l)};\\
{\g1_{u}}_{(\lambda\alpha)(\tau\beta)}=V_{(\lambda\alpha)}^{(i,l)}
V_{(i,l)(\tau\beta)}=g_{(i,l)(m,p)}V_{(\lambda\alpha)}^{(i,l)}
V^{(m,p)}_{(\tau\beta)}, 
\end{array}
\end{equation}
$V^{(\lambda\alpha)}_{(i,l)}(\theta)$ are congruence parameters of curves
(Latin indices refer to tetrad components).
Here, as usual, we let first subscript in the parentheses labels the
pseudo-vector components, when second refers to ordinary-vector components.
At fixed metric the current
${\J1_{u}}^{(\lambda\alpha)}=\bp_{u}(\theta)
{\g1_{u}}^{(\lambda\alpha)}(\theta)\ps1_{u}(\theta)$
is conserved
$\FFr{1}{\sqrt{-g}}{\pr_{u}}_{(\lambda\alpha)}
(\sqrt{-g}{\J1_{u}}^{(\lambda\alpha)})=0$.
Field equations can be derived at once right through eq.(A.1) and variational
principle of least action
\begin{equation}
\label{A5}
\begin{array}{l}
\left[ i{\g1_{u}}^{(\lambda\alpha)}(\theta)
({\pr_{u}}_{(\lambda\alpha)}-{\Gam_{u}}_{(\lambda\alpha)}(\theta))
- m \right]\ps1_{u}(\theta) =0,\\
\bp_{u}(\theta)\left[
i({\lpr_{u}}_{(\lambda\alpha)}-{\bGam_{u}}_{(\lambda\alpha)}(\theta))
{\g1_{u}}^{(\lambda\alpha)}(\theta) - m \right] =0.
\end{array}
\end{equation}


\begin {thebibliography}{99}
\bibitem{\Ter86} Ter-Kazarian G.T., Selected Questions of Theoretical 
and Mathematical Physics, 1986, VINITI, N5322-B86, Moscow.
\bibitem{\Tercom} Ter-Kazarian G.T.,Comm. Byurakan Obs., 1989,v.62, 1.
\bibitem{\Ter92} Ter-Kazarian G.T., Astrophys. and Space Sci., 1992, v.194, 1.
\bibitem{A1} Ter-Kazarian G.T.,IC/94/290, ICTP Preprint, Trieste, 
Italy, 1, 1994.
\bibitem{A2} Ter-Kazarian G.T., hep-th/9510110, 1995.
\bibitem{A43} Weil H., Sitzungsber. d. Berl. Akad., 1918, 465.
\bibitem{\Yang} Yang C.N., Mills R.L., Phys. Rev., 1954, v. 96, 191.
\bibitem{\Utiyama} Utiyama R., Phys. Rev., 1956, v.101, 1597.
\bibitem{\Dubrovin} Dubrovin B.A., Novikov S.P., Fomenko A.T., 
Contemporary Geometry, 1986, Nauka, Moscow.
\bibitem{\Pont} Pontryagin L.S., Continous Groups, 1984, Nauka, Moscow.
\bibitem{\Weinberg72} Weinberg S., 1972, Gravitation and Cosmology, 
J.W. and Sons, New York.
\bibitem{\Dir} Dirac P.A.M., 1975, General Theory of Relativity,
A Wiley-Interscience Publ., New York.
\bibitem{\Fok} Fok V.A., Zeitsch. fur Phys., 1929, v.57, 261.
\bibitem{\Weinberg64} Weinberg S., 1964, Phys. Rev. b133, 1318. 
\bibitem{A11} Di Bartini, Dokl. Akad. Nauk, SSSR, v.163,4, 1965.
\bibitem{A13} Bjorken J.D., Drell S.D., Relativistic Quantum Fields,
\indent Mc Graw-Hill, New York, 1964.
\end {thebibliography}
\end{document}